\documentclass[english,aps,pra,reprint,superscriptaddress]{revtex4-1}
\usepackage[T1]{fontenc}
\usepackage[latin9]{inputenc}
\usepackage{color}
\usepackage{babel}
\usepackage{float}
\usepackage{units}
\usepackage{amsmath}
\usepackage{amssymb}
\usepackage{graphicx}
\usepackage[unicode=true,
 bookmarks=true,bookmarksnumbered=true,bookmarksopen=false,
 breaklinks=false,pdfborder={0 0 0},pdfborderstyle={},backref=false,colorlinks=true]
 {hyperref}
\hypersetup{
 allcolors=blue}

\makeatletter

\newcommand*\LyXThinSpace{\,\hspace{0pt}}

\usepackage{braket}
\usepackage{dsfont}

\makeatother

\begin{document}

\title{Verifying detailed fluctuation relations for discrete feedback-controlled
quantum dynamics}

\author{Patrice A. Camati}
\email{p.camati@ufabc.edu.br}

\selectlanguage{english}%

\affiliation{Centro de Ciências Naturais e Humanas, Universidade Federal do ABC,
Avenida dos Estados 5001, 09210-580 Santo André, São Paulo, Brazil}

\author{Roberto M. Serra}
\email{serra@ufabc.edu.br}

\selectlanguage{english}%

\affiliation{Centro de Ciências Naturais e Humanas, Universidade Federal do ABC,
Avenida dos Estados 5001, 09210-580 Santo André, São Paulo, Brazil}

\affiliation{Department of Physics, University of York, York YO10 5DD, United
Kingdom}
\begin{abstract}
Discrete quantum feedback control consists of a managed dynamics according
to the information acquired by a previous measurement. Energy fluctuations
along such dynamics satisfy generalized fluctuation relations, which
are useful tools to study the thermodynamics of systems far away from
equilibrium. Due to the practical challenge to assess energy fluctuations
in the quantum scenario, the experimental verification of detailed
fluctuation relations in the presence of feedback control remains
elusive. We present a feasible method to experimentally verify detailed
fluctuation relations for discrete feedback control quantum dynamics.
Two detailed fluctuation relations are developed and employed. The
method is based on a quantum interferometric strategy that allows
for the verification of fluctuation relations in the presence of feedback
control. An analytical example to illustrate the applicability of
the method is discussed. The comprehensive technique introduced here
can be experimentally implemented at a microscale with the current
technology in a variety of experimental platforms.
\end{abstract}
\maketitle

\section{INTRODUCTION}

Feedback control is a process which allows the control of the system
dynamics based on information of its present or past states. Maxwell
pioneered the mathematical analysis of control systems in the second
half of the 19th century \cite{Maxwell1867,Kang2016}. Due to the
growing application of automation in industry, control theory plays
a major role in modern engineering and technology \cite{Bechhoefer2005},
e.g., using feedback control for stabilization of dynamical systems.
Feedback control has also been used to control quantum systems, e.g.,
for reducing unwanted noise \cite{Jacobs2006,Zhang2014}. Maxwell's
demon, an important thought experiment envisaged by Maxwell \cite{Leff2002,Maruyama2009,Lutz2015},
may also be recognized as a feedback control protocol at the microscopic
scale. Along with the work of Jaynes \cite{Jaynes1957a,Jaynes1957b}
concerning the informational character of statistical mechanics, they
comprise prominent examples of the interplay between information theory
and thermodynamics. 

One approach to study the nonequilibrium thermodynamics of small quantum
systems is based on fluctuation relations \cite{Jarzynski2011,Klages2013,Vinjanampathy2015,Millen2016,GelbwaserKlimovsky2015}.
Quantities such as work, heat, and entropy production are described
by stochastic variables and establish deep relations between nonequilibrium
and equilibrium thermodynamic quantities \cite{Esposito2009,Seifert2012}.
The Jarzynski equality \cite{Jarzynski1997a,Jarzynski1997b} and Crooks
relation \cite{Crooks1999} connect the nonequilibrium work with the
equilibrium free-energy difference. Despite their classical origin,
their validity has been extended to driven quantum systems \cite{Tasaki2000,Kurchan2001,Campisi2011,Hanggi2015,Ribeiro2016}.
These fundamental relations were also extended for feedback protocols,
providing further tools to study information thermodynamics of classical
and quantum systems \cite{Lloyd1997,Sagawa2012-1,Parrondo2015,Chapman2015,Brandner2015,Girolami2015,Kammerlander2016,Lebedev2016,Kutvonen2016,Weilenmann2016,Elouard2016,Strasberg2017,Elouard2017,Shu2017}.
They imply generalized forms of the second law where thermodynamic
and information quantities are treated on an equal footing \cite{Sagawa2006,Sagawa2008,Sagawa2010,Morikuni2011,Sagawa2012,Lahiri2012,Funo2013}.
There are a few qualitatively different generalized second laws in
the presence of feedback control, involving the information gain \cite{Fuchs2001,Buscemi2008,Buscemi2009,Berta2014,Jacobs2014,Kosloff2013,Camati2016},
the mutual information \cite{Cover2006}, or even the entanglement
of formation \cite{Tajima2013}.

From the experimental point of view, quantum fluctuation relations
have been tested very recently in a nuclear magnetic resonance (NMR)
quantum processor \cite{Batalhao2014,Batalhao2015}, employing interferometric
methods \cite{Mazzola2013,Dorner2013a}, and in a trapped-ions setup
\cite{An2014a}. In classical information thermodynamics, the extension
of the Jarzynski relation \cite{Toyabe2010}, the Landauer's principle
\cite{B=0000E9rut2012}, the Szilard engine \cite{Rold=0000E1n2014,Koski2014,Koski2014-1},
and Maxwell's demon \cite{Koski2015,vedral2016} have been performed
experimentally. On the other hand, in quantum information thermodynamics
\cite{Goold2015} there has been little experimental activity in the
past couple of years, due to the inherent difficulties associated
to the quantum setup. Chronologically, information-to-energy conversion
\cite{Peterson2016} and Maxwell's demon \cite{Camati2016} have been
accomplished in an NMR quantum processor; a photonic architecture
was employed to address the minimum entropy cost necessary for the
implementation of a quantum measurement \cite{Mancino2017}; communication-assisted
games based on a logical version of Maxwell's demon provided an entanglement-separability
criterion in multiphoton optical interferometers \cite{Ciampini2016};
another realization of Maxwell's demon has been done in both the classical
and quantum regimes in circuit quantum electrodynamics (QED) \cite{Cottet2017};
the experimental verification of the integral fluctuation relation
in the presence of feedback control was very recently implemented
both encompassing absolute irreversibility performing nondemolition
quantum measurements \cite{Masuyama2017} and encompassing inefficient
measurements employed by continuous weak measurements \cite{Naghiloo2018}
using transmon qubits in circuit QED.

The integral fluctuation relation for feedback-controlled systems
has been obtained for two different physical scenarios. In the first,
the quantum system is controlled by a noisy classical feedback device
which introduces a control mismatch between the observable measurement
basis and the feedback control basis. In this situation, the mutual
information (density) between the measurement and the feedback control
appears in the integral fluctuation relation \cite{Morikuni2011}.
This setting assumes that the controller implements the measurement
of an observable, i.e., a projective measurement. On the other hand,
another setting was developed in Ref. \cite{Funo2013} in which an
auxiliary quantum system is employed to implement a controller which
performs an inefficient positive operator-valued measure (POVM). In
this case, the information gain (density) of the measurement appears
in the integral fluctuation relation. We note that these settings
are physically distinct, since the information-theoretic quantity
in the first setting is present due to a (classical) noisy controller,
while in the second setting it comes from the inefficient POVM measurement
performed by the controller. 

The detailed fluctuation relation for feedback control characterizes
the irreversibility in controlled systems and  is a generalization
of the integral fluctuation relations. For the first setting mentioned
above, with a noisy controller, the detailed fluctuation relation
was obtained in Ref.~\cite{Lahiri2012}. Up to the present moment,
the detailed fluctuation relation for feedback control has not been
experimentally verified either in the classical or in the quantum
scenario. Here we propose a method to experimentally assess the statistics
of energy fluctuations which allows the verification of the quantum
detailed fluctuation relation for processes driven by discrete-time
feedback control. The method is based on interferometric protocols
that can be put into action with the current technology in a variety
of experimental platforms.

This paper is organized as follows. In Sec.~\ref{sec:Time-reversal-operator-and-backward-processes}\textbf{
}we review the definition of a backward process using the time-reversal
operator. In Sec.~\ref{sec:QUANTUM-DETAILED-NONEQUILIBRIUM} we employ
this formalism to review the detailed fluctuation relations with and
without feedback control. We show that they remain valid without assuming
time-reversal invariance, which is usually assumed in the literature.
Furthermore, we derive two  detailed fluctuation relations which are
instrumental in our method. One applies to feedback without control
mismatch, whereas the other applies to feedback with control mismatch.
In Sec.~\ref{sec:Method} we present a method to experimentally verify
the detailed fluctuation relation for discrete feedback-controlled
quantum dynamics. The method uses the detailed fluctuation relations
for single histories and five quantum interferometric circuits. To
illustrate our method, we study the analytical example of a qubit
controlled by conditional sudden quenches in Sec. \ref{subsec:Example:-Qubit-Controlled}.
We summarize in Sec. \ref{sec:Conclusions}.

\section{TIME-REVERSAL TRANSFORMATION AND THE DEFINITION OF BACKWARD PROCESSES\label{sec:Time-reversal-operator-and-backward-processes}}

Detailed fluctuation relations involve the forward and backward probability
density functions (PDFs) of a protocol. In order to properly discuss
the backward process, we shortly review the formalism of the time-reversal
operator. In the next section we show that the quantum detailed fluctuation
relations (QDFRs) apply for systems without time-reversal symmetry.
This contrasts with some previous approaches where the system dynamics
was restricted to be time-reversal symmetric. We note that a few authors
have also realized this fact \cite{Gong2015,Rana2012,Lahiri2012-1,Liu2012,Liu2014}.

In quantum mechanics, symmetry operations have to preserve probabilities,
which, in turn, are based on inner products. This necessary condition
implies that a symmetry operator $\mathcal{O}$ must satisfy $\left|\left(\mathcal{O}\Ket{\psi},\mathcal{O}\Ket{\varphi}\right)\right|=\left|\left(\Ket{\psi},\Ket{\varphi}\right)\right|$,
where $\left(\cdot,\cdot\right)$ is the Hilbert space inner product
(here we assume the first entry to be antilinear and the second one
to be linear). Wigner showed that any symmetry operator is given by
either a unitary or antiunitary operator \cite{Wigner1959}. By definition,
unitary operators leave the inner product invariant, $\left(\mathcal{O}\Ket{\psi},\mathcal{O}\Ket{\varphi}\right)=\left(\Ket{\psi},\Ket{\varphi}\right)$,
whereas antiunitary operators induce a conjugation, $\left(\mathcal{O}\Ket{\psi},\mathcal{O}\Ket{\varphi}\right)=\left(\Ket{\psi},\Ket{\varphi}\right)^{*}$=$\left(\Ket{\varphi},\Ket{\psi}\right)$.
The application of antilinear operators is subtle and the interested
reader may find Refs.~\cite{Messiah1962,Ballentine2000,Haake2010}
helpful.

The usual symmetry operations, such as translations or rotations in
space, acquire a unitary representation. These operations can be implemented
in the laboratory by either a change of reference frame or a physical
position of the experimental setting (passive and active coordinate
transformations, respectively). On the other hand, the time-reversal
and charge-conjugation operators are two well-known antiunitary symmetries
and they cannot be implemented either by a change of reference frame
or by a change in the physical setup. Therefore, to experimentally
access the time-reversal symmetry, two different, but otherwise related,
experiments should be performed. These two related processes are called
forward and backward and how they are related to each other will be
discussed below. In the quantum case, the system is time-reversal
symmetric when the mathematical condition $\left[H\left(t\right),\Theta\right]=0$
is satisfied, where $H\left(t\right)$ is a time-dependent Hamiltonian
for a driven system and $\Theta$ is the time-reversal operator. For
our purposes, the most interesting cases comprise dynamics which are
not time-reversal symmetric, such as a spin-1/2 particle driven by
an external magnetic field as in the experiments reported in Refs.~\cite{Batalhao2014,Batalhao2015}.

\begin{figure}
\includegraphics[scale=0.5]{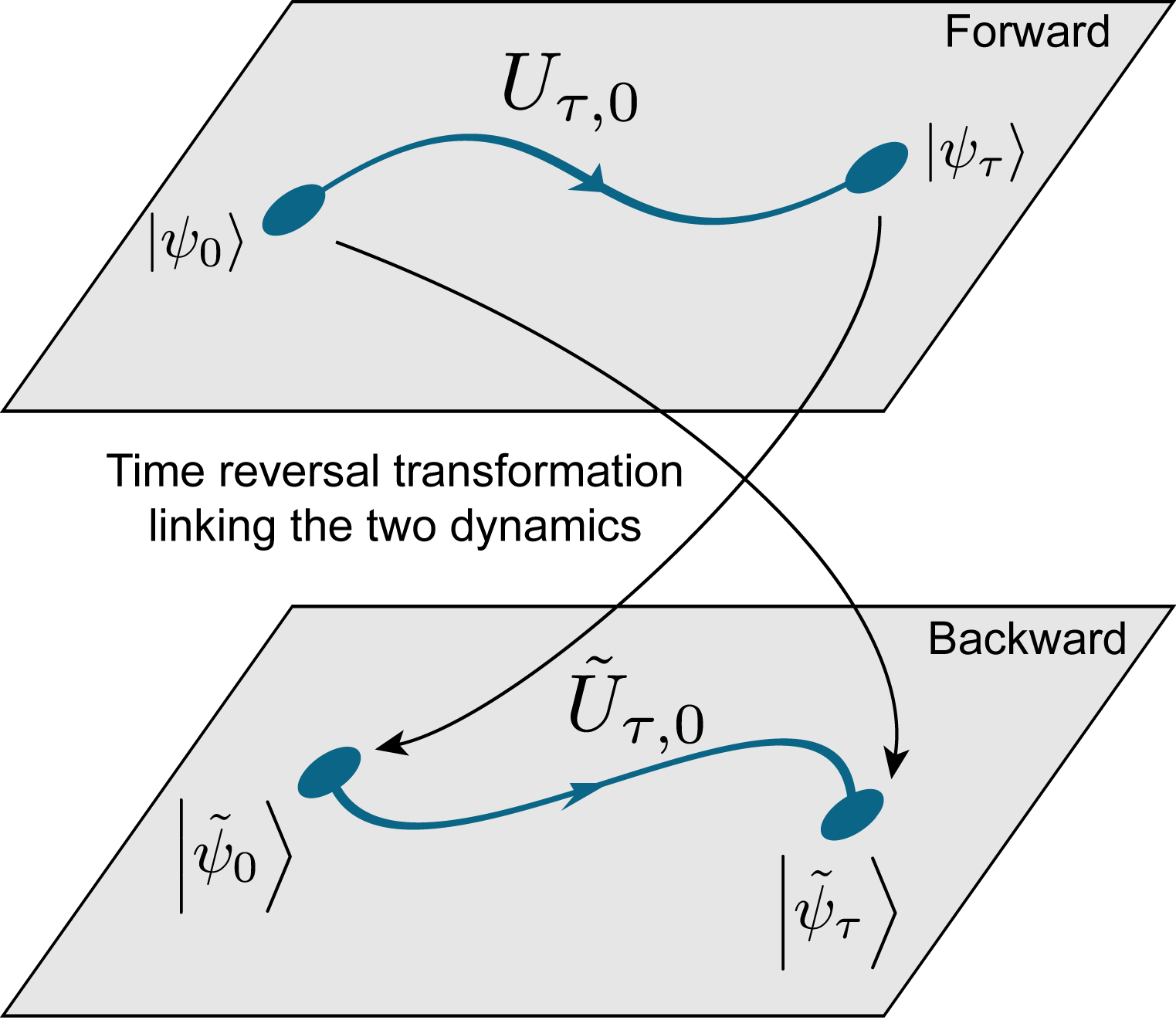}

\caption{Representation of the forward and backward processes. The two planes
represent two parts of the system Hilbert space. The forward and backward
trajectories are mapped by the time-reversal transformation. We emphasize
that from the practical point of view, there are two physically distinct
experiments, one for each history. Each evolution runs forward in
time and both are precisely connected by the time-reversal transformation
for any given time $t\in\left[0,\tau\right]$. \label{fig:forward-backward-trajectories}}

\end{figure}

Let $\Ket{\psi\left(t\right)}$ be the solution of the Schrödinger
equation for a time-dependent Hamiltonian $H\left(t\right)$, driven
by some external parameter, with the initial condition $\Ket{\psi_{0}}$
at $t=0$. Suppose that the total time length for this driven protocol
is $\tau$. This dynamics is called the forward process. The final
state of the forward evolution will be $\Ket{\psi_{\tau}}$ at $t=\tau$.
If we apply the time-reversal operator $\Theta$ and change the time
counting as $t\rightarrow\tau-t$ in the Schrödinger equation of the
forward process a new Schrödinger equation is obtained 

\begin{equation}
i\hbar\frac{d}{dt}\Ket{\tilde{\psi}\left(t\right)}=\tilde{H}\left(t\right)\Ket{\tilde{\psi}\left(t\right)},\label{eq:backward schrodinger equation}
\end{equation}
with $\Ket{\tilde{\psi}\left(t\right)}=\Theta\Ket{\psi\left(\tau-t\right)}$
and $\tilde{H}\left(t\right)=\Theta H\left(\tau-t\right)\Theta^{\dagger}$.
The reverse protocol described by Eq.~(\ref{eq:backward schrodinger equation})
with the initial condition $\Ket{\tilde{\psi}_{0}}=\Theta\Ket{\psi_{\tau}}$
is called the backward process (see Fig.~\ref{fig:forward-backward-trajectories}). 

The evolution operators for the backward and forward processes are
also related to each other. For a nondegenerate discrete Hamiltonian,
the eigenequation of the forward process is $H\left(t\right)\Pi_{n}^{t}=E_{n}^{t}\Pi_{n}^{t}$,
where $E_{n}^{t}$ and $\Pi_{n}^{t}=\Ket{E_{n}^{t}}\Bra{E_{n}^{t}}$
are the instantaneous eigenvalues and eigenprojectors, respectively.
Correspondingly, the eigenequation for the backward process is given
by $\tilde{H}\left(t\right)\tilde{\Pi}_{n}^{t}=\tilde{E}_{n}^{t}\tilde{\Pi}_{n}^{t}$,
where $\tilde{E}_{n}^{t}=E_{n}^{\tau-t}$ and $\tilde{\Pi}_{n}^{t}=\Theta\Pi_{n}^{\tau-t}\Theta^{\dagger}$
are the instantaneous eigenvalues and eigenprojectors of the time-reversed
dynamics, respectively. Furthermore, the evolution operator for the
forward process is $U_{\tau,0}=\mathcal{T}_{>}\exp\left\{ -\frac{i}{\hbar}\int_{0}^{\tau}dt\,H\left(t\right)\right\} $,
where $\mathcal{T}_{>}$ is the time-ordering operator. Using the
time-reversal transformations, one can show that
\begin{eqnarray}
\Theta U_{\tau,0}\Theta^{\dagger} & = & \mathcal{T}_{>}\exp\left\{ -\frac{i}{\hbar}\int_{\tau}^{0}dt\,\Theta H\left(\tau-t\right)\Theta^{\dagger}\right\} \nonumber \\
 & = & \tilde{U}_{\tau,0}^{\dagger}.
\end{eqnarray}
Sometimes this relation is interpreted as follows: The evolution operator
of the forward process between times $0$ and $\tau$ is mapped to
the evolution operator of the backward process between times $\tau$
and $0$; thus, time would run backward.

\section{QUANTUM DETAILED NONEQUILIBRIUM FLUCTUATION RELATIONS\label{sec:QUANTUM-DETAILED-NONEQUILIBRIUM}}

\subsection{Without feedback control\label{sec:Tasaki-Crooks-relation-without-feedback}}

In the light of the formalism introduced in the preceding section,
we review the derivation of the QDFR without feedback and show that
it remains valid for a dynamics which is not symmetric under time
reversal. The QDFR applies to the following scenario. A system starts
in thermal equilibrium in the initial Gibbs state $\rho_{0}^{eq}=e^{-\beta H_{0}}/Z_{0}$
with initial Hamiltonian $H_{0}$, inverse temperature $\beta=\left(k_{B}T\right)^{-1}$,
and partition function $Z_{t}=\text{Tr}\left[e^{-\beta H_{t}}\right]$,
where $k_{B}$ is the Boltzmann constant. The system is driven away
from equilibrium due to the manipulation of controllable parameters
in the Hamiltonian, for instance, the external magnetic field. The
protocol occurs during a time length $\tau$ which drives the system
to the final state $\rho_{\tau}=U_{\tau,0}\rho_{0}^{eq}U_{\tau,0}^{\dagger}$.
For this process, the first law of thermodynamics requires the equality
between the system internal energy difference and the average work,
$\Delta\mathcal{U}=\left\langle W\right\rangle $, with $\mathcal{U}_{t}=\text{Tr}\left[\rho_{t}H_{t}\right]$
being the instantaneous internal energy. The average work is obtained
as the average of the (forward) work PDF, $\left\langle W\right\rangle =\int dW\,W\,P_{F}\left(W\right)$,
where $P_{F}\left(W\right)$ is the work PDF associated with the (forward)
protocol just described. The work PDF $P_{F}\left(W\right)$ can be
obtained through the two-point measurement (TPM) scheme and is given
by \cite{Campisi2011,Talkner2007}
\begin{equation}
P_{F}\left(W\right)=\sum_{m,n}p\left(m,n\right)\delta\left[W-W_{mn}\right],
\end{equation}
where 
\begin{eqnarray}
p\left(m,n\right) & = & p\left(m|n\right)p\left(n\right)\nonumber \\
 & = & \mbox{Tr}\left[\Pi_{m}^{\tau}U_{\tau,0}\Pi_{n}^{0}U_{\tau,0}^{\dagger}\right]\mbox{Tr}\left[\Pi_{n}^{0}\rho_{0}^{eq}\right]\label{eq:forward joint probability distribution-1}
\end{eqnarray}
is the joint probability distribution of obtaining energy outcomes
$E_{m}^{\tau}$ and $E_{n}^{0}$ in the final and initial times, respectively,
$p\left(m|n\right)$ is the conditional probability of obtaining the
energy outcome $E_{m}^{\tau}$ given that $E_{n}^{0}$ was obtained
in the first measurement, $p\left(n\right)$ is the initial probability
to obtain $E_{n}^{0}$, and $W_{mn}=E_{m}^{\tau}-E_{n}^{0}$ is the
value of work in this system history.

Using the relations obtained in the preceding section between forward
and backward processes, one can obtain the QDFR \cite{Tasaki2000,Kurchan2001}
(see \ref{sec:Appendix-A:-Quantum}) 

\textbf{
\begin{equation}
\frac{P_{F}\left(W\right)}{P_{B}\left(-W\right)}=e^{\beta\left(W-\Delta F\right)},
\end{equation}
}where $\Delta F=F_{\tau}-F_{0}$ is the free-energy difference, with
$F_{t}=-\left(\beta\right)^{-1}\ln Z_{t}$, and 
\begin{equation}
P_{B}\left(W\right)=\sum_{n,m}\tilde{p}\left(n,m\right)\delta\left[W-\tilde{W}_{nm}\right]
\end{equation}
is the work PDF associated with the backward process, $\tilde{p}\left(n,m\right)=\tilde{p}\left(n|m\right)\tilde{p}\left(m\right)$
is the joint probability distribution of obtaining energy outcomes
$\tilde{E}_{n}^{\tau}$ and $\tilde{E}_{m}^{0}$ at the end and beginning
of the driven process, respectively, $\tilde{p}\left(n|m\right)=\left|\Bra{\tilde{E}_{n}^{\tau}}\tilde{U}_{\tau,0}\Ket{\tilde{E}_{m}^{0}}\right|^{2}$
is the conditional probability of obtaining the energy outcome $\tilde{E}_{n}^{\tau}$
given that $\tilde{E}_{m}^{0}$ was obtained in the first measurement,
$\tilde{p}\left(m\right)=e^{-\beta\tilde{E}_{m}^{0}}/\tilde{Z}_{0}$
is the initial probability to obtain $\tilde{E}_{m}^{0}$, and $\tilde{W}_{nm}=\tilde{E}_{n}^{\tau}-\tilde{E}_{m}^{0}$
is the value of work. Our calculation is structurally similar to the
others in the literature \cite{Tasaki2000,Kurchan2001,Talkner2007-1,Andrieux2008,Campisi2010-1,Venkatesh2014,Watanabe2014,Campisi2011-1,Campisi2010}.
The distinction of our approach is that the Hamiltonian can be completely
arbitrary, time-reversal symmetric, or otherwise. This has both conceptual
and practical important consequences, which are discussed in the example
below. 

Consider the Zeeman Hamiltonian $H=-\mathbf{\mathbf{\boldsymbol{\mu}}}\cdot\mathbf{B}\left(t\right)$
for a single spin-1/2 particle in a driven magnetic field. In this
case the time-reversal operator satisfies $\Theta^{2}=-\mathds{1}$,
where $\mathds{1}$ denotes the identity operator. For such a system,
a time-reversal symmetric Hamiltonian has a twofold degeneracy called
Kramer's degeneracy \cite{Haake2010}, i.e., if $\Ket{E_{n}}$ is
an eigenvector then so is $\Theta\Ket{E_{n}}$ with same eigenenergy
$E_{n}$. 

Occasionally in the literature \cite{Esposito2009,Campisi2011,Andrieux2008},
in addition to the sign change of the spin operator, the time-reversal
transformation is assumed to change the sign of the magnetic field
as well. If that were the case, then the Zeeman Hamiltonian would
be time-reversal symmetric. This imposed change in the sign of the
magnetic field is artificial, though. If the Zeeman Hamiltonian were
time-reversal symmetric, then Kramer's degeneracy would imply a spectrum
of a single eigenenergy. This would further imply a trivial work PDF,
for a driving spin-1/2, given by a single delta function at $W=0$.
In Ref.~\cite{Batalhao2014} the work PDF for this system was experimentally
obtained and the result is not such a trivial PDF. 

This argument entails that the Zeeman Hamiltonian is not symmetric
under the time-reversal transformation. The (external) magnetic field
is not a property of the system and therefore should not change sign
under the time-reversal transformation. Mathematically, the time-reversal
operator acts on complex numbers, inducing a conjugation, and other
operators, states, or observables. As the magnetic field is a real
vector it does not change under the time-reversal transformation.
However, as we discuss in Sec.~\ref{subsec:Example:-Qubit-Controlled},
the magnetic-field direction can be used to effectively implement
the backward Hamiltonians in an experimental setting \cite{Batalhao2014}.\textcolor{red}{{} }

\begin{figure}[t]
\includegraphics[scale=0.45]{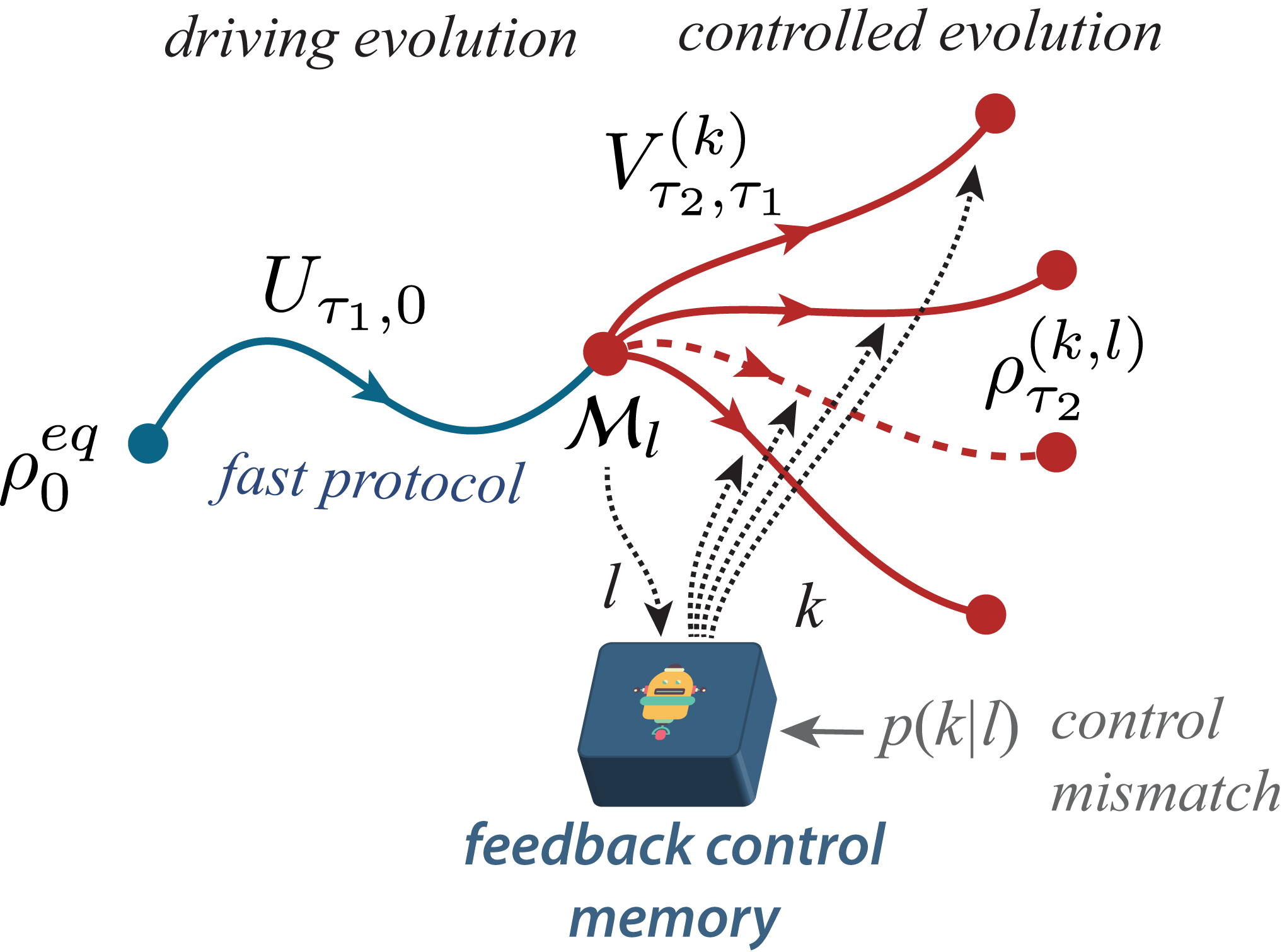}

\caption{Representation of the forward feedback control protocol. The system
starts in a thermal equilibrium state $\rho_{0}^{eq}$ and is driven
by external means to the nonequilibrium state $\rho_{\tau_{1}}$.
An intermediate measurement of some observable $\mathcal{M}$ is performed
at time $\tau_{1}$. The eigenprojectors of the observable are denoted
by $\left\{ \mathcal{M}_{l}\right\} $ and the outcome $l$ is obtained
with probability $p\left(l\right)$. After the measurement, there
might be a misimplementation of the feedback control, i.e., the feedback
operation associated with outcome $k$, $V_{\tau_{2},\tau_{1}}^{\left(k\right)}$,
is implemented when the actual measurement outcome is $l$. This mismatch
may occur due to an error or nonideality in the feedback control memory.
This is modeled by the conditional probability distribution $p\left(k|l\right)$.
For each possible outcome $l$ there will be $k$ different histories
for the system. Each of the $\left(k,l\right)$ histories lead to
the final state $\rho_{\tau_{2}}^{\left(k,l\right)}$, with joint
probability $p\left(k,l\right)$, and final Hamiltonian $H_{\tau_{2}}^{\left(k\right)}$,
with probability $p\left(k\right)$. \label{fig:feedback histories}}
\end{figure}

\subsection{With Feedback Control\label{sec:Tasaki-Crooks-relation-with-feedback-control}}

Although the QDFR for feedback processes was discussed in Ref. \cite{Lahiri2012},
it was not treated in a way which is clearly experimentally accessible.
Hitherto, even in the classical scenario, the generalization of the
Crooks relation for feedback processes has not been tested experimentally.
We recast the discussion in the light of the approach developed in
Sec.~\ref{sec:Time-reversal-operator-and-backward-processes} revisiting
the derivation of the QDFR in the presence of feedback control as
well as introducing two other QDFRs. In the next section we use the
results developed here to present a method to experimentally access
energy fluctuations and verify the QDFR in the presence of feedback
control.

The forward process in the presence of feedback control, illustrated
in Fig.~\ref{fig:feedback histories}, is defined in a similar way
to the forward process without feedback. The system begins in the
Gibbs state $\rho_{0}^{eq}$ and is driven away from equilibrium by
the evolution $U_{\tau_{1},0}$ up to time $\tau_{1}$, $\rho_{\tau_{1}}=U_{\tau_{1},0}\rho_{0}^{eq}U_{\tau_{1},0}^{\dagger}$.
At this time an intermediate measurement of some observable $\mathcal{M}$
is performed, leading to the postmeasurement state $\rho_{\tau_{1}}^{\left(l\right)}=\mathcal{M}_{l}\rho_{\tau_{1}}\mathcal{M}_{l}/p\left(l\right)$
with probability $p\left(l\right)=\mbox{Tr}\left[\mathcal{M}_{l}\rho_{\tau_{1}}\right]$,
where $\left\{ \mathcal{M}_{l}\right\} $ are the observable eigenprojectors. 

After the measurement, a noisy controller implements the feedback.
The feedback operation, described by the unitary operator $V_{\tau_{2},\tau_{1}}^{\left(k\right)}$,
is applied up until time $\tau_{2}$. Note that the index of the feedback
operation $k$ is different from the index of the postmeasurement
states $l$. This is due to the noisy feature of the controller, which
implements the feedback operator $V_{\tau_{2},\tau_{1}}^{\left(k\right)}$
to $\rho_{\tau_{1}}^{\left(l\right)}$ with conditional probability
$p\left(k|l\right)$. This probability quantifies the control mismatch
of the feedback implementation of $V_{\tau_{2},\tau_{1}}^{\left(k\right)}$,
associated with outcome $k$, and the actual measurement outcome $l$.
The control mismatch encompasses possible failures and imperfections
in the feedback mechanism. 

We denote the instantaneous eigenvalues and eigenprojectors of the
$k$th feedback Hamiltonian as $H_{t}^{\left(k\right)}P_{m}^{\left(k\right)t}=E_{m}^{\left(k\right)t}P_{m}^{\left(k\right)t}$.
We note that, rigorously, we should write the indices as $m=m^{\left(k\right)}$
since for each of the $k$th Hamiltonians there are $m^{\left(k\right)}$
possible eigenvalues. We have chosen the compact notation in almost
all of our expressions for the sake of readability. After the measurement,
there will be a number of possible histories for the system, quantified
by the product of the indices $kl$. Each $\left(k,l\right)$ history
will lead to the final state $\rho_{\tau_{2}}^{\left(k,l\right)}=V_{\tau_{2},\tau_{1}}^{\left(k\right)}\rho_{\tau_{1}}^{\left(l\right)}V_{\tau_{2},\tau_{1}}^{\left(k\right)\dagger}$,
with joint probability $p\left(k,l\right)=p\left(k|l\right)p\left(l\right)$,
and the final Hamiltonian $H_{\tau_{2}}^{\left(k\right)}$, with probability
$p\left(k\right)=\sum_{l}p\left(k,l\right)$. 

Due to the control mismatch, the mutual information density (between
the measurement and the controlled evolution) can be introduced \cite{Morikuni2011}
and it is given by $I^{\left(k,l\right)}=\ln p\left(k|l\right)/p\left(k\right)$,
where $p\left(k\right)$ is the probability that the feedback operation
$V_{\tau_{2},\tau_{1}}^{\left(k\right)}$ is implemented. The mutual
information is given by the average of its density, $\left\langle I\right\rangle =\sum_{k,l}p\left(k,l\right)I^{\left(k,l\right)}$,
and quantifies the correlation between the measurement and controlled
evolution. 

The mutual information is a fundamental quantity in information theory
\cite{Cover2006} and it appears explicitly in one of the generalized
forms of the second law in the presence of feedback, $\left\langle \Sigma\right\rangle =\beta\left(\left\langle W\right\rangle -\left\langle \Delta F\right\rangle \right)\geq-\left\langle I\right\rangle $,
where the lower bound for the mean entropy production $\left\langle \Sigma\right\rangle $
is a negative quantity \cite{Morikuni2011}. This inequality reveals
that the feedback control can be used to rectify the mean entropy
production in a nonequilibrium quantum dynamics \cite{Camati2016}.

We introduce the forward mixed joint PDF for a single system history
$\left(k,l\right)$, 
\begin{align}
P_{F}\left(k,l\,;W,\Delta F,I\right)= & \sum_{m,n}p\left(m^{\left(k\right)},k,l,n\right)\nonumber \\
 & \times\delta\left[W-W_{mkn}\right]\delta\left[\Delta F-\Delta F^{\left(k\right)}\right]\nonumber \\
 & \times\delta\left[I-I^{\left(k,l\right)}\right],\label{eq:forward mixed joint pdf single history kl}
\end{align}
where $W_{mkn}=E_{m}^{\left(k\right)\tau_{2}}-E_{n}^{0}$ is the work
in the TPM scheme and $\Delta F^{\left(k\right)}=F_{\tau_{2}}^{\left(k\right)}-F_{0}$
is the free-energy difference associated with the final Hamiltonian
$H_{\tau_{2}}^{\left(k\right)}$. This mixed PDF encompasses all statistical
information of the forward protocol (see \ref{sec:Appendix-B:-Single-History}).
It contains two discrete variables, $k$ and $l$, and three continuous
stochastic variables, $W$, $\Delta F$, and $I$. In Eq.~(\ref{eq:forward mixed joint pdf single history kl}),
$p\left(m^{\left(k\right)},k,l,n\right)=p\left(m^{\left(k\right)}|l\right)p\left(k|l\right)p\left(l|n\right)p\left(n\right)$
is a function of the set of labels $\left(m^{\left(k\right)},k,l,n\right)$,
with $p\left(m^{\left(k\right)}|l\right)=\mbox{Tr}\left[P_{m}^{\left(k\right)\tau_{2}}V_{\tau_{2},\tau_{1}}^{\left(k\right)}\mathcal{M}_{l}V_{\tau_{2},\tau_{1}}^{\left(k\right)\dagger}\right]$
the probability of obtaining $E_{m}^{\left(k\right)\tau_{2}}$ in
the final energy measurement given that the feedback measurement outcome
was $l$, $p\left(l|n\right)=\mbox{Tr}\left[\mathcal{M}_{l}U_{\tau_{1},0}P_{n}^{0}U_{\tau_{1},0}^{\dagger}\right]$
the probability of obtaining the outcome $l$ in the feedback measurement
given that the initial energy outcome was $E_{n}^{0}$, and $p\left(n\right)=e^{-\beta E_{n}^{0}}/Z_{0}$
the probability of obtaining the energy outcome $E_{n}^{0}$ in the
first measurement. Integrating some variables, we can obtain marginal
probability distributions associated with the protocol (see \ref{sec:Appendix-B:-Single-History}).
In particular, a useful quantity in our method is the mixed work PDF
\begin{equation}
P_{F}\left(k,l\,;W\right)=\sum_{m,n}p\left(m^{\left(k\right)},k,l,n\right)\delta\left[W-W_{mkn}\right],\label{eq:single history forward distribution}
\end{equation}
which is obtained from Eq.~(\ref{eq:forward mixed joint pdf single history kl})
by integrating over the free-energy difference and mutual information
density.

\begin{figure}[t]
\includegraphics[scale=0.5]{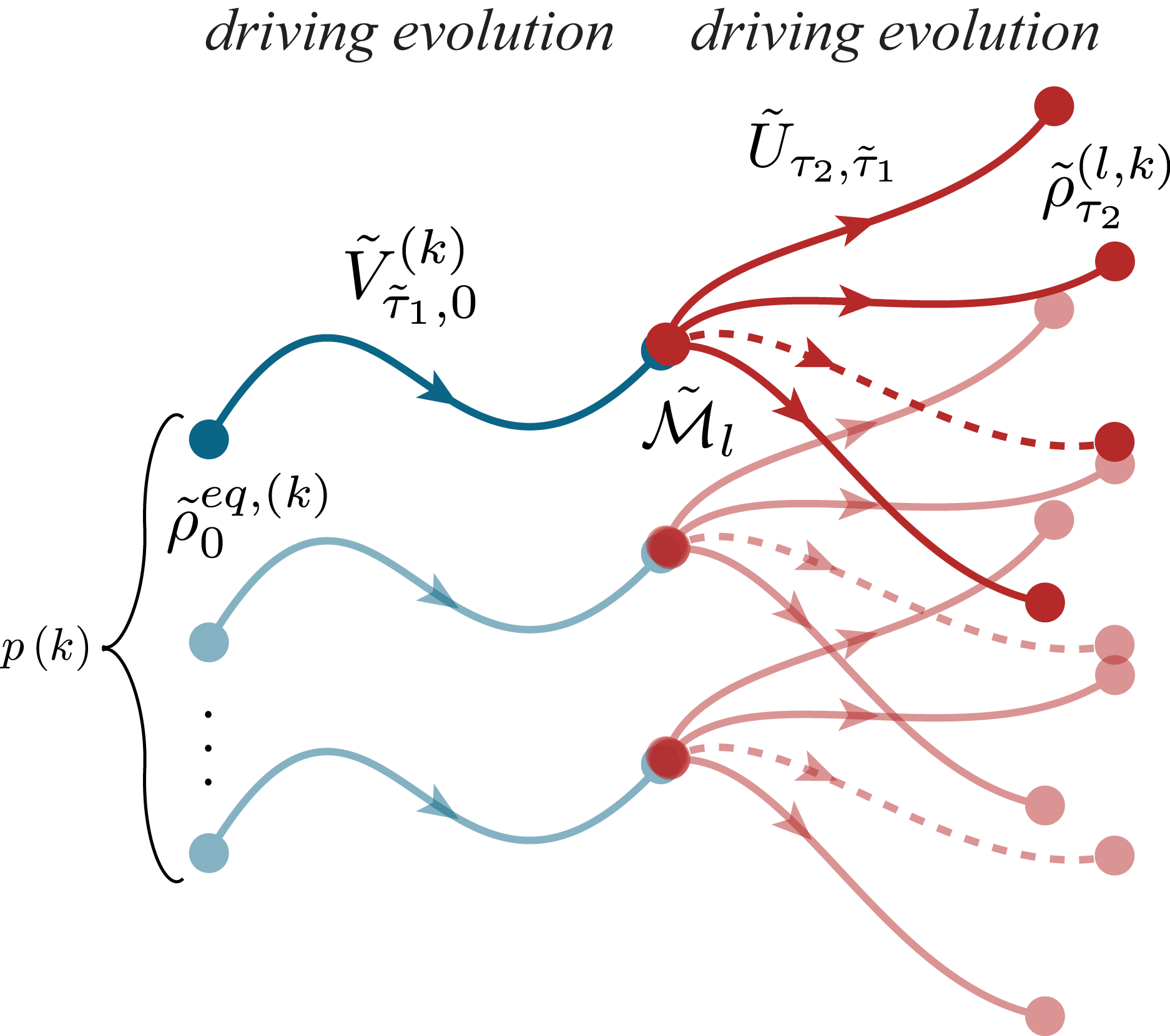}

\caption{Schematic representation of the backward feedback control protocol.
The initial thermal state for the backward protocol $\tilde{\rho}_{0}^{eq,\left(k\right)}=e^{-\beta\tilde{H}_{0}^{\left(k\right)}}/\tilde{Z}_{0}$
(where $\tilde{Z}_{t}=\text{Tr}e^{-\beta\tilde{H}_{t}^{\left(k\right)}}$)
is sampled with the probability $p\left(k\right)$ that the associated
final Hamiltonian of the forward protocol $H_{\tau_{2}}^{\left(k\right)}$
occurs. The first driven evolution is conditioned to the corresponding
feedback operation in the forward process. For each sampled initial
state the appropriate evolution corresponding to the time-reversed
Hamiltonian $\tilde{H}_{t}^{\left(k\right)}=\Theta H_{\tau_{2}-t}^{\left(k\right)}\Theta^{\dagger}$
from $0$ to $\tilde{\tau}_{1}=\tau_{2}-\tau_{1}$ should be implemented.
The measurement of the time-reversed observable $\tilde{\mathcal{M}}$
is performed at time $\tilde{\tau}_{1}$. The set of measurement eigenprojectors
is represented by $\left\{ \tilde{\mathcal{M}}_{l}\right\} $ and
the outcome $l$ occurs with probability $\tilde{p}\left(l|k\right)$.
The conditioning in $k$ is due to the fact that for each initial
state, labeled by $k$, the measurement probabilities will be different.
The final state is designated by $\tilde{\rho}_{\tau_{2}}^{\left(l,k\right)}$.
\label{fig:Backward-feedback-protocol.}}
\end{figure}

Employing the relations discussed in Sec.~\ref{sec:Time-reversal-operator-and-backward-processes},
we obtain a QDFR with feedback for the mixed work PDF {[}see Eq.~(\ref{eq:S21}){]}
\begin{equation}
\frac{P_{F}\left(k,l\,;W\right)}{P_{B}\left(l,k\,;-W\right)}=e^{\beta\left(W-\Delta F^{\left(k\right)}\right)+I^{\left(k,l\right)}},\label{eq:tasaki-crooks single history}
\end{equation}
 where 
\begin{equation}
P_{B}\left(l,k\,;W\right)=\sum_{n,m}\tilde{p}\left(n,l,k,m^{\left(k\right)}\right)\delta\left[W-\tilde{W}_{nkm}\right]\label{eq:single history backward distribution}
\end{equation}
is the mixed work PDF of the backward protocol (see Fig.~\ref{fig:Backward-feedback-protocol.}).
This QDFR applies for each history $\left(k,l\right)$ of the feedback
protocol. In Eq.~(\ref{eq:single history backward distribution}),
$\tilde{p}\left(n,l,k,m^{\left(k\right)}\right)=\tilde{p}\left(n|l\right)\tilde{p}\left(l|m^{\left(k\right)}\right)\tilde{p}\left(m\right)p\left(k\right)$,
with $\tilde{p}\left(n|l\right)=\text{Tr}\left[\tilde{P}_{n}^{\tau_{2}}\tilde{U}_{\tau_{2},\tilde{\tau}_{1}}\tilde{\mathcal{M}}_{l}\tilde{U}_{\tau_{2},\tilde{\tau}_{1}}^{\dagger}\right]$
the probability that the final energy outcome is $\tilde{E}_{n}^{\tau_{2}}$
given that the intermediate measurement outcome was $l$, $\tilde{p}\left(l|m^{\left(k\right)}\right)=\text{Tr}\left[\tilde{\mathcal{M}}_{l}\tilde{V}_{\tilde{\tau}_{1},0}^{\left(k\right)}\tilde{P}_{m}^{\left(k\right)0}\tilde{V}_{\tilde{\tau}_{1},0}^{\left(k\right)\dagger}\right]$
the probability that the intermediate measurement outcome is $l$
given that the initial energy outcome was $\tilde{E}_{m}^{\left(k\right),0}$,
$\tilde{p}\left(m\right)=e^{-\beta\left(\tilde{E}_{m}^{\left(k\right)0}-\tilde{F}_{0}^{\left(k\right)}\right)}$
the probability of the initial energy outcome $\tilde{E}_{m}^{\left(k\right),0}$,
$\tilde{\tau}_{1}=\tau_{2}-\tau_{1}$ the time of the intermediate
measurement, and $\tilde{W}_{nkm}=\tilde{E}_{n}^{\tau_{2}}-\tilde{E}_{m}^{\left(k\right)0}$
the work value. The measurement and feedback breaks a direct relational
symmetry between forward and backward processes. Therefore, the backward
protocol consistent with the fluctuation relations (FRs) is structurally
different from the forward protocol in the presence of feedback (see
Fig.~\ref{fig:Backward-feedback-protocol.}). We also note that an
analogous relation was obtained in the classical setting in Ref. \cite{Horowitz2010}.

First, for each value of $k$ associated with the different feedback
operations of the forward protocol, a different initial thermal state
should be prepared, $\tilde{\rho}_{0}^{eq,\left(k\right)}=e^{-\beta\tilde{H}_{0}^{\left(k\right)}}/\tilde{Z}_{0}^{\left(k\right)}$,
where $\tilde{H}_{0}^{\left(k\right)}=\Theta H_{\tau_{2}}^{\left(k\right)}\Theta^{\dagger}$.
These initial states are sampled with the probability $p\left(k\right)$
that the final Hamiltonian $H_{\tau_{2}}^{\left(k\right)}$ occurs
in the forward protocol. Second, there is no feedback control after
the intermediate measurement in the backward process, and therefore
no control mismatch. Third, we do not assume that the observable is
time reversal \cite{Campisi2011-1}. The observable to be measured
at $\tilde{\tau}_{1}$ should be the transformed observable $\tilde{\mathcal{M}}=\Theta\mathcal{M}\Theta^{\dagger}$
whose eigenprojectors are given by $\left\{ \tilde{\mathcal{M}}_{l}=\Theta\mathcal{M}_{l}\Theta^{\dagger}\right\} $.

When there is no control mismatch we can still define a forward mixed
joint PDF $P_{F}^{\text{wcm}}\left(k\,;W,\Delta F\right)$ that encompasses
the relevant probabilities involved ($\text{wcm}$ stands for ``without
control mismatch''). We derive yet another QDFR with feedback from
the mixed work PDF without mismatch $P_{F}^{\text{wcm }}\left(k\,;W\right)$,
which applies to each history $k$ and is given by 
\begin{equation}
\frac{P_{F}^{\text{wcm}}\left(k\,;W\right)}{P_{B}^{\text{wcm}}\left(k\,;-W\right)}=e^{\beta\left(W-\Delta F^{\left(k\right)}\right)},\label{eq:tasaki-crooks without error}
\end{equation}
where $P_{B}^{\text{wcm}}\left(k\,;-W\right)$ is the associated backward
mixed work PDF. In  \ref{sec:Appendix-C:-Single-History} we discuss
these PDFs in more detail and demonstrate the validity of the QDFRs.
The QDFR given by Eq.~(\ref{eq:tasaki-crooks without error}) could
be expected since it is known that a projective measurement performed
during the dynamics does not change the FRs \cite{Campisi2010}.

Next we discuss the QDFR with feedback control that is valid for the
process as a whole \cite{Lahiri2012}. Within our formalism, we can
obtain the QDFR in a different way. Integrating the discrete variables
in the mixed joint PDF, Eq.~(\ref{eq:forward mixed joint pdf single history kl}),
we obtain the joint PDF of the feedback protocol
\begin{eqnarray}
P_{F}\left(W,\Delta F,I\right) & = & \sum_{m,k,l,n}p\left(m^{\left(k\right)},k,l,n\right)\delta\left[W-W_{mkn}\right]\nonumber \\
 &  & \times\delta\left[\Delta F-\Delta F^{\left(k\right)}\right]\delta\left[I-I^{\left(k,l\right)}\right].\label{eq:whole forward process distribution}
\end{eqnarray}
We can test the consistency of this joint PDF by calculating the averages
of the individual variables. Averaging the work variable we obtain
the average work of the feedback protocol, $\left\langle W\right\rangle =\sum_{k,l}p\left(k,l\right)\left[\mathcal{U}\left(\rho_{\tau_{2}}^{\left(k,l\right)}\right)-\mathcal{U}\left(\rho_{0}^{eq}\right)\right]$,
which is the average, over $p\left(k,l\right)$, of the work of each
history $\left(k,l\right)$. Averaging the free energy, we obtain
$\left\langle \Delta F\right\rangle =\sum_{k}p\left(k\right)\Delta F^{\left(k\right)}$,
which is the weighted sum of the possible values of free-energy variation.
Averaging the mutual information density, we obtain $\left\langle I\right\rangle =\sum_{k,l}p\left(k,l\right)I^{\left(k,l\right)}$,
which is the mutual information. From Eq.~(\ref{eq:whole forward process distribution})
and the relation between forward and backward processes given in Sec.~\ref{sec:Time-reversal-operator-and-backward-processes}
we obtain the QDFR with feedback for the whole process \cite{Lahiri2012}
\begin{equation}
\frac{P_{F}\left(W,\Delta F,I\right)}{P_{B}\left(-W,-\Delta F,I\right)}=e^{\beta\left(W-\Delta F\right)+I},\label{eq:tasaki-crooks whole process}
\end{equation}
where
\begin{eqnarray}
P_{B}\left(W,\Delta F,I\right) & = & \sum_{n,l,k,m}\tilde{p}\left(n,l,k,m^{\left(k\right)}\right)\delta\left[W-\tilde{W}_{nkm}\right]\nonumber \\
 &  & \times\delta\left[\Delta F-\Delta\tilde{F}^{\left(k\right)}\right]\delta\left[I-I^{\left(k,l\right)}\right]\label{eq:whole backward process distribution}
\end{eqnarray}
is the associated joint PDF of the backward protocol (see also \ref{sec:Appendix-D:-Whole-Process}).

The characteristic function of a PDF is given by its Fourier transform
and contains as much information as the PDF itself. Our method to
experimentally verify the QDFR with feedback control relies on the
direct measurement of the characteristic functions associated with
the PDFs appearing in the FRs discussed above. Without the presence
of feedback control, a method to obtain the characteristic functions
of work distributions was introduced in Refs.~\cite{Mazzola2013,Dorner2013a}
and successfully applied in an experimental scenario in Refs.~\cite{Batalhao2014,Batalhao2015}.
Here we extend these ideas for discrete feedback-controlled quantum
processes and, as will be seen, the complexity of the method increases
considerably. We write down explicitly these characteristic functions
below (see \ref{sec:Appendix-E:-Characteristic}). From the forward
and backward mixed work PDFs, Eqs.~(\ref{eq:single history forward distribution})
and (\ref{eq:single history backward distribution}), we obtain \begin{widetext}
\begin{eqnarray}
\chi_{F}^{\left(k,l\right)}\left(u\right) & = & p\left(k|l\right)\mbox{Tr}\left[e^{+iuH_{\tau_{2}}^{\left(k\right)}}V_{\tau_{2},\tau_{1}}^{\left(k\right)}\mathcal{M}_{l}U_{\tau_{1},0}e^{-iuH_{0}}\rho_{0}^{eq}U_{\tau_{1},0}^{\dagger}\mathcal{M}_{l}V_{\tau_{2},\tau_{1}}^{\left(k\right)\dagger}\right],\label{eq:forward characteristic work function single history}\\
\chi_{B}^{\left(l,k\right)}\left(u\right) & = & p\left(k\right)\mbox{Tr}\left[e^{+iu\tilde{H}_{\tau_{2}}}\tilde{U}_{\tau_{2},\tilde{\tau}_{1}}\tilde{\mathcal{M}}_{l}\tilde{V}_{\tilde{\tau}_{1},0}^{\left(k\right)}e^{-iu\tilde{H}_{0}^{\left(k\right)}}\tilde{\rho}_{0}^{eq,\left(k\right)}\tilde{V}_{\tilde{\tau}_{1},0}^{\left(k\right)\dagger}\tilde{\mathcal{M}}_{l}\tilde{U}_{\tau_{2},\tilde{\tau}_{1}}^{\dagger}\right],\label{eq:backward characteristic work function single history}
\end{eqnarray}
\end{widetext} where we take the Fourier transformation of the work
variable $W$ and $u$ is the transformed variable associated with
the quantum work. For further reference throughout this paper, we
denote the trace in Eq.~(\ref{eq:backward characteristic work function single history})
by $\mathcal{A}\left(k,l\right)=\chi_{B}^{\left(l,k\right)}\left(u\right)/p\left(k\right)$.

From the forward and backward joint PDFs, Eqs.~(\ref{eq:whole forward process distribution})
and (\ref{eq:whole backward process distribution}), we obtain
\begin{eqnarray}
\chi_{F}\left(u,v,w\right) & = & \sum_{k,l}e^{iwI^{\left(k,l\right)}}e^{iv\Delta F^{\left(k\right)}}\chi_{F}^{\left(k,l\right)}\left(u\right),\label{eq:forward characteristic work function whole process}\\
\chi_{B}\left(u,v,w\right) & = & \sum_{l,k}e^{iwI^{\left(k,l\right)}}e^{iv\Delta\tilde{F}^{\left(k\right)}}\chi_{B}^{\left(k,l\right)}\left(u\right),\label{eq:backward characteristic work function whole process}
\end{eqnarray}
where $u$, $v$, and $w$ are the transformed variables associated
with the stochastic work $W$, free-energy variation $\Delta F^{\left(k\right)}$,
and mutual information density $I^{\left(k,l\right)}$, respectively.

\begin{figure}[b]
\includegraphics[scale=0.45]{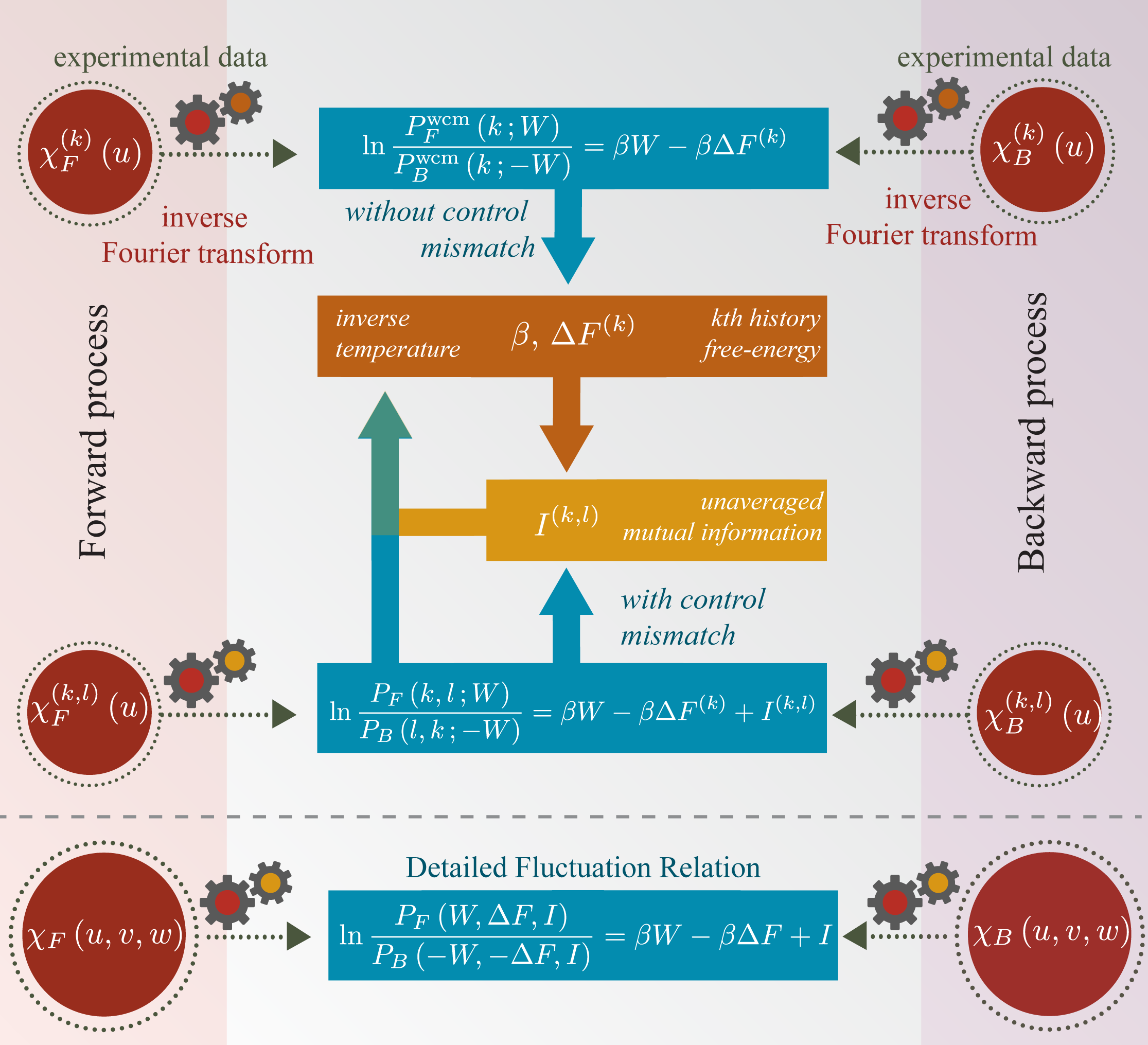}

\caption{Flowchart representing the proposed method to verify the QDFR in the
presence of feedback control (in blue below the dashed line). The
four characteristic functions represented above the dashed line are
the experimental data needed for the verification. They are obtained
from measurements of five quantum algorithms. Taking their inverse
Fourier transform we use the corresponding PDFs in the two QDFRs displayed
in blue. The QDFR without control mismatch (uppermost) is used to
obtain the free energy differences $\Delta F^{\left(k\right)}$. The
QDFR with control mismatch (just above the dashed line) and the free
energies previously obtained are used to obtain the mutual information
density $I^{\left(k,l\right)}$. If wanted, several independent estimations
of the initial inverse temperature $\beta$ can be obtained as explained
in the text. With the free-energy differences, the mutual information
density and the forward, $\chi_{F}^{\left(k,l\right)}\left(u\right)$,
and backward, $\chi_{B}^{\left(l,k\right)}\left(u\right)$, characteristic
work functions, the joint forward and backward characteristic functions
(below the dashed line) can be reconstructed. By inverse Fourier transform
the QDFR for discrete feedback processes can be verified through multiple
linear regression (see Fig. \ref{fig:alternative} in \ref{sec:Appendix-F:-Quantum}
for an additional description of the method). \label{fig:Flowchart}}
\end{figure}

\section{VERIFYING QUANTUM DETAILED FLUCTUATION RELATIONS IN THE PRESENCE
OF FEEDBACK CONTROL\label{sec:Method}}

In this section we present a feasible method to experimentally verify
the QDFR for feedback processes, Eq.~(\ref{eq:tasaki-crooks whole process}).
The method is schematically depicted in Fig.~\ref{fig:Flowchart}
and the strategy is as follows. In order to experimentally obtain
the forward and backward joint PDFs, Eqs.~(\ref{eq:whole forward process distribution})
and (\ref{eq:whole backward process distribution}), we will describe
a way to measure the corresponding characteristic functions, Eqs.~(\ref{eq:forward characteristic work function whole process})
and (\ref{eq:backward characteristic work function whole process}).
The joint PDFs can be obtained from the inverse Fourier transform
of the associated characteristic functions. 

For the full reconstruction of the PDFs in the presence of feedback,
four quantities are required. The characteristic functions of the
forward and backward mixed PDFs, Eqs.~(\ref{eq:forward characteristic work function single history})
and (\ref{eq:backward characteristic work function single history}),
are directly obtained from the interferometric quantum algorithms
that we will introduce in what follows. The remaining two quantities
are the free-energy differences and the mutual information density.
Using the relation between forward and backward processes, it can
be shown that $\Delta\tilde{F}^{\left(k\right)}=-\Delta F^{\left(k\right)}$
and therefore only the forward free-energy differences are really
needed. 

The forward free-energy differences can be obtained from the QDFR
without control mismatch, Eq.~(\ref{eq:tasaki-crooks without error}),
using an adaptation of the methods developed in Refs.~\cite{Batalhao2014,Batalhao2015}
and which we will detail below. The characteristic functions of the
forward and backward mixed work PDF without control mismatch, $P_{F}^{\text{wcm}}\left(k;W\right)$
and $P_{B}^{\text{wcm}}\left(k;W\right)$, which we denote as $\chi_{F}^{\text{wcm}\left(k\right)}\left(u\right)$
and $\chi_{B}^{\text{wcm}\left(k\right)}\left(u\right)$, respectively,
will also be required (see also \ref{sec:Appendix-E:-Characteristic}). 

The mutual information density can be analogously obtained from the
QDFR with control mismatch, Eq.~(\ref{eq:tasaki-crooks single history}),
which involves the characteristic functions $\chi_{F}^{\left(k,l\right)}\left(u\right)$
and $\chi_{B}^{\left(l,k\right)}\left(u\right)$. In summary, it is
necessary to experimentally measure four characteristic functions.
Our method proposes five quantum algorithms which accomplish this
task using an interferometric strategy and auxiliary quantum systems
that will encode the characteristic function to be measured.

\begin{figure}[b]
\includegraphics[width=1\columnwidth]{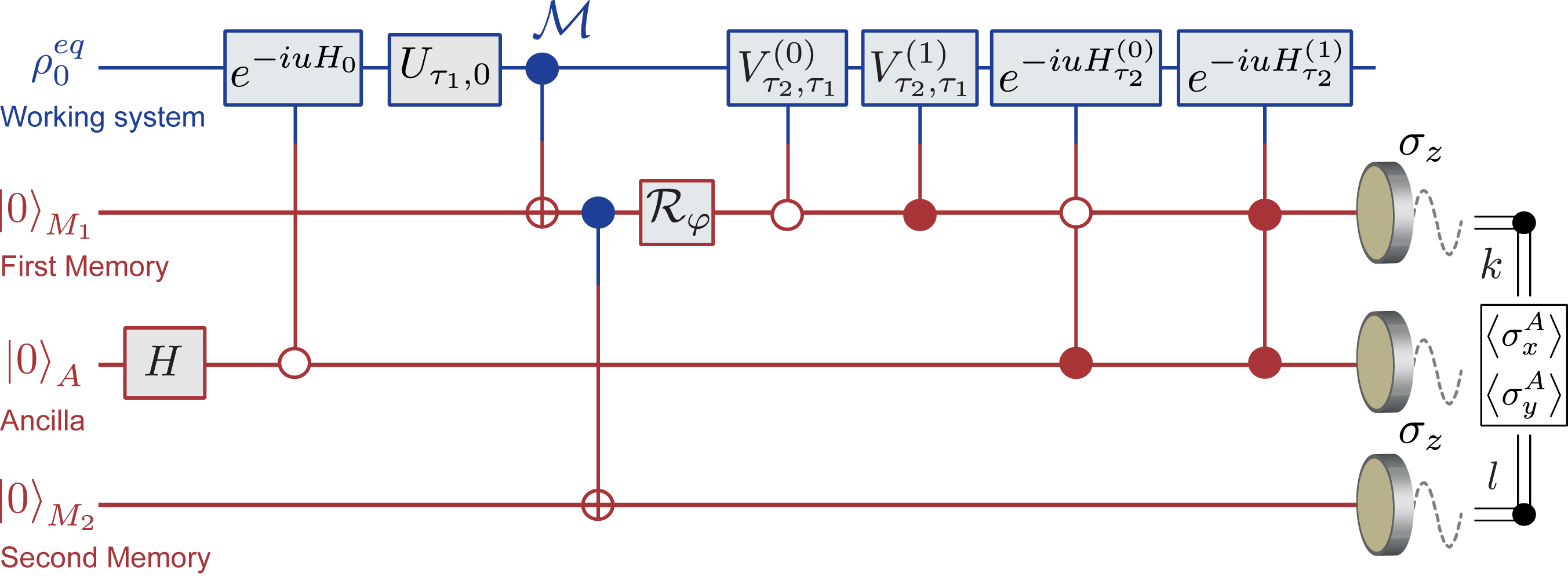}

\caption{Quantum interferometric circuit for the measurement of the forward
work characteristic function with control mismatch, $\chi_{F}^{\left(k,l\right)}\left(u\right)$.
The nonselective measurement is implemented with a $\mathcal{M}$-CNOT
gate. The rotation $\mathcal{R}_{\varphi}$ changes the memory states
so that the feedback is implemented with control mismatch. After a
measurement in the computational basis of the two memories (of the
feedback protocol), $M_{1}$ and $M_{2}$, whose outcomes $\left(k,l\right)$
occur with probability $p^{M_{1}M_{2}}\left(k,l\right)$, the ancilla
$A$ encodes the characteristic function in its coherence elements
in the computational basis. It can be retrieved from the average of
the Pauli observables in the auxiliary system $\left\langle \sigma_{x}^{A}\right\rangle =\mbox{Re}\left[\chi_{F}^{\left(k,l\right)}\left(u\right)\right]/p^{M_{1}M_{2}}\left(k,l\right)$
and $\left\langle \sigma_{y}^{A}\right\rangle =\mbox{Im}\left[\chi_{F}^{\left(k,l\right)}\left(u\right)\right]/p^{M_{1}M_{2}}\left(k,l\right)$.
\label{fig:Single-history-forward-with-error}}
\end{figure}

Here we employ a full quantum perspective for the feedback protocol
illustrated in Fig.~\ref{fig:feedback histories}, which is expressed
by the fact that feedback controller is regarded as a quantum system.
The characteristic function of the forward mixed work PDF, $\chi_{F}^{\left(k,l\right)}\left(u\right)$,
has two fixed parameters, $k$ and $l$. Our algorithm employs two
quantum memories to store the information concerning these parameters.
These memories should have a state space dimension at least as large
as the ranges of the indices $k$ and $l$ so as to encode them in
orthogonal states. Moreover, an ancilla qubit is used to encode the
information of the characteristic function \cite{Mazzola2013}. We
assume from now on, without loss of generality, that the system of
interest is a qubit. Hence, the two memories, which we designate as
the feedback controller memories, are two qubits. We stress that the
ancilla system (which encodes the characteristic function) can always
be a qubit, irrespective of the dimension of the system of interest.
Notwithstanding, the ideas presented here can be extended to other
scenarios: where the control has a fixed probability to fail or to
a semiclassical feedback device where the feedback memory and control
mechanism are classical objects.

The quantum interferometric circuit depicted in Fig.~\ref{fig:Single-history-forward-with-error}
enables the measurement of the characteristic function $\chi_{F}^{\left(k,l\right)}\left(u\right)$.
The measurement stage of the feedback is implemented by an emulated
nonselective measurement. This is accomplished by an $\mathcal{M}$-CNOT
gate (right after the unitary gate $U_{\tau_{1},0}$) in Fig.~\ref{fig:Single-history-forward-with-error}.
This gate is similar to the usual CNOT gate, except that the control
basis is not the computational basis, $\{\ket{0},\ket{1}\}$, but
the basis of the eigenvectors of the observable $\mathcal{M}$ to
be measured in the feedback protocol {[}see Eq.~(\ref{eq:S41}){]}.
We consider one extra rotation ($\mathcal{R}_{\varphi}=e^{-i\varphi\sigma_{x}}$
in Fig.~\ref{fig:Single-history-forward-with-error}) that plays
the role of switching the control mismatch in the feedback protocol.
The feedback stage is implemented, in a quantum way, by controlled
unitary gates to properly perform the correct feedback operations
$V_{\tau_{2},\tau_{1}}^{\left(k\right)}$ using the encoded information
in the quantum memory. 

When the feedback device is described as a quantum system, which is
our case here, the state of the composite system at the end of the
interferometric circuit may be highly entangled. At the end of the
circuit a measurement in the computational basis of the two memory
qubits is performed giving outcomes $\left(k,l\right)$ with probability
$p^{M_{1}M_{2}}\left(k,l\right)$ {[}for details see Eq.~(\ref{eq:S46}){]}.
After such a measurement, the ancillary qubit $A$ encodes the information
of the forward work characteristic function associated with the corresponding
outcome and feedback control operation $\left(k,l\right)$, $\chi_{F}^{\left(k,l\right)}\left(u\right)$. 

The characteristic function and its conjugate are encoded in the coherence
elements of the reduced density matrix of the ancilla in the computational
basis. It can be extracted from the averages $\left\langle \sigma_{x}^{A}\right\rangle =\mbox{Re}\left[\chi_{F}^{\left(k,l\right)}\left(u\right)\right]/p^{M_{1}M_{2}}\left(k,l\right)$
and $\left\langle \sigma_{y}^{A}\right\rangle =\mbox{Im}\left[\chi_{F}^{\left(k,l\right)}\left(u\right)\right]/p^{M_{1}M_{2}}\left(k,l\right)$
of the Pauli observables $\sigma_{x}^{A}$ and $\sigma_{y}^{A}$ in
the ancillary qubit $A$. 

In the actual implementation of the interferometric circuit, the conjugate
variable $u$ will be associated with the time of a suitable interaction
between the qubits implementing the controlled unitaries. Therefore,
each run of the algorithm actually measures a single value of the
characteristic function, i.e., for a given interaction time and thus
a given value of $u$. This means that the quantum algorithm has to
run several times to obtain a discretized characteristic function
\cite{Batalhao2014}. The sampling rate of the parametrization of
$u$, in an actual experiment, will be also related to the accuracy
of the inverse Fourier transform of the acquired data in order to
reconstruct the work PDFs. 

\begin{figure}
\includegraphics[width=1\columnwidth]{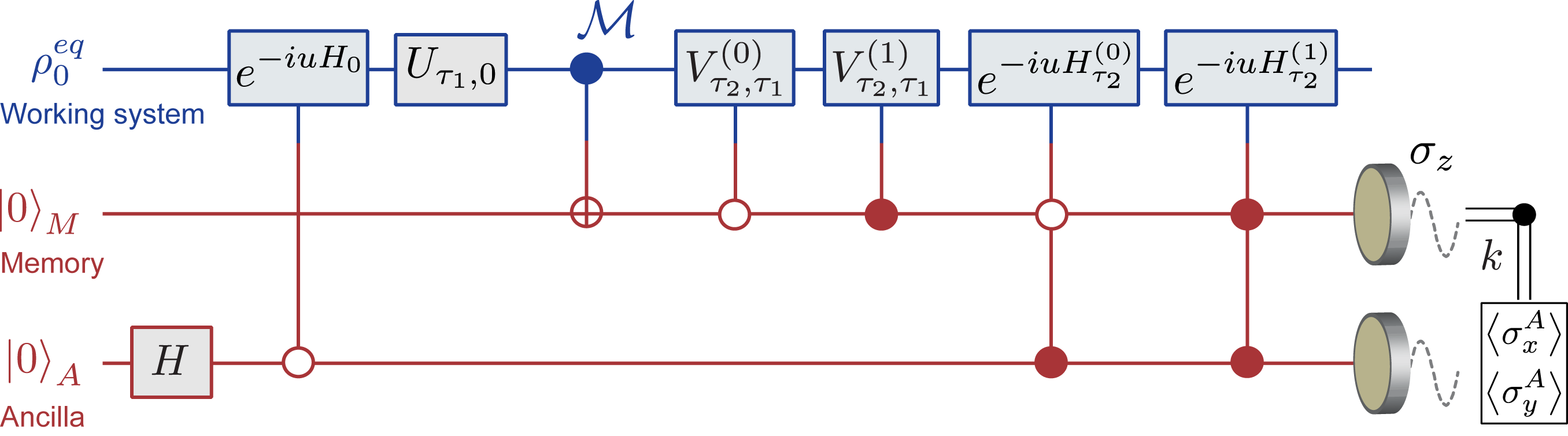}

\caption{Quantum interferometric circuit for the measurement of the forward
characteristic work function without control mismatch $\chi_{F}^{\left(k\right)}\left(u\right)$.
After a measurement in the computational basis of the memory $M$
{[}whose outcomes $k$ occur with probability $p^{M}\left(k\right)${]},
the ancilla $A$ encodes the characteristic function in its coherence
elements in the computational basis. It can be retrieved from the
average of the Pauli observables $\left\langle \sigma_{x}^{A}\right\rangle =\mbox{Re}\left[\chi_{F}^{\left(k\right)}\left(u\right)\right]/p^{M}\left(k\right)$
and $\left\langle \sigma_{y}^{A}\right\rangle =\mbox{Im}\left[\chi_{F}^{\left(k\right)}\left(u\right)\right]/p^{M}\left(k\right)$.
\label{fig:single-history without error}}
\end{figure}

When there is no feedback control mismatch, one memory to encode the
value of $k$ suffices. The quantum circuit which measures the characteristic
work function $\chi_{F}^{\text{wcm}\left(k\right)}\left(u\right)$
is shown in Fig.~\ref{fig:single-history without error}. As before,
$\chi_{F}^{\text{wcm}\left(k\right)}\left(u\right)$ is encoded in
the coherence elements of the ancilla in the computational basis when
the (feedback) memory measurement outcome $k$ is obtained. This occurs
with probability $p^{M}\left(k\right)$ {[}see also Eq.~(\ref{eq:S47}){]}.
In this case, the average of the Pauli observables $\left\langle \sigma_{x}^{A}\right\rangle =\mbox{Re}\left[\chi_{F}^{\text{wcm}\left(k\right)}\left(u\right)\right]/p^{M}\left(k\right)$
and $\left\langle \sigma_{y}^{A}\right\rangle =\mbox{Im}\left[\chi_{F}^{\text{wcm}\left(k\right)}\left(u\right)\right]/p^{M}\left(k\right)$
provides the real and imaginary parts of the characteristic function.

\begin{figure}[b]
\includegraphics[width=0.9\columnwidth]{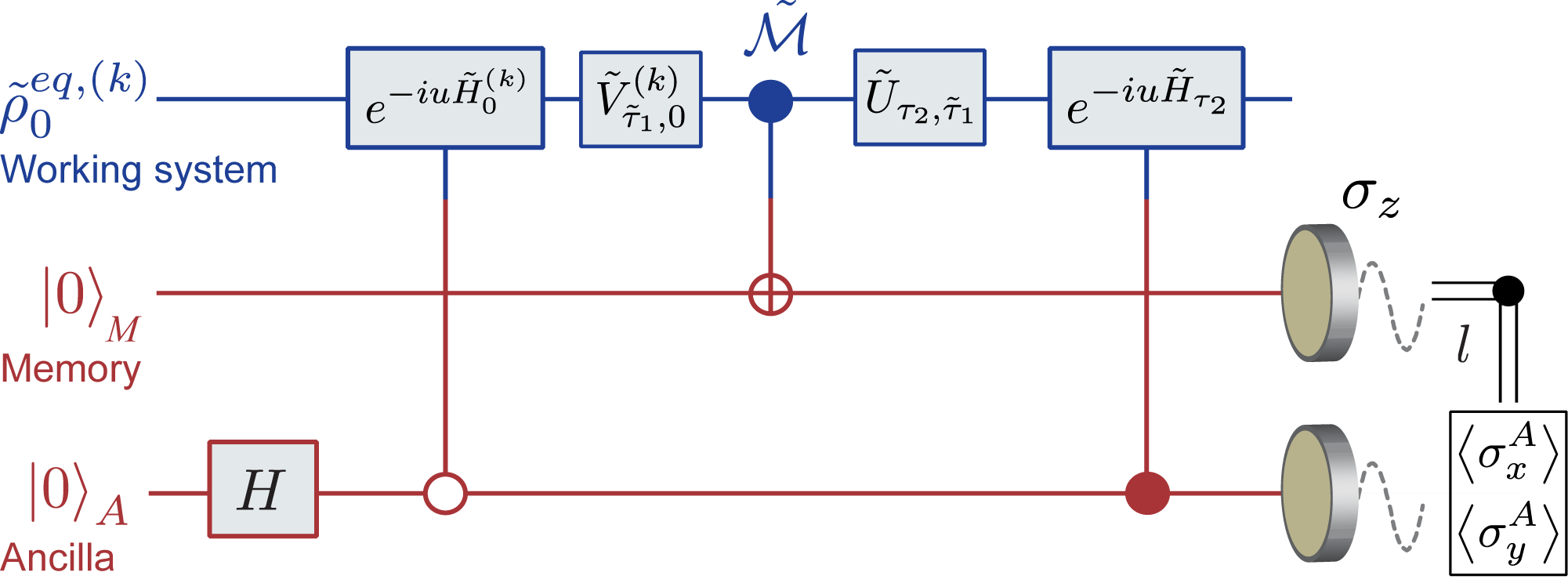}

\caption{Quantum interferometric circuit for the measurement of the backward
characteristic work functions with and without mismatch, $\chi_{B}^{\left(l,k\right)}\left(u\right)$
and $\chi_{B}^{\text{wcm}\left(k\right)}\left(u\right)$, respectively.
In order to measure $\chi_{B}^{\text{wcm}\left(k\right)}\left(u\right)$
the ancilla $A$ is measured only when the outcome of the memory measurement
$M$ is $l=k$. After a measurement of the feedback memory in the
computational basis, whose outcomes $l$ occur with probability $p_{B}^{M}\left(l\right)$,
the ancilla encodes the information of $\mathcal{A}\left(k,l\right)$
{[}defined after Eq.~(\ref{eq:backward characteristic work function single history}){]}
in its coherence elements in the computational basis. The information
can be retrieved from the average of the Pauli observables $\left\langle \sigma_{x}^{A}\right\rangle =\mbox{Re}\left[\mathcal{A}\left(k,l\right)\right]/p^{M}\left(l\right)$
and $\left\langle \sigma_{y}^{A}\right\rangle =\mbox{Im}\left[\mathcal{A}\left(k,l\right)\right]/p^{M}\left(l\right)$.
For the case where feedback occurs without control mismatch the equality
$\chi_{B}^{\text{wcm}\left(k\right)}\left(u\right)=\mathcal{A}\left(k,k\right)$
holds. Therefore, in that case, the ancilla encodes the characteristic
function as $\left\langle \sigma_{x}^{A}\right\rangle =\mbox{Re}\left[\chi_{B}^{\text{wcm}\left(k\right)}\left(u\right)\right]/p_{B}^{M}\left(l=k\right)$
and $\left\langle \sigma_{y}^{A}\right\rangle =\mbox{Im}\left[\chi_{B}^{\text{wcm}\left(k\right)}\left(u\right)\right]/p_{B}^{M}\left(l=k\right)$.
With control mismatch the equality $\chi_{B}^{\left(l,k\right)}\left(u\right)=p\left(k\right)\mathcal{A}\left(k,l\right)$
holds. Therefore, the average of the ancilla Pauli observables provides
part of the backward characteristic function. The probability $p\left(k\right)$
is measured with the quantum circuit in Fig.~\ref{fig:joint probability distribution}.
\label{fig:backward process}}
\end{figure}

To measure the backward characteristic function $\chi_{B}^{\text{wcm}\left(k\right)}\left(u\right)$
without control mismatch the quantum algorithm in Fig.~\ref{fig:backward process}
will be employed . The strategy of the interferometric protocol is
as follows. Prepare the thermal initial state of the backward protocol,
$\tilde{\rho}_{0}^{eq,\left(k\right)}=e^{-\beta\tilde{H}_{0}}/\tilde{Z}_{0}^{\left(k\right)}$.
When the measurement outcome of the feedback memory measurement is
$l\neq k$, the data are discarded and the initial state is prepared
again. When the measurement outcome of the feedback memory is $l=k$,
the ancilla $A$ encodes the information of $\chi_{B}^{\text{wcm}\left(k\right)}\left(u\right)$
in its coherences in the computational basis. Again, the average of
the Pauli observables $\left\langle \sigma_{x}^{A}\right\rangle =\mbox{Re}\left[\chi_{B}^{\text{wcm}\left(k\right)}\left(u\right)\right]/p_{B}^{M}\left(l=k\right)$
and $\left\langle \sigma_{y}^{A}\right\rangle =\mbox{Im}\left[\chi_{B}^{\text{wcm}\left(k\right)}\left(u\right)\right]/p_{B}^{M}\left(l=k\right)$
is used to construct the characteristic function. The probability
$p_{B}^{M}\left(l\right)$ is the probability that the memory measurement
provides the outcome $l$ {[}see Eq.~(\ref{eq:S48}){]}.

\begin{figure}
\includegraphics[scale=0.5]{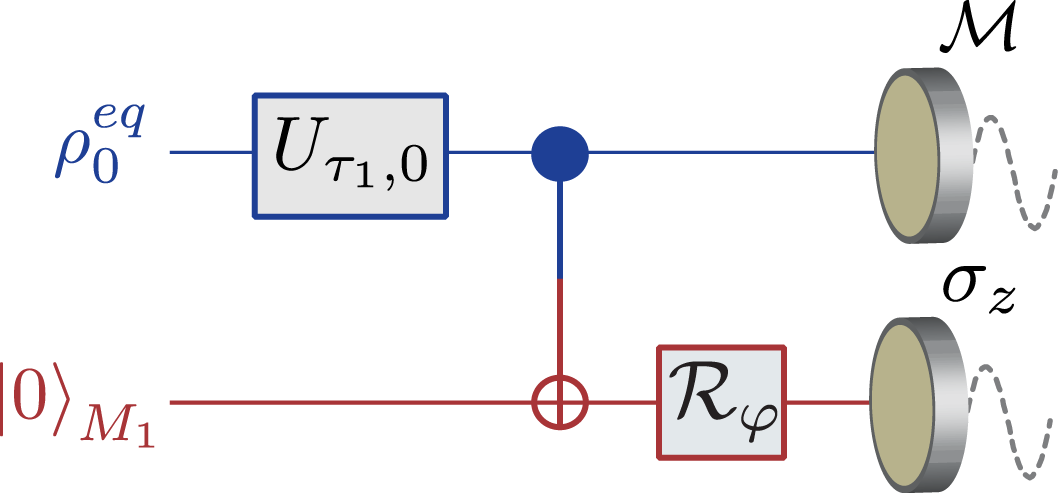}

\caption{Quantum circuit to characterizing the feedback mechanism. It may be
employed for the determination of the joint probability distribution
$p\left(k,l\right)$ of the $\left(k,l\right)$ history. A composite
measurement of the joint observable $\mathcal{M}^{S}\otimes\sigma_{z}^{M}$
gives outcomes $\left(k,l\right)$ with probability $p\left(k,l\right)$.\label{fig:joint probability distribution}}
\end{figure}

The backward characteristic work function with mismatch, $\chi_{B}^{\left(l,k\right)}\left(u\right)$,
requires two quantum circuits: the one we just considered in Fig.~\ref{fig:backward process}
and the one depicted in Fig.~\ref{fig:joint probability distribution}.
This latter circuit measures the joint probability $p\left(k,l\right)$
from the statistics of the composite measurement $\mathcal{M}^{S}\otimes\sigma_{z}^{M}$.
This is necessary to obtain the distribution $p\left(k\right)=\sum_{k,l}p\left(k,l\right)$
that appears in Eq.~(\ref{eq:backward characteristic work function single history}).
The trace in Eq.~(\ref{eq:backward characteristic work function single history}),
denoted by $\mathcal{A}\left(k,l\right)$, is obtained by the measurement
of the ancilla qubit in Fig.~\ref{fig:backward process}. The average
of the Pauli observables $\left\langle \sigma_{x}^{A}\right\rangle =\mbox{Re}\left[\mathcal{A}\left(k,l\right)\right]/p_{B}^{M}\left(l\right)$
and $\left\langle \sigma_{y}^{A}\right\rangle =\mbox{Im}\left[\mathcal{A}\left(k,l\right)\right]/p_{B}^{M}\left(l\right)$
enables the reconstruction of $\mathcal{A}\left(k,l\right)$ (see
\ref{sec:Appendix-F:-Quantum}). With $p\left(k\right)$ and $\mathcal{A}\left(k,l\right)$
one can reconstruct $\chi_{B}^{\left(l,k\right)}\left(u\right)=p\left(k\right)\mathcal{A}\left(k,l\right)$
{[}see Eq.~(\ref{eq:backward characteristic work function single history}){]}.

We have hitherto shown how to obtain the four characteristic work
functions: $\chi_{F}^{\text{wcm}\left(k\right)}\left(u\right)$ for
the forward process without mismatch, $\chi_{B}^{\text{wcm}\left(k\right)}\left(u\right)$
for the backward process without mismatch, $\chi_{F}^{\left(k,l\right)}\left(u\right)$
for the forward process with mismatch, and $\chi_{B}^{\left(l,k\right)}\left(u\right)$
for the backward process with mismatch. Thus, by means of the inverse
Fourier transform, one can obtain the four corresponding mixed work
PDFs: $P_{F}^{\text{wcm}}\left(k\,;W\right)$ for the forward process
without mismatch, $P_{B}^{\text{wcm}}\left(k\,;W\right)$ for the
backward process without mismatch, $P_{F}\left(k,l\,;W\right)$ for
the forward process with mismatch, and $P_{B}\left(l,k\,;W\right)$
for the backward process with mismatch. 

For the moment, let us consider the mixed work PDFs without mismatch,
$P_{F}^{\text{wcm}}\left(k\,;W\right)$ and $P_{B}^{\text{wcm}}\left(k\,;W\right)$.
For a system with discrete energy spectrum these PDFs should be theoretically
given by a sum of delta functions with all the possible transitions
from the eigenstates of the initial Hamiltonian to the eigenstates
of the final Hamiltonian. For a feedback-driven qubit dynamics in
which the initial and final Hamiltonians have different energy spectra,
each of these PDFs has four different peaks for each $k$ \cite{Batalhao2014}.
The peaks of the forward and backward PDFs will coincide because $E_{m}^{\left(k\right)\tau_{2}}-E_{n}^{0}=-\left(\tilde{E}_{n}^{\tau_{2}}-\tilde{E}_{m}^{\left(k\right)0}\right)$
from the relations discussed in Sec. \ref{sec:Time-reversal-operator-and-backward-processes}.
The peaks will differ only by multiplicative factors, i.e., the factors
multiplying the delta functions {[}see Eqs.~(\ref{eq:20}) and (\ref{eq:22})
of the next section{]}. 

Taking the ratio of the two PDFs and then the logarithm, one obtains
$\ln P_{F}^{\text{wcm}}\left(k\,;W\right)/P_{B}^{\text{wcm}}\left(k\,;W\right)=\beta W-\beta\Delta F^{\left(k\right)}$,
which is the QDFR without mismatch, Eq.~(\ref{eq:tasaki-crooks without error}).
For each $k$, a logarithmic plot of this ratio will provide a set
of points which ideally lie in a straight line. From each straight
line, one can infer the inverse temperature $\beta$ from the slope
and the free-energy difference $\Delta F^{\left(k\right)}$ for each
history from the interception of the line with the vertical axis. 

The same reasoning applies to the forward, $P_{F}\left(k,l\,;W\right)$,
and backward, $P_{B}\left(l,k\,;W\right)$, PDFs with control mismatch.
In that case, the logarithm of the ratio is $\ln P_{F}\left(k,l\,;W\right)/P_{B}\left(l,k\,;W\right)=\beta W-\beta\Delta F^{\left(k\right)}+I^{\left(k,l\right)}$,
which is the QDFR with mismatch, Eq.~(\ref{eq:tasaki-crooks single history}).
For each history $\left(k,l\right)$ the logarithmic plot of the ratio
will provide a set of points which ideally lie in a straight line.
From each straight line one can independently obtain the inverse temperature
$\beta$ from the slopes. The points of interception with the vertical
axis give the quantities $I^{\left(k,l\right)}-\beta\Delta F^{\left(k\right)}$.
Since $\beta$ was determined from the corresponding slope and $\Delta F^{\left(k\right)}$
from the QDFR without control mismatch, one can infer the values of
the mutual information density $I^{\left(k,l\right)}$ from the interception
points. The nonidealities inevitably associated with actual experiments
are discussed in the next section when we consider a concrete example.

From our discussion at the beginning of this section, we explained
how to obtain (via an interferometric strategy) the four quantities
required to reconstruct the joint forward and backward characteristic
functions, $\chi_{F}\left(u,w,v\right)$ and $\chi_{B}\left(u,w,v\right)$.
One can infer the free energies $\Delta F^{\left(k\right)}$ and the
mutual information density $I^{\left(k,l\right)}$ and directly measure
$\chi_{F}^{\left(k,l\right)}\left(u\right)$ and $\chi_{B}^{\left(l,k\right)}\left(u\right)$
from the quantum circuits as explained. The forward, $P_{F}\left(W,\Delta F,I\right)$,
and backward, $P_{B}\left(W,\Delta F,I\right)$, joint PDFs can be
obtained from the inverse Fourier transform of the corresponding characteristic
functions. 

The logarithm of the ratio of the two PDFs $\ln P_{F}\left(W,\Delta F,I\right)/P_{B}\left(W,\Delta F,I\right)=\beta W-\beta\Delta F+I$,
which is the logarithm of the QDFR introduced in Eq.~(\ref{eq:tasaki-crooks whole process}).
As these PDFs depend on three stochastic variables $W$, $\Delta F$,
and $I$, the logarithmic plot of their ratio will be related to a
set of points in four dimensions. Moreover, denoting by $z=f\left(W,\Delta F,I\right)$
the logarithm of the ratio, these points should ideally lie in a hyperplane
given by the equation $\beta W-\beta\Delta F+I-z=0$. The parameters
$\beta$, $-\beta$, and $1$ that multiply the variables $W$, $\Delta F$,
and $I$, respectively, can be obtained using multiple linear regression. 

Further details are discussed in the illustrative example provided
in the next section. We note that the coefficient of the mutual information
density $I$ is independent of any parameter of the protocols. Consequently,
the analysis of how much this coefficient differs from $1$ using
the experimental data is a robust consistency test of the QDFR for
feedback control protocols.

\section{Illustrative Example: Qubit Controlled by Conditional Quenches\label{subsec:Example:-Qubit-Controlled}}

\begin{figure}
\includegraphics[scale=0.5]{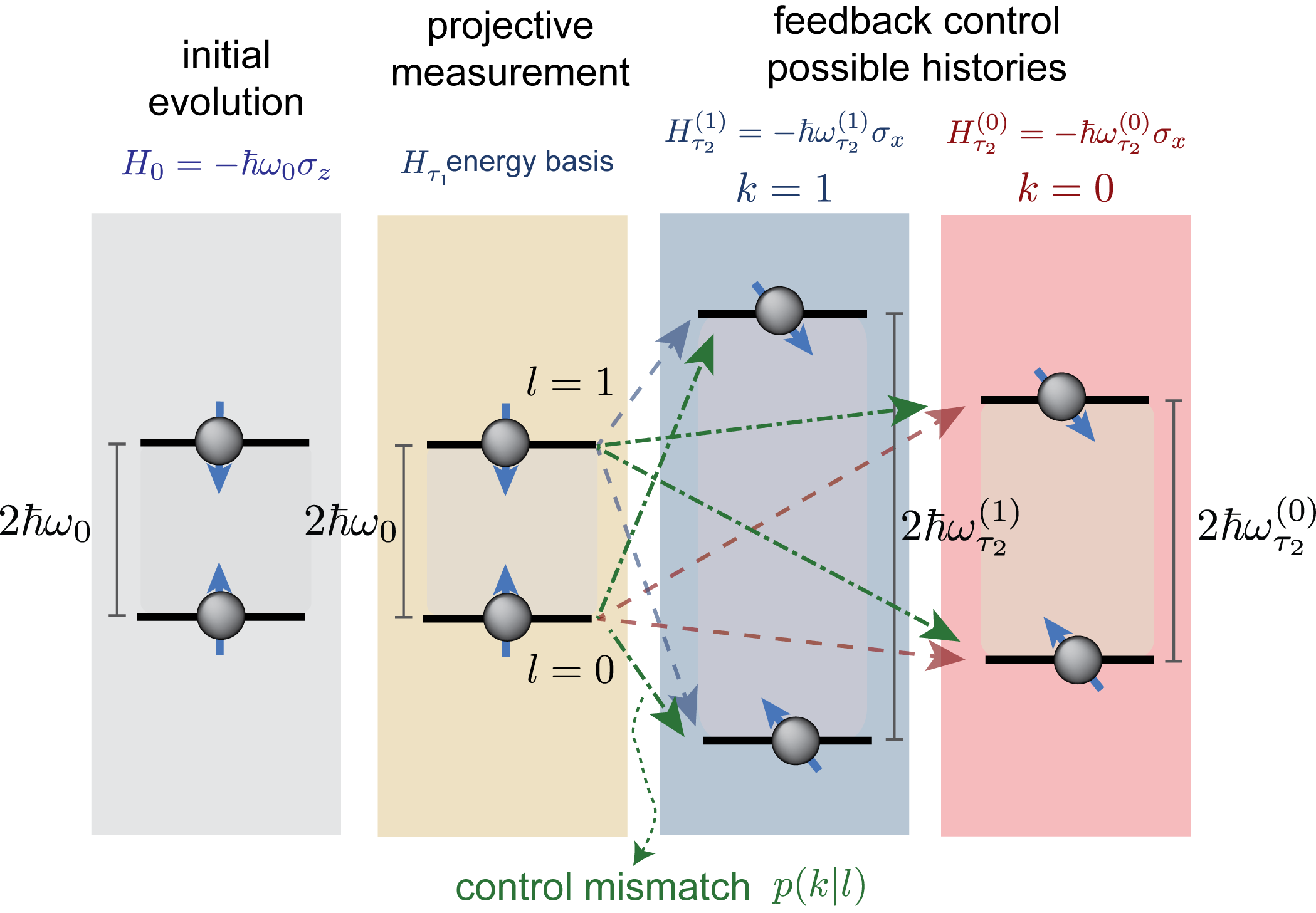}

\caption{Representation of the feedback control protocol considered in the
illustrative example. The system starts in the thermal state with
Hamiltonian $H_{0}=-\hbar\omega_{0}\sigma_{z}$ and inverse temperature
$\beta$. The Hamiltonian is kept constant up to time $\tau_{1}$
along a free evolution, $H_{\tau_{1}}=H_{0}$. At time $\tau_{1}$,
the feedback measurement is performed. We considered the instantaneous
Hamiltonian as the measurement observable. The feedback operation
corresponding to $k=0$ ($k=1$) is a sudden change of the Hamiltonian
from $H_{0\ensuremath{}}$ to $H_{\tau_{1}}^{\left(0\right)}=-\hbar\omega_{\tau_{2}}^{\left(0\right)}\sigma_{x}$
($H_{\tau_{1}}^{\left(1\right)}=-\hbar\omega_{\tau_{2}}^{\left(1\right)}\sigma_{x}$).
Whichever the case, the Hamiltonian is kept constant up to time $\tau_{2}$.
If there is no control mismatch, the red (blue) arrows correspond
to the possible transitions among the instantaneous eigenstates in
the TPM scheme corresponding to $l=k=0$ ($l=k=1$). The control mismatch
will be modeled by an $x$-rotation on the feedback memory changing
the reference basis for the feedback implementation from the computational
basis. The control mismatch probability is parametrized by a single
angle $\varphi$ through $p\left(k|l\right)=\Bra{k}R_{x}\left(\varphi\right)\Ket{l}$
and the transitions are depicted by the green arrows. \label{fig:Feedback-control-protocol.}}
\end{figure}

Let us consider a qubit which is controlled by a conditional sudden
quench, in a similar way to what was experimentally implemented in
a spin-1/2 system in Ref.~\cite{Camati2016}. After the measurement
stage of the feedback protocol, the system is driven in two different
ways, depending on the measurement outcome. Elements of the protocol
are outlined in Fig.~\ref{fig:Feedback-control-protocol.}. The qubit
starts in the Gibbs state with initial Hamiltonian $H_{0}=-\hbar\omega_{0}\sigma_{z}$.
In the pre-measurement part of the protocol, the Hamiltonian remains
constant up to time $\tau_{1}$, hence the evolution operator is given
by $U_{\tau_{1},0}=\exp\left\{ +i\omega_{0}\tau_{1}\sigma_{z}\right\} $
and the population of the systems state does not change. The associated
work of this process is zero since there is no external drive. Then,
a measurement of the Hamiltonian $H_{\tau_{1}}=H_{0}$ is performed
at time $\tau_{1}$. The measurement projectors are denoted by $\mathcal{M}_{l}=\Ket{l}\Bra{l}$,
where $\Ket{0}$ and $\Ket{1}$ are the eigenstates of $\sigma_{z}$
with positive and negative eigenvalues, respectively. 

The feedback control mismatch is modeled by a single parameter $\varphi$,
i.e., $p\left(k|l\right)=\Bra{k}R_{x}\left(\varphi\right)\Ket{l}$,
where $R_{x}\left(\varphi\right)=\exp\left\{ -i\varphi\sigma_{x}\right\} $
is an $x$ rotation. When $\varphi=0$ there is no mismatch between
the measurement basis and the reference basis which applies the feedback
operations. When the mismatch angle is different from zero, the measurement
basis (computational basis) will be different from the basis which
implements the feedback. In the latter case the reference basis for
the feedback is the basis obtained by performing the $x$ rotation
with an angle $\varphi$ onto the computational basis. 

The index $k$ represents the feedback operation being performed on
the postmeasurement state whereas $l$ is the actual measurement outcome.
Without control mismatch we would have $k=l$. If $k=0$, the Hamiltonian
is suddenly changed to $H_{\tau_{1}}^{\left(0\right)}=-\hbar\omega_{\tau_{2}}^{\left(0\right)}\sigma_{x}$
and then kept constant up to time $\tau_{2}$. Hence, the evolution
operator is $V_{\tau_{2},\tau_{1}}^{\left(0\right)}=\exp\left\{ +i\omega_{\tau_{2}}^{\left(0\right)}\left(\tau_{2}-\tau_{1}\right)\sigma_{x}\right\} $
and the final Hamiltonian is $H_{\tau_{2}}^{\left(0\right)}=H_{\tau_{1}}^{\left(0\right)}$.
If $k=1$, the Hamiltonian is suddenly changed to $H_{\tau_{1}}^{\left(1\right)}=-\hbar\omega_{\tau_{2}}^{\left(1\right)}\sigma_{x}$
and then kept constant up to time $\tau_{2}$. Hence, the evolution
operator is $V_{\tau_{2},\tau_{1}}^{\left(1\right)}=\exp\left\{ +i\omega_{\tau_{2}}^{\left(1\right)}\left(\tau_{2}-\tau_{1}\right)\sigma_{x}\right\} $
and the final Hamiltonian is $H_{\tau_{2}}^{\left(1\right)}=H_{\tau_{1}}^{\left(0\right)}$.
The difference between the two feedback operations is the energy gap
between the qubit eigenstates ($\hbar\omega_{\tau_{2}}^{\left(0\right)}$
and $\hbar\omega_{\tau_{2}}^{\left(1\right)}$). For a spin-1/2 particle
these sudden changes correspond to fast switches in the magnetic-field
direction and intensity. Furthermore, for the sake of illustration,
we assume that $\omega_{\tau_{2}}^{\left(0\right)}=2\omega_{0}$,
$\omega_{\tau_{2}}^{\left(1\right)}=3\omega_{0}$, and $5\hbar\omega_{0}=k_{B}T$. 

With the protocol properly defined, the quantum algorithms introduced
in the preceding section have to be performed in order to obtain the
quantities of interest. In the case in which there is no mismatch,
the characteristic work function for a single history is obtained
by performing the algorithm in Fig.~\ref{fig:single-history without error}.
In other words, the characteristic function will be encoded in the
auxiliary system $A$, and reads
\begin{align}
\chi_{F}^{\text{wcm}\left(k\right)}\left(u\right)= & \frac{1}{2}\frac{e^{-\beta E_{k}^{0}}}{Z_{0}}[e^{iu\left(E_{1}^{\left(k,\tau_{2}\right)}-E_{k}^{0}\right)}\nonumber \\
 & +e^{iu\left(E_{0}^{\left(k,\tau_{2}\right)}-E_{k}^{0}\right)}].\label{eq:19}
\end{align}
This characteristic function leads to the mixed work PDF 
\begin{align}
P_{F}^{\text{wcm}}\left(k\,;W\right)= & \frac{1}{2}\frac{e^{-\beta E_{k}^{0}}}{Z_{0}}[\delta\left[W-\left(E_{1}^{\left(k,\tau_{2}\right)}-E_{k}^{0}\right)\right]\nonumber \\
 & +\delta\left[W-\left(E_{0}^{\left(k,\tau_{2}\right)}-E_{k}^{0}\right)\right]].\label{eq:20}
\end{align}
In actual experimental realizations, the oscillatory complex functions
in Eq.~(\ref{eq:19}) may be restricted to some time interval and
the values of the variable $u$ are discrete with some suitable sampling
rate \cite{Batalhao2014}. This, in turn, will lead to a set of finite
width distributions for the PDFs (actual experimental data), in this
case Lorentz distributions, instead of delta functions (as in the
idealized theory). In what follows, we plot representative values
for the quantities with mismatch.

The backward process will be defined by the corresponding time-reversal
operator for a spin-1/2 particle \cite{Haake2010}, $\Theta=i\sigma_{y}K$,
where $K\left[\cdot\right]$ is the conjugation map. Therefore, $\tilde{H}\left(t\right)=i\sigma_{y}K\left[H\left(t\right)iK\left[\sigma_{y}\right]\right]=\sigma_{y}H\left(t\right)^{*}\sigma_{y}$,
where $H\left(t\right)$ is assumed to be written in the computation
basis and we used $\Theta^{\dagger}=iK\left[\sigma_{y}\right]=-i\sigma_{y}$.
Observe that $K\left[\cdot\right]$ has to be interpreted as a map
applied to everything after it. The backward Hamiltonians turn out
to be $\tilde{H}_{0}^{\left(0\right)}=\hbar\omega_{\tau_{2}}^{\left(0\right)}\sigma_{x}$,
$\tilde{H}_{0}^{\left(1\right)}=\hbar\omega_{\tau_{2}}^{\left(1\right)}\sigma_{x}$,
and $\tilde{H}_{\tau_{2}}=\hbar\omega_{0}\sigma_{z}$. 

Let us emphasize an important detail discussed at the end of Sec.~\ref{sec:Tasaki-Crooks-relation-without-feedback}.
Consider the qubit system is encoded in a spin-$1/2$ particle immersed
in an external magnetic field. A change in the magnetic-field direction
would change the sign of the Larmor frequency $\omega$. Therefore,
one could interpret the minus sign difference between the forward
$H$ and backward $\tilde{H}$ Hamiltonians to be a change in the
direction of the magnetic field. However, the time-reversal transformation
changes the sign of the spin operator, a property of the system, and
not the sign of the magnetic field, which is a property of the environment
where the system is immersed (the control system). Therefore, we conclude
that the time-reversal transformation has the same dynamical effect
on the spin as would be a change in the external magnetic-field direction.
Thus, in the actual experiment, the backward Hamiltonian can be implemented
by the reversal of the magnetic-field direction and its time modulation.
We emphasize that this is possible due to the dynamical equivalence
of the time-reversed Hamiltonian and the Hamiltonian with magnetic-field
direction reversed. This dynamical equivalence was exploited in Refs.~\cite{Batalhao2014,Batalhao2015}
to experimentally implement the backward protocol. 

The evolution operators of the two possible backward process are $\tilde{V}_{\tilde{\tau}_{1},0}^{\left(0\right)}=\exp\left\{ -i\omega_{\tau_{2}}^{\left(0\right)}\tilde{\tau}_{1}\sigma_{x}\right\} $,
$\tilde{V}_{\tilde{\tau}_{1},0}^{\left(1\right)}=\exp\left\{ -i\omega_{\tau_{2}}^{\left(1\right)}\tilde{\tau}_{1}\sigma_{x}\right\} $,
and $\tilde{U}_{\tau_{2},\tilde{\tau}_{1}}=\exp\left\{ -i\omega_{0}\tau_{1}\sigma_{z}\right\} $,
where $\tilde{\tau}_{1}=\tau_{2}-\tau_{1}$. To obtain the backward
characteristic work function without control mismatch the quantum
algorithm in Fig. \ref{fig:backward process} has to be implemented.
In the preceding section we discussed that, in the absence of control
mismatch, one has to consider only the situations when $l=k$. This
means that if the Gibbs state $\tilde{\rho}_{0}^{eq,\left(0\right)}$,
associated with the time-reversed initial Hamiltonian $\tilde{H}_{0}^{\left(0\right)}$,
is prepared, then only the values obtained for the ancilla observables
corresponding to the outcome of the memory measurement $l=k=0$ has
to be considered. Conversely, when the initial state is the Gibbs
state $\tilde{\rho}_{0}^{eq,\left(1\right)}$, associated with the
Hamiltonian $\tilde{H}_{0}^{\left(1\right)}$, the statistics of the
measurements on the ancilla when the memory measurement outcome is
$l=k=1$ have to be considered. In summary, an outcome-dependent postprocessing
of the data has to be done. This will provide the characteristic function
\begin{align}
\chi_{B}^{\text{wcm}\left(k\right)}\left(u\right)= & \frac{1}{2}\frac{e^{-\beta\tilde{E}_{0}^{\left(k\right)0}}}{\tilde{Z}_{0}^{\left(k\right)}}e^{iu\left(\tilde{E}_{k}^{\tau_{2}}-\tilde{E}_{0}^{\left(k\right)0}\right)}\nonumber \\
 & +\frac{1}{2}\frac{e^{-\beta\tilde{E}_{1}^{\left(k\right)0}}}{\tilde{Z}_{0}^{\left(k\right)}}e^{iu\left(\tilde{E}_{k}^{\tau_{2}}-\tilde{E}_{1}^{\left(k\right)0}\right)}.
\end{align}
This characteristic function leads to the mixed work PDF
\begin{align}
P_{B}^{\text{wcm}}\left(k\,;W\right)= & \frac{1}{2}\frac{e^{-\beta\tilde{E}_{0}^{\left(k\right)0}}}{\tilde{Z}_{0}^{\left(k\right)}}\delta\left[W-\left(\tilde{E}_{k}^{\tau_{2}}-\tilde{E}_{0}^{\left(k\right)0}\right)\right]\nonumber \\
 & +\frac{1}{2}\frac{e^{-\beta\tilde{E}_{1}^{\left(k\right)0}}}{\tilde{Z}_{0}^{\left(k\right)}}\delta\left[W-\left(\tilde{E}_{k}^{\tau_{2}}-\tilde{E}_{1}^{\left(k\right)0}\right)\right].\label{eq:22}
\end{align}

The next step is to take the ratio of the forward {[}Eq.~(\ref{eq:20}){]}
and backward {[}Eq.~(\ref{eq:22}){]} PDFs. In our theoretical example
this is simply done by taking the ratio of the coefficients preceding
the corresponding delta functions. However, the actual data would
consist, typically, of a sum of Lorentz distributions which possess
finite width. Therefore, an estimator for the best representative
value of the heights would have to be considered. For instance, an
estimator would simply to consider Lorentz distributions that fit
the experimental data. 

\begin{figure}
\includegraphics[scale=0.56]{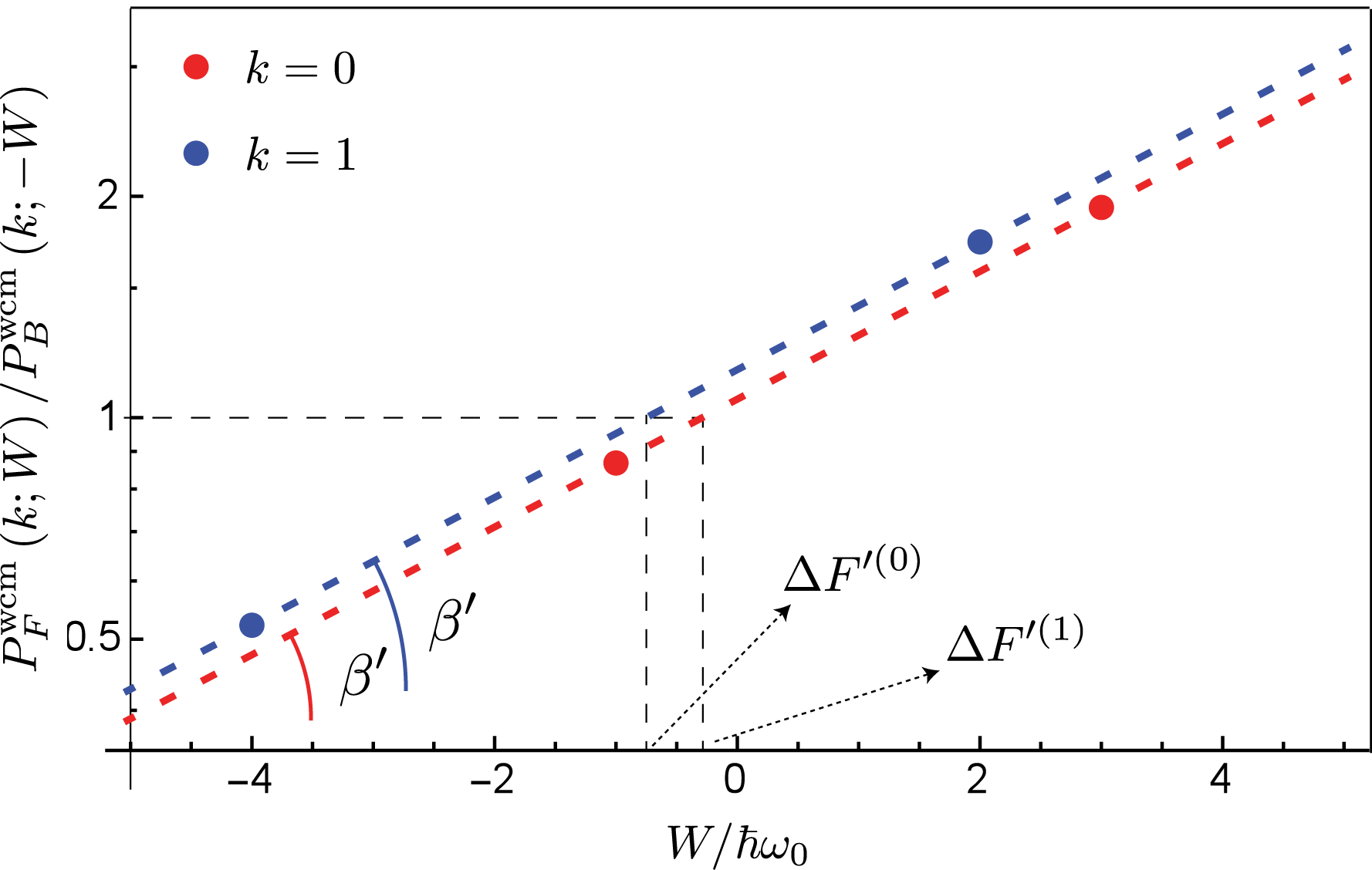}

\caption{Logarithmic plot of the QDFR for a single history without mismatch.
The four points show the possible transitions which may occur in the
forward process if the outcome of the measurement is $k=0$ (red)
or $k=1$ (blue). The straight dashed lines represent the linear fit
which should be related to the two functions $\beta W^{\left(k\right)}-\beta\Delta F^{\left(k\right)}$,
for both outcomes, $k=0$ (red) or $k=1$ (blue). From each curve,
one can obtain $\beta$ from the slope and $\Delta F^{\left(k\right)}$
from the work value whose image is $1$. Since in our analytical example
we plotted the normalized work variable $W/\hbar\omega_{0}$, the
slope actually provides $\beta'=\hbar\omega_{0}\beta$ and the values
whose image is $1$ are $\Delta F'^{\left(k\right)}=\Delta F^{\left(k\right)}/\hbar\omega_{0}$.
We assumed that $\omega_{\tau_{2}}^{\left(0\right)}=2\omega_{0}$,
$\omega_{\tau_{2}}^{\left(1\right)}=3\omega_{0}$, and $5\hbar\omega_{0}=k_{B}T$.
\label{fig:single-history-Tasaki-Crooks-withour-error}}

\end{figure}

In our example, the PDFs have two terms, so the ratio of the PDFs
provides two points for each value of $k$. In Fig.~\ref{fig:single-history-Tasaki-Crooks-withour-error}
we show the result of this ratio plotted on a logarithmic scale. From
the QDFR without mismatch for each $k$, the straight lines connecting
the corresponding pair of points should be equal to
\begin{equation}
\ln\frac{P_{F}^{\text{wcm}}\left(k\,;W\right)}{P_{B}^{\text{wcm}}\left(k\,;-W\right)}=\beta W-\beta\Delta F^{\left(k\right)}.\label{eq:23}
\end{equation}
The interception of the graph with the vertical axis should approximately
be $-\beta\Delta F^{\left(k\right)}$. Therefore, the free-energy
variation can be estimated, given that $\beta$ was obtained from
the slope of the best line fitting the data. However, the value of
$\beta$ may contain experimental errors which also propagate to this
estimation of the free energies $\Delta F^{\left(k\right)}$. A better
strategy to estimate the free energies is the following. When the
vertical axis is equal to one, the left-hand side of Eq.~(\ref{eq:23})
implies that $W=\Delta F^{\left(k\right)}$ (since we are using a
logarithmic scale). Therefore, for each $k$, the corresponding value
of $W$ when the image is $1$ provides an estimation of the corresponding
free-energy difference $\Delta F^{\left(k\right)}$. Such estimation
would encompass the experimental errors of the linear fit avoiding
the error propagation from the inverse temperature in the direct determination
of the linear coefficient, $-\beta\Delta F^{\left(k\right)}$, of
the fitting curve. 

Let us consider now the case with control mismatch. The forward characteristic
work function is obtained from the quantum algorithm in Fig.~\ref{fig:Single-history-forward-with-error}.
For our example this will give

\begin{figure}[b]
\includegraphics[scale=0.55]{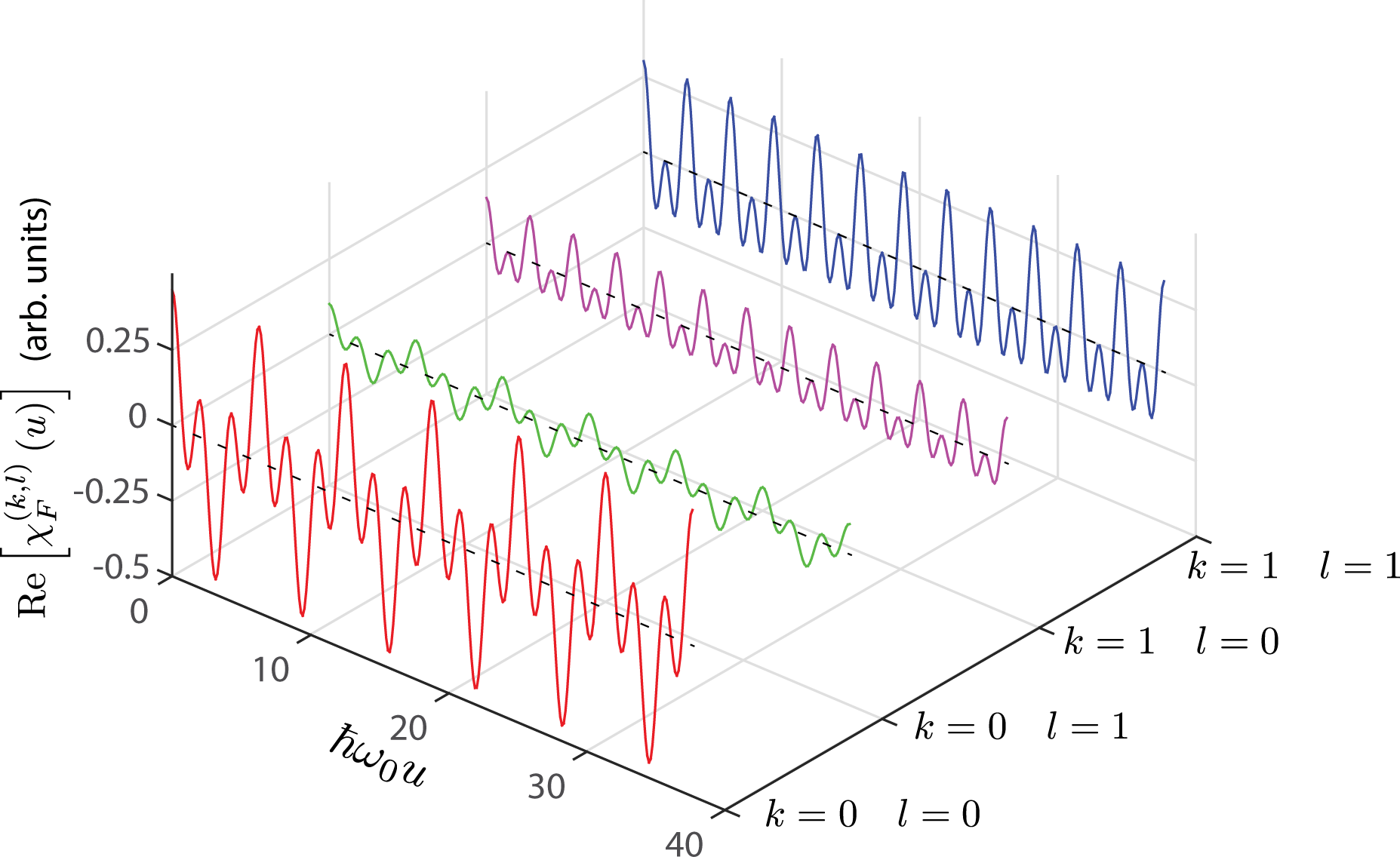}

\caption{Real part of the forward work characteristic function with control
mismatch, $\chi_{F}^{\left(k,l\right)}\left(u\right)$, for each history
$\left(k,l\right)$. We assumed that $\omega_{\tau_{2}}^{\left(0\right)}=2\omega_{0}$,
$\omega_{\tau_{2}}^{\left(1\right)}=3\omega_{0}$, $5\hbar\omega_{0}=k_{B}T$,
and $\varphi=\pi/3$.\label{fig:forward characteristic function }}
\end{figure}
\begin{align}
\chi_{F}^{\left(k,l\right)}\left(u\right)= & \frac{1}{2}p\left(k|l\right)\frac{e^{-\beta E_{l}^{0}}}{Z_{0}}[e^{iu\left(E_{1}^{\left(k,\tau_{2}\right)}-E_{l}^{0}\right)}\nonumber \\
 & +e^{iu\left(E_{0}^{\left(k,\tau_{2}\right)}-E_{l}^{0}\right)}].\label{eq:24}
\end{align}

\begin{figure}
\includegraphics[scale=0.55]{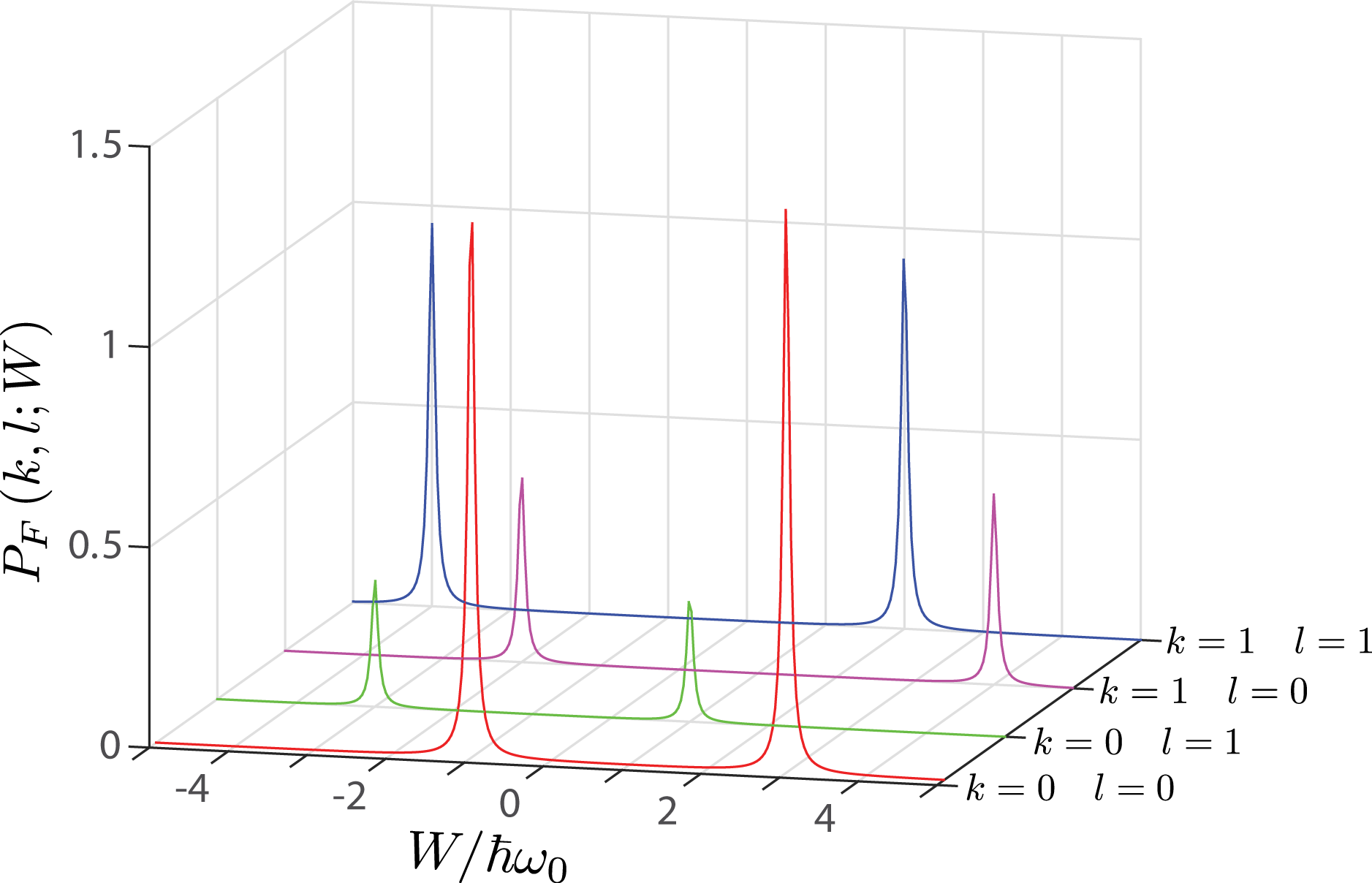}

\caption{Mixed work PDF with control mismatch, $P_{F}\left(k,l;W\right)$.
We plotted Lorentz distributions representing the inverse Fourier
transformation of some actual experimental data. We used $\omega_{\tau_{2}}^{\left(0\right)}=2\omega_{0}$,
$\omega_{\tau_{2}}^{\left(1\right)}=3\omega_{0}$, $5\hbar\omega_{0}=k_{B}T$,
and $\varphi=\pi/3$ in this example.\label{fig:Mixed-work-pdf} }
\end{figure}

In Fig.~\ref{fig:forward characteristic function } we show its real
part when the mismatch angle is $\varphi=\pi/3$. The corresponding
mixed work PDF is obtained by the inverse Fourier transform of the
measured forward characteristic work function and should be related
to 
\begin{align}
P_{F}\left(k,l\,;W\right)= & \frac{1}{2}p\left(k|l\right)\frac{e^{-\beta E_{l}^{0}}}{Z_{0}}[\delta\left[W-\left(E_{1}^{\left(k,\tau_{2}\right)}-E_{l}^{0}\right)\right]\nonumber \\
 & +\delta\left[W-\left(E_{0}^{\left(k,\tau_{2}\right)}-E_{l}^{0}\right)\right]].
\end{align}
In Fig.~\ref{fig:Mixed-work-pdf} we plot the four mixed work PDFs
when $\varphi=\pi/3$. 

Recall that we need two quantum algorithms to obtain the backward
characteristic work function with mismatch, $\chi_{B}^{\left(l,k\right)}\left(u\right)$.
The probability $p\left(k\right)$ is obtained from the circuit displayed
in Fig.~\ref{fig:joint probability distribution}, while the trace
$\mathcal{A}\left(k,l\right)$ in $\chi_{B}^{\left(l,k\right)}\left(u\right)$
{[}introduced in Eq.~(\ref{eq:backward characteristic work function single history}){]}
is obtained from the circuit in Fig.~\ref{fig:backward process}.
Together they will give
\begin{align}
\chi_{B}^{\left(l,k\right)}\left(u\right)= & \frac{1}{2}p\left(k\right)\frac{e^{-\beta\tilde{E}_{0}^{\left(k\right)0}}}{\tilde{Z}_{0}^{\left(k\right)}}e^{iu\left(\tilde{E}_{l}^{\tau_{2}}-\tilde{E}_{0}^{\left(k\right)0}\right)}\nonumber \\
 & +\frac{1}{2}p\left(k\right)\frac{e^{-\beta\tilde{E}_{1}^{\left(k\right)0}}}{\tilde{Z}_{0}^{\left(k\right)}}e^{iu\left(\tilde{E}_{l}^{\tau_{2}}-\tilde{E}_{1}^{\left(k\right)0}\right)}.\label{eq:26}
\end{align}
 The corresponding mixed work PDF is
\begin{align}
P_{B}\left(l,k\,;W\right)= & \frac{1}{2}p\left(k\right)\frac{e^{-\beta\tilde{E}_{0}^{\left(k\right)0}}}{\tilde{Z}_{0}^{\left(k\right)}}\delta\left[W-\left(\tilde{E}_{l}^{\tau_{2}}-\tilde{E}_{0}^{\left(k\right)0}\right)\right]\nonumber \\
 & +\frac{1}{2}p\left(k\right)\frac{e^{-\beta\tilde{E}_{1}^{\left(k\right)0}}}{\tilde{Z}_{0}^{\left(k\right)}}\delta\left[W-\left(\tilde{E}_{l}^{\tau_{2}}-\tilde{E}_{1}^{\left(k\right)0}\right)\right].
\end{align}

\begin{figure}
\includegraphics[scale=0.6]{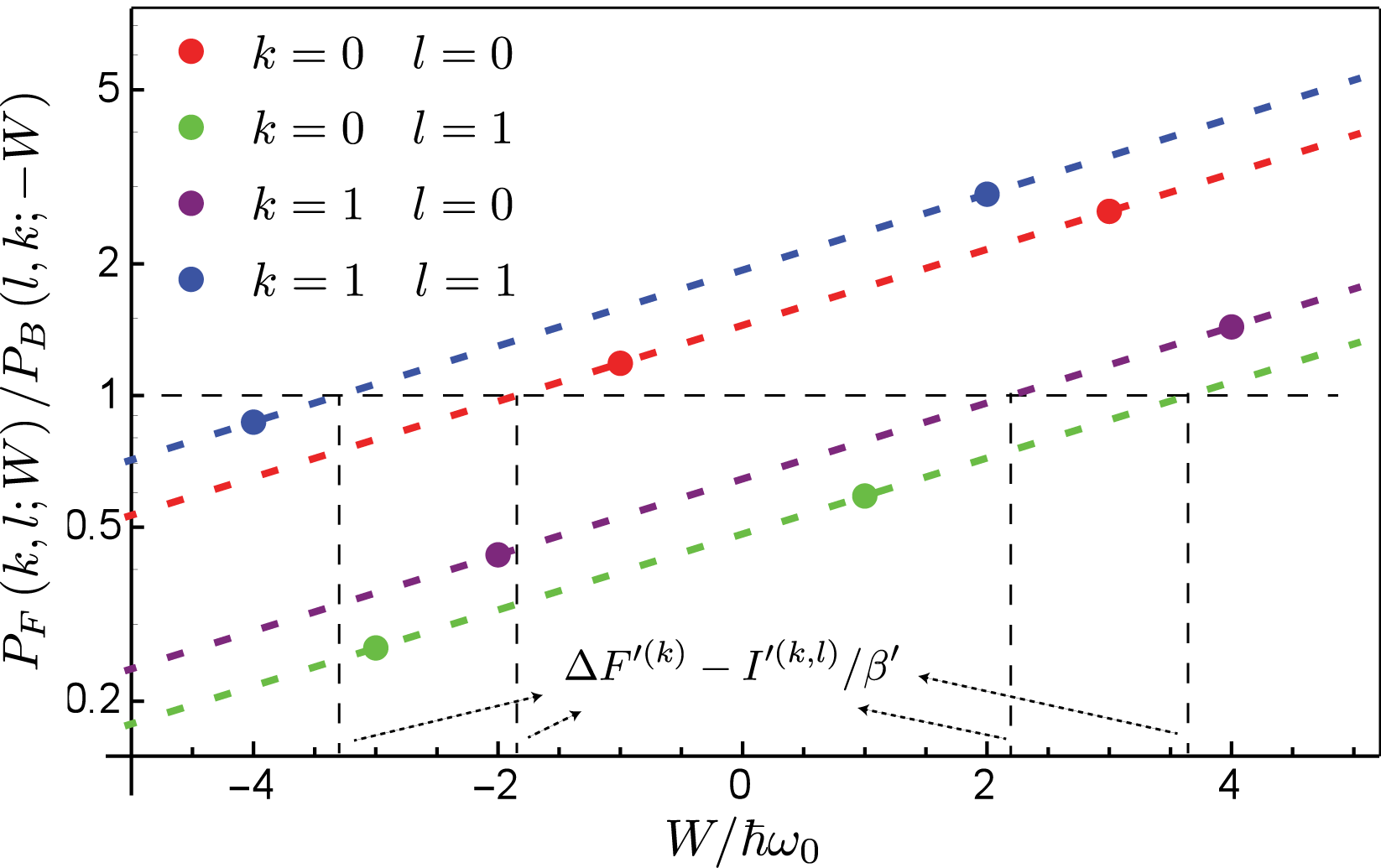}

\caption{Logarithmic plot of the QDFR for a single history with mismatch. The
eight points show the possible transitions which may occur in the
forward process if the outcome of the measurement and value of the
feedback implemented are, respectively, ($k=0$, $l=0$) (red), ($k=0$,
$l=1$) (green), ($k=1$, $l=0$) (purple), and ($k=1$, $l=1$) (blue).
The straight lines represent the functions $\beta W^{\left(k\right)}-\beta\Delta F^{\left(k\right)}+I^{\left(k,l\right)}$.
For each graph one can obtain another estimate for $\beta$ from their
slope. Assuming the knowledge of $\beta$ and $\Delta F^{\left(k\right)}$,
the mutual information density $I^{\left(k,l\right)}$ is obtained
from the work value whose image is $1$. Since in our analytical example
we plotted the normalized work variable $W/\hbar\omega_{0}$, the
slope actually gives $\beta'=\hbar\omega_{0}\beta$ and the values
whose image is $1$ are $\Delta F'^{\left(k\right)}+I'^{\left(k,l\right)}/\beta'$,
where $\Delta F'^{\left(k\right)}=\Delta F^{\left(k\right)}/\hbar\omega_{0}$
and $I'^{\left(k,l\right)}/\hbar\omega_{0}$. We assumed that $\omega_{\tau_{2}}^{\left(0\right)}=2\omega_{0}$,
$\omega_{\tau_{2}}^{\left(1\right)}=3\omega_{0}$, $5\hbar\omega_{0}=k_{B}T$,
and $\varphi=\pi/3$. \label{fig:single-history-Tasaki-Crooks-with-error}}
\end{figure}

To employ the QDFR with control mismatch for a single history $\left(k,l\right)$
{[}Eq.~(\ref{eq:tasaki-crooks single history}){]} the ratio of forward
and backward mixed work PDFs has to be taken. As we did before, we
take the ratio of the factors preceding the delta functions. Since
in our example the PDF of each $\left(k,l\right)$ history has only
two terms, two points are obtained. As there are four possible histories
there will be four straight lines as presented in Fig.~\ref{fig:single-history-Tasaki-Crooks-with-error}.
The QDFR with mismatch implies that the straight lines are given by
\begin{equation}
\ln\frac{P_{F}\left(k,l\,;W\right)}{P_{B}\left(l,k\,;-W\right)}=\beta W-\beta\Delta F^{\left(k\right)}+I^{\left(k,l\right)}.\label{eq:28}
\end{equation}

From the slope of these straight lines the inverse temperature $\beta$
can be estimated again (that can be used as a consistency test for
the measured data). This estimation is independent of the previous
one because the algorithm employed is different. From the same reasoning
as before, the work value $W=\Delta F^{\left(k\right)}-\beta^{-1}I^{\left(k,l\right)}$
corresponds to the image $1$ in the logarithmic plot. If the free-energy
differences $\Delta F^{\left(k\right)}$ have been obtained from the
analysis without mismatch, the mutual information density $I^{\left(k,l\right)}$
can be estimated from Fig.~\ref{fig:single-history-Tasaki-Crooks-with-error}.
An independent estimation for the mutual information density could
be realized using the circuit in Fig.~\ref{fig:joint probability distribution}.
That circuit provides the joint distribution $p\left(k,l\right)$
from which the mutual information density $I^{\left(k,l\right)}=\ln\nicefrac{p\left(k,l\right)}{p\left(k\right)p\left(l\right)}$
can be directly computed.

\begin{figure}[b]
\includegraphics[scale=0.6]{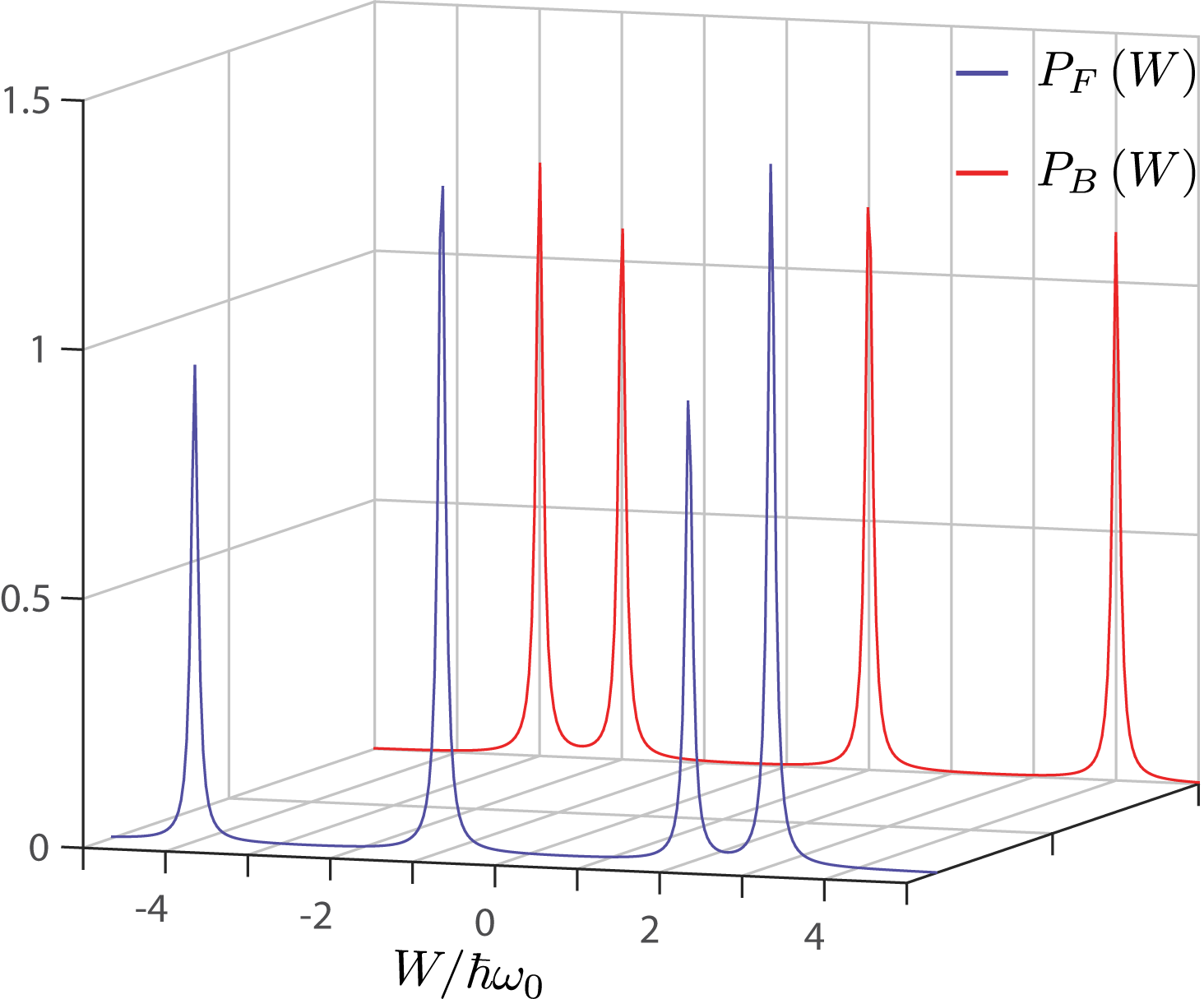}

\caption{Plot of the forward (red) and backward (blue) work PDFs for the feedback
process with control mismatch. We assumed that $\omega_{\tau_{2}}^{\left(0\right)}=2\omega_{0}$,
$\omega_{\tau_{2}}^{\left(1\right)}=3\omega_{0}$, $5\hbar\omega_{0}=k_{B}T$,
and $\varphi=\pi/3$.\label{fig:forward and backward work pdf}}
\end{figure}

Before we proceed with the method, we plotted the forward, $P_{F}\left(W\right)$,
and backward, $P_{B}\left(W\right)$, work PDF in Fig.~\ref{fig:forward and backward work pdf}.
Without feedback control these PDFs characterize the irreversibility
of the process. Here they only characterize the work distribution
and thus can be used to obtain the moments of the work stochastic
variable, such as the average work. The irreversibility for processes
with feedback is characterized by the joint PDFs, $P_{F}\left(W,\Delta F,I\right)$
and $P_{B}\left(W,\Delta F,I\right)$, which are the quantities that
our method is able to obtain from the experimental data.

Hitherto, the method provided several estimates of the inverse temperature
$\beta$: two possible ways to estimate each free-energy difference
$\Delta F^{\left(k\right)}$, one from the QDFR without mismatch for
a single history $k$ and one in the presence of mismatch (since we
have another way to estimate $I^{\left(k,l\right)}$), and two estimations
of the mutual information density $I^{\left(k,l\right)}$, one from
the QDFR with mismatch for a single history $\left(k,l\right)$ and
the other from direct computation using the probability distribution
$p\left(k,l\right)$. With all this information, the final quantities
for the verification of the QDFR with feedback control can be computed.
Different routes can be chosen depending on which kind of control
and information are accessed in a particular experimental setup. 

As discussed in the preceding section, the quantities obtained so
far are enough to reconstruct the forward and backward joint characteristic
functions, $\chi_{F}\left(W,\Delta F,I\right)$ and $\chi_{B}\left(W,\Delta F,I\right)$.
The multidimensional inverse Fourier transform of these quantities
will give the joint PDFs $P_{F}\left(W,\Delta F,I\right)$ and $P_{B}\left(W,\Delta F,I\right)$,
respectively. Each joint PDF will be a sum of delta functions (or
Lorentz distributions typically in the experimental analysis) {[}see
Eqs.~(\ref{eq:S47}) and (\ref{eq:S49}){]}. In our example, they
contain eight terms. The logarithm of the ratio of these PDFs will
provide eight points in a four-dimensional space whose axes are comprised
of the work, free-energy difference, and mutual information density
random variables. When plotted in logarithmic scale, these points
should lie in a hyperplane in four dimensions given by 
\begin{equation}
\ln\frac{P_{F}\left(W,\Delta F,I\right)}{P_{B}\left(-W,-\Delta F,I\right)}=\beta\left(W-\Delta F\right)+I.\label{eq:eq29}
\end{equation}
From the experimental data, one would have to use statistical methods
to estimate the best hyperplane representing these points.

The coefficients preceding the variables $W$, $\Delta F$, and $I$
in Eq.~(\ref{eq:eq29}) can be obtained from a multiple linear regression.
These values should be $\beta$, $-\beta$, and $1$, respectively.
These values for inverse temperature could be compared to all the
independent estimations inferred previously. The fact that the slope
in the $I$ direction is independent of any parameter of the system
or the protocol provides a robust consistency test to verify the QDFR. 

\begin{figure}[H]
\includegraphics[scale=0.6]{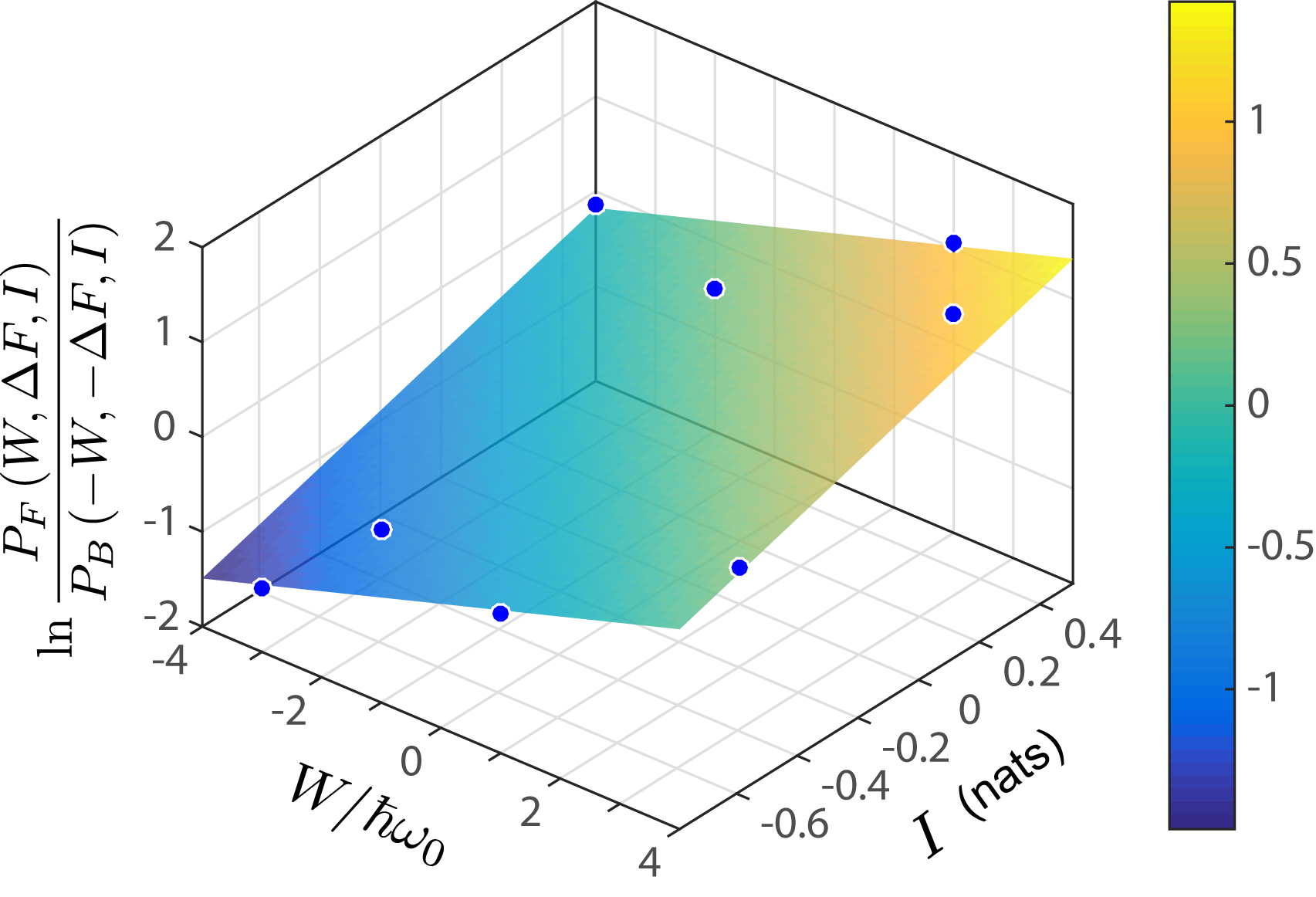}

\caption{Logarithmic plot of the QDFR for feedback processes. For this plot
we assumed $\omega_{\tau_{2}}^{\left(0\right)}=\omega_{\tau_{2}}^{\left(1\right)}=2\omega_{0}$,
$5\hbar\omega_{0}=k_{B}T$, and $\varphi=\pi/3$. With these parameters
the free energy is independent of the feedback control index $k$,
$\Delta F^{\left(0\right)}=\Delta F^{\left(1\right)}=\Delta F$. In
this case, the QDFR can be verified by a linear regression technique.
\label{fig:whole-process-Tasaki-Crooks}}
\end{figure}

Some information would necessarily be lost when we plot a projection
of this hyperplane associated with Eq.~(\ref{eq:eq29}) in three
dimensions. Therefore, just for the next plot of our example, we assumed
that the two gaps in quenched feedback control evolutions are equal,
i.e., $\omega_{\tau_{2}}^{\left(0\right)}=\omega_{\tau_{2}}^{\left(1\right)}=2\omega_{0}$.
In this way, the free-energy difference becomes independent of the
feedback $k$th protocol, $\Delta F^{\left(0\right)}=\Delta F^{\left(1\right)}=\Delta F$,
and the whole feedback process can be fully characterized by a plane
in three dimensions. The result of this choice is illustrated in Fig.~\ref{fig:whole-process-Tasaki-Crooks}.
In that case, where the final gaps are equal but the final Hamiltonians
may possibly be different, the coefficients can be estimated as follows.
The slope with respect to $W$ estimates the value of the inverse
temperature $\beta$. The value of $\Delta F$ can be obtained by
the point of interception of the plane with the $z$ axis, $-\beta\Delta F$,
since $\beta$ was already obtained. The slope with respect to $I$
should be equal to $1$, i.e., independent of any parameter of the
protocols. Therefore, the deviation of the slope of $I$ from $1$
is a robust way to verify the QDFR with feedback control.

\section{CONCLUSIONS\label{sec:Conclusions}}

We have presented a detailed discussion of the backward protocol and
showed that the QDFRs remain valid even when the dynamics is not time-reversal
invariant. Our discussion emphasizes that system's external parameters,
such as an external magnetic field, should not change under the time-reversal
transformation. Furthermore, we derived two QDFRs in the presence
of feedback control which apply to each single history after the measurement
stage of the control protocol. 

Employing the discussed formalism, we also provided a derivation of
the QDFR for the whole discrete feedback process. Additionally, we
introduced forward and backward mixed joint PDFs. These PDFs incorporate
the relevant statistical information of the forward and backward protocols
which are instrumental for an experimental verification of QDFRs.
The PDFs appearing in the QDFRs can be obtained from these mixed PDFs
as marginal distributions. In this sense, the mixed PDFs introduced
here provide a unified way to obtain the complete set of PDFs related
to the QDFRs in the presence of feedback control. These theoretical
developments were applied to obtain our main result. 

We proposed a systematic method to experimentally verify the QDFR
for discrete feedback-controlled quantum dynamics. The method employs
quantum interferometric circuits to measure characteristic functions
associated with work distribution in the presence of feedback control.
The general strategy involves one ancillary qubit that encodes information
of the characteristic function related to a PDF. The feedback mechanism
is modeled as a quantum system that carries two quantum memories.
Nonetheless, the general ideas introduced throughout the paper can
be applied to different feedback control mechanism or to a semiclassical
device where the feedback control is processed by classical means. 

The QDFRs for feedback processes provide a valuable contribution to
quantum information thermodynamics. They quantify the irreversibility
of such processes, allowing for generalizations of the second law
in the presence of feedback. The algorithm introduced here is experimentally
feasible with the current technology for quantum control, for example,
in the NMR setup \cite{Camati2016,Batalhao2014,Batalhao2015,Kaonan-2017,Peterson2018}
or in a circuit QED setting \cite{Cottet2017,Masuyama2017,Naghiloo2018}.
We hope that our findings inspire new experimental efforts towards
a detailed description of the role played by information in the feedback-controlled
nonequilibrium quantum dynamics. 
\begin{acknowledgments}
We acknowledge financial support from UFABC, CNPq, CAPES, and FAPESP.
R.M.S. gratefully acknowledges financial support from the Royal Society
through the Newton Advanced Fellowship scheme (Grant no.~NA140436).
This research was performed as part of the Brazilian National Institute
of Science and Technology for Quantum Information. 

\setcounter{section}{0}
\global\long\def\thesection{APPENDIX \Alph{section}}
\end{acknowledgments}

\section{Quantum Detailed Fluctuation Relation Without Feedback\label{sec:Appendix-A:-Quantum}}

\setcounter{figure}{0}
\setcounter{equation}{0}
\global\long\def\thefigure{A\arabic{figure}}
\global\long\def\theequation{A\arabic{equation}}

The work PDF in the driven forward protocol without feedback control
may be written as \cite{Campisi2011,Talkner2007}
\begin{equation}
P_{F}\left(W\right)=\sum_{m,n}p\left(m,n\right)\delta\left[W-W_{mn}\right],\label{eq:forward work pdf}
\end{equation}
where $p\left(m,n\right)=\mbox{Tr}\left[P_{m}^{\tau}U_{\tau,0}P_{n}^{0}U_{\tau,0}^{\dagger}\right]\mbox{Tr}\left[P_{n}^{0}\rho_{0}^{eq}\right]$,
with $\rho_{0}^{eq}=e^{-\beta\left(H_{0}-F_{0}\right)}$. Introducing
identity operators represented in terms of the time-reversal operator,
$\Theta^{\dagger}\Theta=\mathds{1}$, in the expression for the joint
distribution $p\left(m,n\right)$, one obtains 
\begin{align}
p\left(m,n\right) & =\mbox{Tr}\left[\Theta P_{m}^{\tau}\Theta^{\dagger}\Theta U_{\tau,0}\Theta^{\dagger}\Theta P_{n}^{0}\Theta^{\dagger}\Theta U_{\tau,0}^{\dagger}\Theta^{\dagger}\right]\nonumber \\
 & \times e^{-\beta\left(E_{n}^{0}-F_{0}\right)},\label{eq:S2}
\end{align}
where we also used $\mbox{Tr}\left[P_{n}^{0}\rho_{0}^{eq}\right]=e^{-\beta\left(E_{n}^{0}-F_{0}\right)}$.
Applying the relations between forward and backward protocols discussed
in Sec.~II, Eq.~(\ref{eq:S2}) can be rewritten as
\begin{align}
p\left(m,n\right) & =\mbox{Tr}\left[\tilde{P}_{n}^{\tau}\tilde{U}_{\tau,0}\tilde{P}_{m}^{0}\tilde{U}_{\tau,0}^{\dagger}\right]\text{Tr}\left[\tilde{P}_{m}^{0}\tilde{\rho}_{0}^{eq}\right]\nonumber \\
 & \times e^{-\beta\left(E_{n}^{0}-F_{0}\right)}e^{+\beta\left(\tilde{E}_{m}^{0}-\tilde{F}_{0}\right)},
\end{align}
where we have multiplied and divided by the occupation probability
of the initial eigenstates in the backward protocol, $\tilde{p}\left(m\right)=\text{Tr}\left[\tilde{\Pi}_{m}^{0}\tilde{\rho}_{0}^{eq}\right]=e^{-\beta\left(\tilde{E}_{m}^{0}-\tilde{F}_{0}\right)}$.
One can identify the product of the traces in the above equation as
the joint probability $\tilde{p}\left(n,m\right)$ of the backward
process. Using the relations of the eigenenergies from Sec.~II, $\tilde{E}_{m}^{0}=E_{n}^{\tau}$,
and $\tilde{F}_{0}=F_{\tau}$, one can write
\begin{equation}
p\left(m,n\right)=e^{+\beta\left(W_{mn}-\Delta F\right)}\tilde{p}\left(n,m\right).
\end{equation}
With the above result, Eq.~(\ref{eq:forward work pdf}) becomes
\begin{align}
P_{F}\left(W\right)= & \sum_{m,n}e^{+\beta\left(W_{mn}-\Delta F\right)}\tilde{p}\left(m,n\right)\nonumber \\
 & \times\delta\left[W-W_{mn}\right].
\end{align}
Removing the exponential from the sum and using the relation of the
eigenenergies again, $\tilde{W}_{nm}=-W_{mn}$, we obtain the QDFR
\begin{align}
P_{F}\left(W\right) & =e^{+\beta\left(W-\Delta F\right)}\sum_{m,n}\tilde{p}\left(m,n\right)\delta\left[\left(-W\right)-\tilde{W}_{nm}\right]\nonumber \\
 & =e^{+\beta\left(W-\Delta F\right)}P_{B}\left(-W\right),
\end{align}
without assuming time-reversal invariance of the Hamiltonian. 

\section{Single-History QDFR with control mismatch\label{sec:Appendix-B:-Single-History}}

\setcounter{figure}{0}
\setcounter{equation}{0}
\global\long\def\thefigure{B\arabic{figure}}
\global\long\def\theequation{B\arabic{equation}}

The forward and backward protocol with control mismatch was discussed
in the main text. Here, we describe the steps to demonstrate the QDFR
with control mismatch, Eq.~(\ref{eq:tasaki-crooks single history}).
The forward mixed joint PDF (introduced in the main text) is given
by 

\begin{align}
P_{F}\left(k,l\,;W,\Delta F,I\right) & =\sum_{m,n}p\left(m^{\left(k\right)},k,l,n\right)\nonumber \\
 & \times\delta\left[W-W_{mkn}\right]\delta\left[\Delta F-\Delta F^{\left(k\right)}\right]\nonumber \\
 & \times\delta\left[I-I^{\left(k,l\right)}\right],\label{eq:mixed joint pdf}
\end{align}
where 
\begin{equation}
p\left(m^{\left(k\right)},k,l,n\right)=p\left(m^{\left(k\right)}|l\right)p\left(k|l\right)p\left(l|n\right)p\left(n\right)
\end{equation}
are quantities defined previously in the main text. This mixed PDF
contains all the relevant statistical information of the forward protocol.
By integrating over the proper variables this mixed joint PDF encompasses
\begin{align}
P_{F}\left(k,l\right) & =\iiint dWd\left(\Delta F\right)dI\,P_{F}\left(k,l;W,\Delta F,I\right)\nonumber \\
 & =p\left(k,l\right),
\end{align}
the joint probability distribution of the $\left(k,l\right)$ history,
and
\begin{align}
P_{F}\left(\Delta F\right) & =\sum_{k,l}\iint dWdI\,P_{F}\left(k,l;W,\Delta F,I\right)\nonumber \\
 & =\sum_{k}p\left(k\right)\delta\left[\Delta F-\Delta F^{\left(k\right)}\right],
\end{align}
the free-energy difference PDF, consistently showing that the probability
that the $k$th value of the free-energy difference $\Delta F^{\left(k\right)}$
occurs is $p\left(k\right)$, i.e., the probability that the $k$th
final Hamiltonian, $H_{\tau_{2}}^{\left(k\right)}$, occurs in the
feedback protocol. The mutual information density PDF
\begin{equation}
P_{F}\left(I\right)=\sum_{k,l}p\left(k,l\right)\delta\left[I-I^{\left(k,l\right)}\right]
\end{equation}
also consistently shows that the probability that the mutual information
density, $I^{\left(k,l\right)}$, occurs is the probability of the
$\left(k,l\right)$ history takes place and finally we can write 
\begin{equation}
P_{F}\left(W\right)=\sum_{m,k,l,n}p\left(m^{\left(k\right)},k,l,n\right)\delta\left[W-W_{mkn}\right],
\end{equation}
the probability of obtaining the instantaneous eigenenergies difference
$W_{mkn}=E_{m}^{\left(k\right)\tau_{2}}-E_{n}^{0}$ in the TPM scheme.
This last PDF is consistent with the forward protocol depicted in
Fig.~\ref{fig:feedback histories}, since the average work associated
with it is given by 
\begin{align}
\left\langle W\right\rangle  & =\int dW\,WP_{F}\left(W\right)\nonumber \\
 & =\sum_{k,l}p\left(k,l\right)\left[\mathrm{\mathcal{U}}\left(\rho_{\tau_{2}}^{\left(k,l\right)}\right)-\mathrm{\mathcal{U}}\left(\rho_{0}^{eq}\right)\right].
\end{align}
The average work for each history $\left(k,l\right)$ is given by
the internal energy difference, $\mathrm{\mathcal{U}}\left(\rho_{\tau_{2}}^{\left(k,l\right)}\right)-\mathrm{\mathcal{U}}\left(\rho_{0}^{eq}\right)$
{[}with $\mathrm{\mathcal{U}}\left(\rho_{\tau_{2}}^{\left(k,l\right)}\right)=\text{Tr}\left(\rho_{\tau_{2}}^{\left(k,l\right)}H_{\tau_{2}}^{\left(k\right)}\right)$
and $\mathrm{\mathcal{U}}\left(\rho_{0}^{eq}\right)=\text{Tr}\left(\rho_{0}^{eq}H_{0}\right)${]},
and it occurs with probability $p\left(k,l\right)$. Therefore, the
above average work is the mean value over all histories $\left(k,l\right)$
of each single-history average work.

To show the QDFR we employ the forward mixed work PDF 
\begin{align}
P_{F}\left(k,l\,;W\right)= & \sum_{m,n}p\left(m^{\left(k\right)},k,l,n\right)\nonumber \\
 & \times\delta\left[W-W_{mkn}\right].\label{eq:forward mixed work pdf}
\end{align}
 Inserting identities, $\Theta^{\dagger}\Theta=\mathds{1}$, in the
expression for the conditional probability for obtaining the $m$th
instantaneous energy state of $H_{\tau_{2}}^{\left(k\right)}$ in
the $k$th history given the $l$th outcome in the feedback measurement,
$p\left(m^{\left(k\right)}|l\right)$, and the conditional probability
for obtaining the $l$th outcome in the feedback measurement given
the $n$th initial energy state, $p\left(l|n\right)$, one obtains
the relation between the forward and backward conditional probabilities

\begin{align}
p\left(m^{\left(k\right)}|l\right) & =\mbox{Tr}\left[\Theta P_{m}^{\left(k\right)\tau_{2}}\Theta^{\dagger}\Theta V_{\tau_{2},\tau_{1}}^{\left(k\right)}\Theta^{\dagger}\Theta\mathcal{M}_{l}\Theta^{\dagger}\Theta V_{\tau_{2},\tau_{1}}^{\left(k\right)\dagger}\Theta^{\dagger}\right]\nonumber \\
 & =\mbox{Tr}\left[\tilde{\mathcal{M}}_{l}\tilde{V}_{\tilde{\tau}_{1},0}^{\left(k\right)}\tilde{P}_{m}^{\left(k\right)0}\tilde{V}_{\tilde{\tau}_{1},0}^{\left(k\right)\dagger}\right]\nonumber \\
 & =\tilde{p}\left(l|m^{\left(k\right)}\right)
\end{align}
and 

\begin{align}
p\left(l|n\right) & =\mbox{Tr}\left[\Theta\mathcal{M}_{l}\Theta^{\dagger}\Theta U_{\tau_{1},0}\Theta^{\dagger}\Theta P_{n}^{0}\Theta^{\dagger}\Theta U_{\tau_{1},0}^{\dagger}\Theta\right]\nonumber \\
 & =\mbox{Tr}\left[\tilde{P}_{n}^{\tau_{2}}\tilde{U}_{\tau_{2},\tilde{\tau}_{1}}\tilde{\mathcal{M}}_{l}\tilde{U}_{\tau_{2},\tilde{\tau}_{1}}^{\dagger}\right]\nonumber \\
 & =\tilde{p}\left(n|l\right),
\end{align}
where $\tilde{\tau}_{1}=\tau_{2}-\tau_{1}$ and $\tilde{\mathcal{M}}=\Theta\mathcal{M}\Theta^{\dagger}$
is the time-reversed observable. Therefore, 
\begin{align}
p\left(m^{\left(k\right)},k,l,n\right) & =e^{I^{\left(k,l\right)}}\frac{p\left(n\right)}{\tilde{p}\left(m^{\left(k\right)}\right)}\tilde{p}\left(l|m^{\left(k\right)}\right)\nonumber \\
 & \times\tilde{p}\left(n|l\right)\tilde{p}\left(m^{\left(k\right)}\right)p\left(k\right),
\end{align}
where we introduced the probabilities $\tilde{p}\left(m^{\left(k\right)}\right)=e^{-\beta\left(\tilde{E}_{m}^{\left(k\right)0}-\tilde{F}_{0}^{\left(k\right)}\right)}$
and the mutual information density $I^{\left(k,l\right)}=\ln\nicefrac{p\left(k|l\right)}{p\left(k\right)}$.
Using the relation between forward and backward instantaneous eigenenergies,
$\tilde{E}_{n}^{\left(k\right)t}=E_{n}^{\left(k\right)\tau_{2}-t}$
and $\tilde{F}_{t}^{\left(k\right)}=F_{\tau_{2}-t}^{\left(k\right)}$,
the ratio between the initial energy state probabilities will be given
by $\nicefrac{p\left(n\right)}{\tilde{p}\left(m\right)}=e^{\beta\left(W_{mkn}-\Delta F^{\left(k\right)}\right)}$.
One can identify the remaining probabilities as associated with some
backward protocol so that 
\begin{align}
\tilde{p}\left(n,l,m^{\left(k\right)},k\right)= & \tilde{p}\left(l|m^{\left(k\right)}\right)\tilde{p}\left(n|l\right)\tilde{p}\left(m^{\left(k\right)}\right)\nonumber \\
 & \times p\left(k\right).
\end{align}
Therefore, 
\begin{align}
p\left(m^{\left(k\right)},k,l,n\right) & =e^{\beta\left(W_{mkn}-\Delta F^{\left(k\right)}\right)+I^{\left(k,l\right)}}\nonumber \\
 & \times\tilde{p}\left(n,l,m^{\left(k\right)},k\right).\label{eq:forward backward relation}
\end{align}
Substituting Eq.~(\ref{eq:forward backward relation}) into Eq.~(\ref{eq:forward mixed work pdf}),
we obtain 
\begin{align}
P_{F}\left(k,l\,;W\right) & =\sum_{m,n}e^{\beta\left(W_{mkn}-\Delta F^{\left(k\right)}\right)+I^{\left(k,l\right)}}\tilde{p}\left(n,l,m^{\left(k\right)},k\right)\nonumber \\
 & \times\delta\left[W-W_{mkn}\right].
\end{align}
Using the delta function to extract the exponential of work from the
sum and identifying the backward mixed work PDF, using $W_{mkn}=-\tilde{W}_{nkm}$,
we end up with the QDFR written as
\begin{equation}
P_{F}\left(k,l\,;W\right)=e^{\beta\left(W-\Delta F^{\left(k\right)}\right)+I^{\left(k,l\right)}}P_{B}\left(l,k\,;-W\right),\label{eq:S21}
\end{equation}
where 
\begin{align}
P_{B}\left(l,k\,;W\right)= & \sum_{n,m}\tilde{p}\left(n,l,m^{\left(k\right)},k\right)\nonumber \\
 & \times\delta\left[W-\tilde{W}_{nkm}\right].\label{eq:backward mixed work pdf}
\end{align}
The mixed work PDF in Eq. (\ref{eq:backward mixed work pdf}) gives
the average work for the backward protocol, which can be written as
\begin{align}
\left\langle W\right\rangle = & \sum_{l,k}\int dW\,WP_{B}\left(l,k;W\right)\nonumber \\
= & \sum_{k}p\left(k\right)\left\{ \sum_{l}\tilde{p}\left(l|k\right)\left[\tilde{\mathrm{\mathcal{U}}}\left(\tilde{\rho}_{\tau_{2}}^{\left(l,k\right)}\right)\right.\right.\nonumber \\
 & -\left.\left.\tilde{\mathrm{\mathcal{U}}}\left(\tilde{\rho}_{0}^{eq,\left(k\right)}\right)\right]\right\} .\label{eq:average backward protocol with mismatch}
\end{align}
This average work is consistent with the physical description of the
backward protocol illustrated in Fig.~\ref{fig:Backward-feedback-protocol.}.
The difference in the internal energies $\tilde{\mathrm{\mathcal{U}}}\left(\tilde{\rho}_{\tau_{2}}^{\left(l,k\right)}\right)-\tilde{\mathrm{\mathcal{U}}}\left(\tilde{\rho}_{0}^{eq,\left(k\right)}\right)$
provides the average work for a single history $l$ associated with
the backward protocol with the $k$th initial thermal state $\tilde{\rho}_{0}^{eq,\left(k\right)}$.
Then, an average over all histories $l$ is performed, for each $k$,
by the conditional probability $\tilde{p}\left(l|k\right)=\text{Tr}\left[\tilde{\mathcal{M}}_{l}\tilde{V}_{\tilde{\tau}_{1},0}^{\left(k\right)}\tilde{\rho}_{0}^{eq,\left(k\right)}\tilde{V}_{\tilde{\tau}_{1},0}^{\left(k\right)\dagger}\right]$
that the outcome $l$ occurs given that the system began in $\tilde{\rho}_{0}^{eq,\left(k\right)}$.
Another average over the initial sampling probability $p\left(k\right)$
is performed in Eq. (\ref{eq:average backward protocol with mismatch}).
This last average corresponds to an average of the $k$ distinct protocols
which begin with the corresponding Gibbs state $\tilde{\rho}_{0}^{eq,\left(k\right)}$.
Equation~(\ref{eq:average backward protocol with mismatch}) gives
the mean value for the work over all possible histories of the backward
protocol.

The backward mixed joint PDF is obtained from the backward mixed work
PDF, Eq.~(\ref{eq:backward mixed work pdf}), by introducing the
free-energy difference and mutual information density delta functions
as
\begin{align}
P_{B}\left(l,k\,;W,\Delta F,I\right)= & \sum_{n,m}\tilde{p}\left(n,l,m^{\left(k\right)},k\right)\delta\left[W-\tilde{W}_{nkm}\right]\nonumber \\
 & \times\delta\left[\Delta F-\Delta\tilde{F}^{\left(k\right)}\right]\nonumber \\
 & \times\delta\left[I-I^{\left(k,l\right)}\right].\label{eq:backward mixed joint pdf}
\end{align}

\begin{figure}[t]
\includegraphics[scale=0.5]{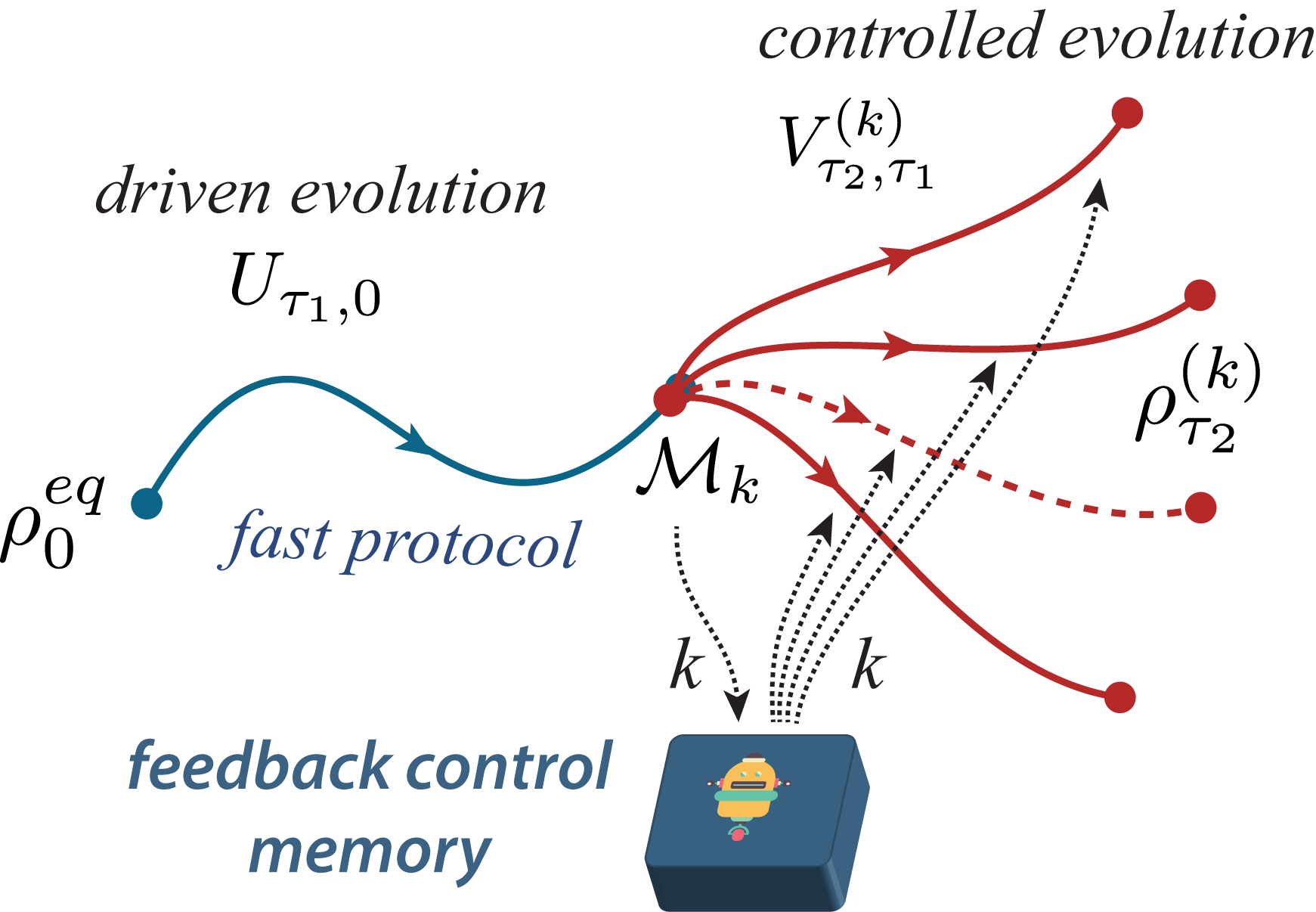}

\caption{Representation of the forward feedback control protocol without control
mismatch. The system starts in thermal equilibrium $\rho_{0}^{eq}$
and is driven by external means to the nonequilibrium state $\rho_{\tau_{1}}$.
The corresponding evolution operator is $U_{\tau_{1},0}$. At time
$\tau_{1}$ an intermediate measurement of some observable $\mathcal{M}$
is performed. The eigenprojectors of the observable are $\left\{ \mathcal{M}_{k}\right\} $
and the outcome $k$ is obtained with probability $q\left(k\right)$.
After the measurement the feedback operation $V_{\tau_{2},\tau_{1}}^{\left(k\right)}$
is implemented leading to the Hamiltonian $H_{\tau_{2}}^{\left(k\right)}$
and final state $\rho_{\tau_{2}}^{\left(k\right)}$.\label{fig:asd.forward protocol wihout mismatch}}
\end{figure}

\section{Single-History QDFR without control mismatch\label{sec:Appendix-C:-Single-History}}

\setcounter{figure}{0}
\setcounter{equation}{0}
\global\long\def\thefigure{C\arabic{figure}}
\global\long\def\theequation{C\arabic{equation}}

Here we will discuss the protocol without control mismatch in detail.
The forward feedback control protocol is very similar to the case
with control mismatch and is depicted in Fig.~\ref{fig:asd.forward protocol wihout mismatch}.
The only difference is that, without control mismatch, the feedback
operation $V_{\tau_{2},\tau_{1}}^{\left(k\right)}$ is always applied
to the corresponding outcome $k$. The system begins in the thermal
state $\rho_{0}^{eq}=e^{-\beta H_{0}}/Z_{0}$ and evolves under the
unitary evolution $U_{\tau_{1},0}=\mathcal{T}_{>}\exp\left\{ -\frac{i}{\hbar}\int_{0}^{\tau_{1}}dt\,H\left(t\right)\right\} $
up to time $\tau_{1}$ and the state evolves to $\rho_{\tau_{1}}=U_{\tau_{1},0}\rho_{0}^{eq}U_{\tau_{1},0}^{\dagger}$.
Then, a measurement of the observable $\mathcal{M}$ is performed,
whose eigenprojectors are $\left\{ \mathcal{M}_{k}\right\} $. After
the measurement, the state changes to $\rho_{\tau_{1}}^{\left(k\right)}=\mathcal{M}_{k}\rho_{\tau_{1}}\mathcal{M}_{k}/q\left(k\right)$
with probability $q\left(k\right)=\text{Tr}\left[\mathcal{M}_{k}\rho_{\tau_{1}}\right]$.
The feedback operation $V_{\tau_{2}\tau_{1}}^{\left(k\right)}=\mathcal{T}_{>}\exp\left\{ -\frac{i}{\hbar}\int_{\tau_{1}}^{\tau_{2}}dt\,H^{\left(k\right)}\left(t\right)\right\} $
leads to the final state $\rho_{\tau_{2}}^{\left(k\right)}=V_{\tau_{2}\tau_{1}}^{\left(k\right)}\rho_{\tau_{1}}^{\left(k\right)}V_{\tau_{2}\tau_{1}}^{\left(k\right)\dagger}$
and final Hamiltonian $H_{\tau_{2}}^{\left(k\right)}$ with probability
$q\left(k\right)$. 

Now we introduce the forward mixed PDF 
\begin{align}
P_{F}^{\text{wcm}}\left(k\,;W,\Delta F\right)= & \sum_{m,n}q\left(m^{\left(k\right)},k,n\right)\delta\left[W-W_{mkn}\right]\nonumber \\
 & \times\delta\left[\Delta F-\Delta F^{\left(k\right)}\right],\label{eq:single-history forward work pdf}
\end{align}
where $q\left(m^{\left(k\right)},k,n\right)=q\left(m^{\left(k\right)}|k\right)q\left(k|n\right)p\left(n\right)$,
with $q\left(m^{\left(k\right)}|k\right)=\mbox{Tr}\left[P_{m}^{\left(k\right)\tau_{2}}V_{\tau_{2},\tau_{1}}^{\left(k\right)}\mathcal{M}_{k}V_{\tau_{2},\tau_{1}}^{\left(k\right)\dagger}\right]$,
$q\left(k|n\right)=\mbox{Tr}\left[\mathcal{M}_{k}U_{\tau_{1},0}P_{n}^{0}U_{\tau_{1},0}^{\dagger}\right]$,
and $p\left(n\right)=e^{-\beta E_{n}^{0}}/Z_{0}$, is a function of
the set of labels $\left(m^{\left(k\right)},k,n\right)$ and $W_{mkn}=E_{m}^{\left(k\right)\tau_{2}}-E_{n}^{0}$.
This PDF contains the relevant information about energy fluctuations
in the protocol. By integrating over all the stochastic variables
but one, we consistently obtain the correct marginal probabilities
\begin{align}
P_{F}^{\text{wcm}}\left(k\right)= & \iint dWd\left(\Delta F\right)P_{F}^{\text{wcm}}\left(k\,;W,\Delta F\right)\nonumber \\
= & q\left(k\right),
\end{align}
which is the probability of obtaining outcome $k$ in the intermediate
measurement;
\begin{align}
P_{F}^{\text{wcm}}\left(\Delta F\right) & =\sum_{k}\int dW\,P_{F}^{\text{wcm}}\left(k\,;W,\Delta F\right)\nonumber \\
 & =\sum_{k}q\left(k\right)\delta\left[\Delta F-\Delta F^{\left(k\right)}\right],
\end{align}
which correctly reflects the fact that the free energy difference
$\Delta F^{\left(k\right)}$ occurs with probability $q\left(k\right)$;
and 
\begin{align}
P_{F}^{\text{wcm}}\left(W\right) & =\sum_{m,n}q\left(m^{\left(k\right)},k,n\right)\delta\left[W-W_{mkn}\right]\nonumber \\
 & =\sum_{k}q\left(k\right)\left[\mathrm{\mathcal{U}}\left(\rho_{\tau_{2}}^{\left(k\right)}\right)-\mathrm{\mathcal{U}}\left(\rho_{0}^{eq}\right)\right],
\end{align}
which correctly gives the average over histories of the internal energy
variation of each history. 

We use the same strategy as we did in \ref{sec:Appendix-B:-Single-History}
to connect forward and backward protocols. Taking the conditional
probabilities $q\left(m^{\left(k\right)}|k\right)$ and $q\left(k|n\right)$
and inserting identities, one obtains 
\begin{align}
q\left(m^{\left(k\right)}|k\right) & =\mbox{Tr}\left[\Theta P_{m}^{\left(k\right)\tau_{2}}\Theta^{\dagger}\Theta V_{\tau_{2},\tau_{1}}^{\left(k\right)}\Theta^{\dagger}\Theta\mathcal{M}_{k}\Theta^{\dagger}\Theta V_{\tau_{2},\tau_{1}}^{\left(k\right)\dagger}\Theta^{\dagger}\right]\nonumber \\
 & =\mbox{Tr}\left[\tilde{\mathcal{M}}_{k}\tilde{V}_{\tilde{\tau}_{1},0}^{\left(k\right)}\tilde{P}_{m}^{\left(k\right)0}\tilde{V}_{\tilde{\tau}_{1},0}^{\left(k\right)\dagger}\right]\nonumber \\
 & =\tilde{q}\left(k|m^{\left(k\right)}\right)
\end{align}
 and 
\begin{align}
q\left(k|n\right) & =\mbox{Tr}\left[\Theta\mathcal{M}_{k}\Theta^{\dagger}\Theta U_{\tau_{1},0}\Theta^{\dagger}\Theta P_{n}^{0}\Theta^{\dagger}\Theta U_{\tau_{1},0}^{\dagger}\Theta\right]\nonumber \\
 & =\mbox{Tr}\left[\tilde{P}_{n}^{\tau_{2}}\tilde{U}_{\tau_{2},\tilde{\tau}_{1}}\tilde{\mathcal{M}}_{k}\tilde{U}_{\tau_{2},\tilde{\tau}_{1}}^{\dagger}\right]\nonumber \\
 & =\tilde{q}\left(n|k\right),
\end{align}
where $\tilde{\tau}_{1}=\tau_{2}-\tau_{1}$ and $\tilde{\mathcal{M}}=\Theta\mathcal{M}\Theta^{\dagger}$
is the time-reversed observable. We also have multiplied and divided
by the probability $\tilde{p}\left(m^{\left(k\right)}\right)=\text{Tr}\left[\tilde{P}_{m}^{\left(k\right)0}\tilde{\rho}_{0}^{eq,\left(k\right)}\right]=e^{-\beta\left(\tilde{E}_{m}^{\left(k\right)0}-\tilde{F}_{0}^{\left(k\right)}\right)}$
of having the initial eigenenergies of the backward process with initial
Gibbs state $\tilde{\rho}_{0}^{eq,\left(k\right)}=e^{-\beta\tilde{H}_{0}^{\left(k\right)}}/\tilde{Z}_{0}^{\left(k\right)}$.
All together, these calculations will give
\begin{align}
q\left(m^{\left(k\right)},k,n\right) & =\tilde{q}\left(n|k\right)\tilde{q}\left(k|m^{\left(k\right)}\right)\tilde{p}\left(m^{\left(k\right)}\right)\nonumber \\
 & \times e^{-\beta\left(E_{n}^{0}-F_{0}\right)}e^{+\beta\left(\tilde{E}_{m}^{\left(k\right)0}-\tilde{F}_{0}^{\left(k\right)}\right)}.\label{eq:10}
\end{align}
We identify the product in the first line of Eq.~(\ref{eq:10}) as
$\tilde{q}\left(n,k,m^{\left(k\right)}\right)=\tilde{q}\left(n|k\right)\tilde{q}\left(k|m^{\left(k\right)}\right)\tilde{p}\left(m^{\left(k\right)}\right)$,
which will be associated to the backward protocol in a moment. The
term in the second line of Eq.~(\ref{eq:10}) can be written as $e^{+\beta\left(W_{mkn}-\Delta F^{\left(k\right)}\right)}$.
Therefore, the mixed work PDF, obtained by integrating over $\Delta F$
in Eq.~(\ref{eq:single-history forward work pdf}), can be written
as
\begin{align}
P_{F}^{\text{wcm}}\left(k\,;W\right)= & \sum_{m,n}e^{+\beta\left(W_{mkn}-\Delta F^{\left(k\right)}\right)}\tilde{q}\left(n,k,m^{\left(k\right)}\right)\nonumber \\
 & \times\delta\left[W-W_{mkn}\right].
\end{align}
Using the delta function and identifying the associated backward work
PDF, we arrive at the single-history QDFR without control mismatch
\begin{equation}
P_{F}^{\text{wcm}}\left(k\,;W\right)=e^{+\beta\left(W-\Delta F^{\left(k\right)}\right)}P_{B}^{\text{wcm}}\left(k\,;-W\right),
\end{equation}
where 
\begin{align}
P_{B}^{\text{wcm}}\left(k\,;W\right)= & \sum_{n,m}\tilde{q}\left(n,k,m^{\left(k\right)}\right)\nonumber \\
 & \times\delta\left[W-\tilde{W}_{nkm}\right].\label{eq:backward mixed work pdf without mismatch}
\end{align}

The backward work PDF $P_{B}^{\text{wcm}}\left(W\right)=\sum_{k}P_{B}^{\text{wcm}}\left(k\,;W\right)$
does not provide another backward protocol as one would have expected.
The average work of the backward work PDF, $\int dW\,WP_{B}^{\text{wcm}}\left(W\right)$,
cannot be interpreted as the average over histories of the work of
some TPM protocol. This could call into question the validity of $P_{B}^{\text{wcm}}\left(W\right)$
as a PDF. However, $P_{B}^{\text{wcm}}\left(W\right)$ is always nonnegative
and normalized, i.e., $\int dW\,P_{B}^{\text{wcm}}\left(W\right)=1$.
Based on these properties, we have chosen to still call it a PDF,
although it lacks a proper protocol associated with it in the absence
of control mismatch. 

From a pragmatic point of view, the function $P_{B}^{\text{wcm}}\left(k\,;W\right)$
provides a fluctuation relation and its Fourier transform can be experimentally
measured. Even if one choses to not call it a PDF, this quantity can
be measured and it is instrumental in our method. 

\section{Whole-Process QDFR for feedback control\label{sec:Appendix-D:-Whole-Process}}

\setcounter{figure}{0}
\setcounter{equation}{0}
\global\long\def\thefigure{D\arabic{figure}}
\global\long\def\theequation{D\arabic{equation}}

From the mixed joint PDF, Eq.~(\ref{eq:mixed joint pdf}), we obtain
the joint PDF by summing over the discrete variables $k$ and $l$.
This gives
\begin{align}
P_{F}\left(W,\Delta F,I\right) & =\sum_{k,l}P_{F}\left(k,l;W,\Delta F,I\right)\nonumber \\
 & =\sum_{m,k,l,n}p\left(m^{\left(k\right)},k,l,n\right)\delta\left[W-W_{mkn}\right]\nonumber \\
 & \times\delta\left[\Delta F-\Delta F^{\left(k\right)}\right]\delta\left[I-I^{\left(k,l\right)}\right].
\end{align}
We use the relation between forward and backward protocols and introduce
conveniently identities inside the definition of $p\left(m^{\left(k\right)},k,l,n\right)$.
This this calculation was already performed and the result was given
in Eq.~(\ref{eq:forward backward relation}). Therefore, 
\begin{align}
P_{F}\left(W,\Delta F,I\right) & =\sum_{m,k,l,n}e^{\beta\left(W_{mkn}-\Delta F^{\left(k\right)}\right)+I^{\left(k,l\right)}}\nonumber \\
 & \times\tilde{p}\left(n,l,m^{\left(k\right)},k\right)\delta\left[W-W_{mkn}\right]\nonumber \\
 & \times\delta\left[\Delta F-\Delta F^{\left(k\right)}\right]\delta\left[I-I^{\left(k,l\right)}\right].
\end{align}
Using the delta functions to remove the exponential from the summation
symbol and identifying the remainder term as the backward joint PDF,
we arrive at the QDFR
\begin{equation}
P_{F}\left(W,\Delta F,I\right)=e^{\beta\left(W-\Delta F\right)+I}P_{B}\left(-W,-\Delta F,I\right),
\end{equation}
where the three variables $W$, $\Delta F$, and $I$ are stochastic
and 
\begin{align}
P_{B}\left(W,\Delta F,I\right) & =\sum_{n,l,m,k}\tilde{p}\left(n,l,m^{\left(k\right)},k\right)\delta\left[W-\tilde{W}_{nkm}\right]\nonumber \\
 & \times\delta\left[\Delta F-\Delta\tilde{F}^{\left(k\right)}\right]\delta\left[I-I^{\left(k,l\right)}\right]
\end{align}
is the backward joint PDF, which can be obtained from Eq.~(\ref{eq:backward mixed joint pdf})
by summing over $k$ and $l$.

\section{Characteristic functions\label{sec:Appendix-E:-Characteristic}}

\setcounter{figure}{0}
\setcounter{equation}{0}
\global\long\def\thefigure{E\arabic{figure}}
\global\long\def\theequation{E\arabic{equation}}

As mentioned in the main text, the four characteristic functions relevant
for our method are: $\chi_{F}^{\left(k,l\right)}\left(u\right)$,
$\chi_{B}^{\left(l,k\right)}\left(u\right)$, $\chi_{F}^{\text{wcm}\left(k\right)}\left(u\right)$,
and $\chi_{B}^{\text{wcm}\left(k\right)}\left(u\right)$, which are
the Fourier transform of the work variable of the mixed work PDFs,
$P_{F}\left(k,l;W\right)$, $P_{B}\left(l,k;W\right)$, $P_{F}^{\text{wcm}}\left(k;W\right)$,
and $P_{B}^{\text{wcm}}\left(k;W\right)$, respectively. These calculations
are reasonably easy to perform. We explicitly wrote the result of
the forward and backward characteristic functions with control mismatch
in the main text {[}see Eqs.~(\ref{eq:forward characteristic work function single history})
and (\ref{eq:backward characteristic work function single history}){]}.
For the sake of completeness, in this appendix we write explicitly
the characteristic functions without control mismatch. The forward
and backward characteristic functions without control mismatch are
respectively given by \begin{widetext} 
\begin{align}
\chi_{F}^{\text{wcm}\left(k\right)}\left(u\right) & =\int dW\,P_{F}^{\text{wcm}}\left(k;W\right)e^{iuW}\nonumber \\
 & =\mbox{Tr}\left[e^{+iuH_{\tau_{2}}^{\left(k\right)}}V_{\tau_{2},\tau_{1}}^{\left(k\right)}\mathcal{M}_{k}U_{\tau_{1},0}e^{-iuH_{0}}\rho_{0}^{eq}U_{\tau_{1},0}^{\dagger}\mathcal{M}_{k}V_{\tau_{2},\tau_{1}}^{\left(k\right)\dagger}\right],
\end{align}
\begin{align}
\chi_{B}^{\text{wcm}\left(k\right)}\left(u\right) & =\int dW\,P_{B}^{\text{wcm}}\left(k;W\right)e^{iuW}\nonumber \\
 & =\mbox{Tr}\left[e^{+iu\tilde{H}_{\tau_{2}}}\tilde{U}_{\tau_{2},\tilde{\tau}_{1}}\tilde{\mathcal{M}}_{k}\tilde{V}_{\tilde{\tau}_{1},0}^{\left(k\right)}e^{-iu\tilde{H}_{0}^{\left(k\right)}}\tilde{\rho}_{0}^{eq,\left(k\right)}\tilde{V}_{\tilde{\tau}_{1},0}^{\left(k\right)\dagger}\tilde{\mathcal{M}}_{k}\tilde{U}_{\tau_{2},\tilde{\tau}_{1}}\right].
\end{align}
\end{widetext} 

\section{Quantum Interferometric Circuits\label{sec:Appendix-F:-Quantum}}

\setcounter{figure}{0}
\setcounter{equation}{0}
\global\long\def\thefigure{F\arabic{figure}}
\global\long\def\theequation{F\arabic{equation}}

Each interferometric circuit in the main text was designed to encode,
in an auxiliary qubit, all the information necessary to reconstruct
the characteristic function of interest, providing a full characterization
of energy fluctuations in a quantum feedback protocol. Let $\mathcal{M}$
be the observable in the feedback protocol and $\left\{ \mathcal{M}_{k}\right\} $
its eigenprojectors set. The first CNOT gate in Fig.~\ref{fig:Single-history-forward-with-error}
and the CNOT gate in Fig.~\ref{fig:single-history without error}
are not the usual CNOT gates. The control basis is not the computational
basis but the basis composed by the eigenprojectors of the observable
of the feedback protocol. The $\mathcal{M}$-CNOT gate will be defined
as
\begin{equation}
\mathcal{M}_{0}^{S}\otimes\mathds{1}^{M}+\mathcal{M}_{1}^{S}\otimes\sigma_{x}^{M}.\label{eq:S41}
\end{equation}
The target qubit $M$ is flipped when the system's state is associated
with the outcome $1$ (projector $\mathcal{M}_{1}^{S}$). As the memory
starts in the reference state $\Ket{0^{M}}$, the memory state will
remain the same if the system is in state $\mathcal{M}_{0}$ and will
flip to $\Ket{1^{M}}$ if the system is in state $\mathcal{M}_{1}$.
Therefore, this gate correlates the system and memory states; two
of the diagonal elements of the resulting state in the $\mathcal{M}^{S}\otimes\sigma_{z}^{M}$
basis will be $\mathcal{M}_{k}^{S}\otimes\Ket{k^{M}}\Bra{k^{M}}$
weighted by the corresponding probability of measuring the outcome
$k$. It is important to note that this protocol can be generalized
for a system with higher dimensions; in this case the feedback memory
system should have the same dimension as the system of interest. 

The algorithm in Fig.~\ref{fig:Single-history-forward-with-error}
is used to obtain the characteristic function $\chi_{F}^{\left(k,l\right)}\left(u\right)$.
The measurement of the two memories is necessary, i.e., one is related
to the outcome of the feedback measurement and the other is related
to the feedback operation implemented. The probability of obtaining
the outcome pair $\left(k,l\right)$ is given by 
\begin{align}
p^{M_{1}M_{2}}\left(k,l\right)= & \frac{1}{2}p\left(k|l\right)\sum_{i=1,2}\text{Tr}[B_{ki}V_{\tau_{2},\tau_{1}}^{\left(k\right)}\mathcal{M}_{l}U_{\tau_{1},0}W_{i}\nonumber \\
 & \times\rho_{0}^{eq}W_{i}^{\dagger}U_{\tau_{1},0}^{\dagger}\mathcal{M}_{l}V_{\tau_{2},\tau_{1}}^{\left(k\right)\dagger}B_{ki}^{\dagger}],
\end{align}
where $W_{0}=e^{-iuH_{0}}$, $W_{1}=\mathds{1}$, $B_{00}=B_{10}=\mathds{1}$,
$B_{k1}=e^{-iuH_{\tau_{2}}^{\left(k\right)}}$, and $p\left(k|l\right)=\left|\Bra{k}R_{x}\left(\varphi\right)\Ket{l}\right|^{2}$.

The algorithm in Fig.~\ref{fig:single-history without error} is
used to measure the characteristic function $\chi_{F}^{\text{wcm}\left(k\right)}\left(u\right)$.
In this case, the measurement of a single memory, related to the outcome
of the feedback measurement, is necessary. The probability of obtaining
the outcome $k$ is given by
\begin{align}
p^{M}\left(k\right)= & \frac{1}{2}\sum_{i=1,2}\text{Tr}[B_{ki}V_{\tau_{2},\tau_{1}}^{\left(k\right)}\mathcal{M}_{k}U_{\tau_{1},0}W_{i}\nonumber \\
 & \times\rho_{0}^{eq}W_{i}^{\dagger}U_{\tau_{1},0}^{\dagger}\mathcal{M}_{k}V_{\tau_{2},\tau_{1}}^{\left(k\right)\dagger}B_{ki}^{\dagger}],
\end{align}
where $W_{i}$ and $B_{kl}$ were defined in the previous paragraph.

The algorithm in Fig.~\ref{fig:joint probability distribution} is
used to measure the trace $\mathcal{A}\left(k,l\right)$ defined in
Sec.~\ref{sec:Tasaki-Crooks-relation-with-feedback-control} {[}see
Eq.~(\ref{eq:backward characteristic work function single history}){]}.
This trace comprised part of the characteristic function with mismatch
$\chi_{B}^{\left(l,k\right)}\left(u\right)=p\left(k\right)\mathcal{A}\left(k,l\right)$
and gives the characteristic function without mismatch $\chi_{B}^{\text{wcm}\left(k\right)}\left(u\right)=\mathcal{A}\left(k,k\right)$.
The probability of the memory measurement is given by 
\begin{align}
p_{B}^{M}\left(l\right)= & \frac{1}{2}\sum_{i=1,2}\text{Tr}[A_{i}\tilde{U}_{\tau_{2},\tilde{\tau}_{1}}\tilde{\mathcal{M}}_{l}\tilde{V}_{\tilde{\tau}_{1},0}^{\left(k\right)}W_{i}^{\left(k\right)}\nonumber \\
 & \times\tilde{\rho}_{0}^{eq,\left(k\right)}W_{i}^{\left(k\right)\dagger}\tilde{V}_{\tilde{\tau}_{1},0}^{\left(k\right)\dagger}\tilde{\mathcal{M}}_{l}\tilde{U}_{\tau_{2},\tilde{\tau}_{1}}^{\dagger}A_{i}^{\dagger}],
\end{align}
where $W_{0}^{\left(k\right)}=e^{-i\tilde{H}_{0}^{\left(k\right)}}$,
$W_{1}^{\left(k\right)}=A_{0}=\mathds{1}$, and $A_{1}=e^{-iu\tilde{H}_{\tau_{2}}}$.
After the memory measurement with an outcome $l$ the matrix element
$\Bra{0^{A}}\rho_{l}^{A}\Ket{1^{A}}$ of the ancilla state at the
end of the circuit is given by 
\begin{equation}
\Bra{0^{A}}\rho_{l}^{A}\Ket{1^{A}}=\frac{1}{2p^{M}\left(l\right)}\mathcal{A}\left(k,l\right).
\end{equation}
Therefore, we can see that if the outcome is $l=k$, then $\Bra{0^{A}}\rho_{l=k}^{A}\Ket{1^{A}}=\chi_{B}^{\text{wcm}\left(k\right)}\left(u\right)/2p_{B}^{M}\left(l=k\right)$.
The averages $\left\langle \sigma_{x}^{A}\right\rangle =\text{Re}\left[\chi_{B}^{\text{wcm}\left(k\right)}\left(u\right)\right]/p_{B}^{M}\left(l=k\right)$
and $\left\langle \sigma_{y}^{A}\right\rangle =\text{Im}\left[\chi_{B}^{\text{wcm}\left(k\right)}\left(u\right)\right]/p_{B}^{M}\left(l=k\right)$
will provide the real and imaginary parts of the work characteristic
function, respectively. For an arbitrary $l$, the averages of the
ancilla Pauli operators will be $\left\langle \sigma_{x}^{A}\right\rangle =\text{Re}\left[\mathcal{A}\left(k,l\right)\right]/p_{B}^{M}\left(l\right)$
and $\left\langle \sigma_{y}^{A}\right\rangle =\text{Im}\left[\mathcal{A}\left(k,l\right)\right]/p_{B}^{M}\left(l\right)$.
Hence, $\mathcal{A}\left(k,l\right)$ can be constructed and part
of $\chi_{B}^{\left(l,k\right)}\left(u\right)$ is obtained.

\begin{widetext}

\begin{figure}
\includegraphics[scale=0.82]{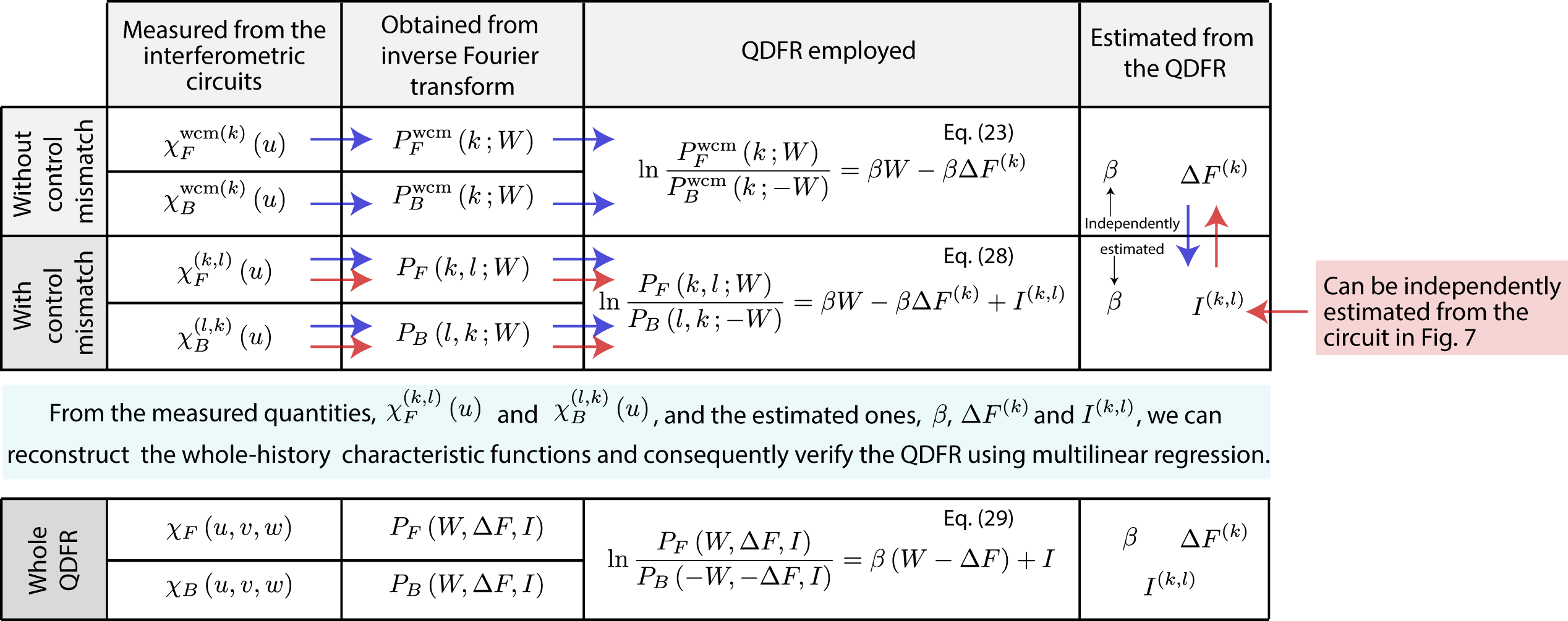}

\caption{In order to verify the QDFR {[}Eq.~(\ref{eq:tasaki-crooks whole process})
or (\ref{eq:eq29}){]} the two measured characteristic functions,
$\chi_{F}^{\left(k,l\right)}\left(u\right)$ and $\chi_{B}^{\left(l,k\right)}\left(u\right)$,
and the three estimated quantities, $\beta$, $\Delta F^{\left(k\right)}$,
and $I^{\left(k,l\right)}$, are necessary. Given the estimated quantities,
the procedure to verify the QDFR is the same: Constructing the characteristic
functions $\chi_{F}\left(u,v,w\right)$ and $\chi_{B}\left(u,v,w\right)$
from the measured data, one obtains the corresponding joint PDFs by
inverse Fourier transform and then applies multilinear regression
to the set of points corresponding to the ratio of the PDFs in logarithmic
scale. On the other hand, there are two pathways to obtain the estimated
quantities ($\beta$, $\Delta F^{\left(k\right)}$, and $I^{\left(k,l\right)}$).
These two pathways are indicated by blue and red arrows in the sketched
table. In any case, the inverse temperature $\beta$ is always estimated
independently. From the blue pathway one obtains first the free-energy
differences $\Delta F^{\left(k\right)}$ from the QDFR given by Eq.~(\ref{eq:tasaki-crooks without error})
or (\ref{eq:23}) and uses such quantities to obtain the mutual information
density from the other QDFR in Eq.~(\ref{eq:tasaki-crooks single history})
or (\ref{eq:28}). This pathway was employed in the analytical example
discussed in Sec.~\ref{subsec:Example:-Qubit-Controlled}. In the
red pathway, we obtain the mutual information density first, through
the measurement of the joint probability $p\left(k,l\right)$ employing
the circuit in Fig.~\ref{fig:joint probability distribution}. With
the joint probability $p\left(k,l\right)$, the mutual information
density can be calculated and then employed to estimate the free-energy
differences using the QDFR of Eq.~(\ref{eq:tasaki-crooks single history})
or (\ref{eq:28}). \label{fig:alternative}}
\end{figure}

\end{widetext}

For the sake of clarity, Fig.~\ref{fig:alternative} displays an
alternative way to understand the whole strategy to verify the detailed
fluctuation relation for feedback-controlled quantum systems.

\section{Details About the Analytical Example\label{sec:Appendix-G:-Details}}

\setcounter{figure}{0}
\setcounter{equation}{0}
\global\long\def\thefigure{G\arabic{figure}}
\global\long\def\theequation{G\arabic{equation}}

In this appendix we show the explicit expressions for the forward
and backward joint PDFs $P_{F}\left(W,\Delta F,I\right)$ and $P_{B}\left(W,\Delta F,I\right)$,
respectively. The forward joint characteristic function, $\chi_{F}\left(u,v,w\right)$,
is given by Eq.~(\ref{eq:forward characteristic work function single history}),
whereas the forward work characteristic function, $\chi_{F}^{\left(k,l\right)}\left(u\right)$,
for our example is given by Eq.~(\ref{eq:24}). The PDF is obtain
from the inverse Fourier transform of the respective characteristic
function

\begin{align}
P_{F}\left(W,\Delta F,I\right)= & \frac{1}{\left(2\pi\right)^{3}}\iiint_{-\infty}^{\infty}du\,dv\,dw\nonumber \\
 & \times e^{-i\left(uW+v\Delta F+wI\right)}\chi_{F}\left(u,v,w\right).\label{eq:S46}
\end{align}
Substituting $\chi_{F}^{\left(k,l\right)}\left(u\right)$ into $\chi_{F}\left(u,v,w\right)$
and evaluating the integrals, we obtain\begin{widetext}
\begin{align}
P_{F}\left(W,\Delta F,I\right)= & \sum_{k,l}\frac{1}{2}p\left(k|l\right)\frac{e^{-\beta E_{l}^{0}}}{Z_{0}}\left\{ \delta\left[W-\left(E_{1}^{\left(k,\tau_{2}\right)}-E_{l}^{0}\right)\right]+\delta\left[W-\left(E_{0}^{\left(k,\tau_{2}\right)}-E_{l}^{0}\right)\right]\right\} \nonumber \\
 & \times\delta\left[\Delta F-\Delta F^{\left(k\right)}\right]\delta\left[I-I^{\left(k,l\right)}\right].\label{eq:S47}
\end{align}
\end{widetext}Since $k$ and $l$ have two possible values, there
are eight peaks in the stochastic variable space $\left(W,\Delta F,I\right)$.
The forward work PDF, plotted in Fig.~\ref{fig:forward and backward work pdf},
can be easily obtain by integrating over the free-energy difference
and mutual information density.

The backward joint characteristic function, $\chi_{B}\left(u,v,w\right)$,
is given by Eq.~(\ref{eq:backward characteristic work function whole process}),
whereas the backward work characteristic function, $\chi_{B}^{\left(l,k\right)}\left(u\right)$,
is given by Eq.~(\ref{eq:26}). The backward joint PDF is obtained
from the inverse Fourier transform of the respective characteristic
function
\begin{align}
P_{B}\left(W,\Delta F,I\right)= & \frac{1}{\left(2\pi\right)^{3}}\iiint_{-\infty}^{\infty}du\,dv\,dw\nonumber \\
 & \times e^{-i\left(uW+v\Delta F+wI\right)}\chi_{B}\left(u,v,w\right).\label{eq:S48}
\end{align}
Substituting $\chi_{B}^{\left(l,k\right)}\left(u\right)$ into $\chi_{B}\left(u,v,w\right)$
and evaluating the integrals, we obtain\begin{widetext}
\begin{align}
P_{B}\left(W,\Delta F,I\right)= & \sum_{l,k}\frac{1}{2}p\left(k\right)\frac{e^{-\beta\tilde{E}_{0}^{\left(k\right)0}}}{\tilde{Z}_{0}^{\left(k\right)}}\left\{ \delta\left[W-\left(\tilde{E}_{l}^{\tau_{2}}-\tilde{E}_{0}^{\left(k\right)0}\right)\right]+\delta\left[W-\left(\tilde{E}_{l}^{\tau_{2}}-\tilde{E}_{1}^{\left(k\right)0}\right)\right]\right\} \nonumber \\
 & \times\delta\left[\Delta F-\Delta\tilde{F}^{\left(k\right)}\right]\delta\left[I-I^{\left(k,l\right)}\right].\label{eq:S49}
\end{align}
\end{widetext}The backward work PDF plotted in Fig.~\ref{fig:forward and backward work pdf}
can be easily obtain by integrating over the free-energy difference
and mutual information density.


\begin{thebibliography}{10}
\bibitem{Maxwell1867}J. C. Maxwell, On Governors, Proc. R. Soc. \textbf{16},
270 (1867).

\bibitem{Kang2016}C.-G. Kang, Origin of Stability Analysis: ``On
Governors'' by J.C. Maxwell {[}Historical Perspectives{]}, IEEE Control
Systems \textbf{36}, 77 (2016).

\bibitem{Bechhoefer2005}J. Bechhoefer, Feedback for physicists: A
tutorial essay on control, Rev. Mod. Phys.\textbf{ 77}, 783 (2005).

\bibitem{Jacobs2006}K. Jacobs, Applications of Feedback Control in
Quantum Systems, arXiv:quant-ph/0605015v1 (2006).

\bibitem{Zhang2014}J. Zhang, Y.-X. Liu, R.-B. Wu, K. Jacobs, and
F. Nori, Quantum feedback: theory, experiments, and applications,
Phys. Rep. \textbf{679}, 1 (2017).

\bibitem{Leff2002}H. S. Leff and A. F. Rex, \emph{Maxwell's Demon
2 Entropy, Classical and Quantum Information, Computing} (IOP Publishing,
Bristol, 2003), 2nd edition.

\bibitem{Maruyama2009}K. Maruyama, F. Nori, and V. Vedral, Colloquium:
The physics of Maxwell\textquoteright s demon and information, Rev.
Mod. Phys. \textbf{81}, 1 (2009).

\bibitem{Lutz2015}E. Lutz and S. Ciliberto, Information: From Maxwell's
demon to Landauer's eraser, Physics Today \textbf{68}, 30 (2015).

\bibitem{Jaynes1957a}E. Jaynes, Information Theory and Statistical
Mechanics, Phys. Rev. \textbf{106}, 620 (1957).

\bibitem{Jaynes1957b}E. Jaynes, Information Theory and Statistical
Mechanics II, Phys. Rev. \textbf{108}, 171 (1957).

\bibitem{Jarzynski2011}C. Jarzynski, Equalities and inequalities:
Irreversibility and the second law of thermodynamics at the nanoscale,
Annu. Rev. Condens. Matter Phys. \textbf{2}, 329 (2011).

\bibitem{Klages2013}R. Klages, W. Just and C. Jarzsynksi, \emph{Nonequilibrium
Statistical Physics of Small Systems, Fluctuation Relations and Beyond}
(Wiley-VCH, Weinheim, 2013).

\bibitem{Vinjanampathy2015}S. Vinjanampathy and J. Anders, Quantum
thermodynamics, Contemp. Phys. \textbf{57}, 545 (2016).

\bibitem{Millen2016}J. Millen and A. Xuereb, Perspective on quantum
thermodynamics, New J. Phys. \textbf{18}, 011002 (2016).

\bibitem{GelbwaserKlimovsky2015}D. Gelbwaser-Klimovsky, W. Niedenzu,
and G. Kurizki, Chapter Twelve - Thermodynamics of Quantum Systems
Under Dynamical Control, Adv. At. Mol. Opt. Phy. \textbf{64}, 329
(2015).

\bibitem{Esposito2009}M. Esposito, Nonequilibrium fluctuations, fluctuation
theorems, and counting statistics in quantum systems, Rev. Mod. Phys.
\textbf{81}, 1665 (2009).

\bibitem{Seifert2012}U. Seifert, Stochastic thermodynamics, fluctuation
theorems and molecular machines, Rep. Prog. Phys. \textbf{75}, 126001
(2012).

\bibitem{Jarzynski1997a}C. Jarzynski, Nonequilibrium Equality for
Free Energy Differences, Phys. Rev. Lett. \textbf{78}, 2690 (1997).

\bibitem{Jarzynski1997b}C. Jarzynski, Equilibrium free-energy differences
from nonequilibrium measurements: A master-equation approach, Phys.
Rev. E \textbf{56}, 5018 (1997).

\bibitem{Crooks1999}G. E. Crooks, Entropy production fluctuation
theorem and the nonequilibrium work relation for free energy differences,
Phys. Rev. E \textbf{60}, 2721 (1999).

\bibitem{Tasaki2000}H. Tasaki, Jarzynski Relations for Quantum Systems
and Some Applications, arXiv:cond-mat/0009244v2 (2000).

\bibitem{Kurchan2001}J. Kurchan, A Quantum Fluctuation Theorem, arXiv:cond-mat/0007360v2
(2001).

\bibitem{Campisi2011}M. Campisi, P. Hänggi, and P. Talkner, Colloquium:
Quantum fluctuation relations: Foundations and applications, Rev.
Mod. Phys. \textbf{83}, 771 (2011).

\bibitem{Hanggi2015}P. Hänggi and P. Talkner, The other QFT, Nat.
Phys. \textbf{11}, 108 (2015).

\bibitem{Ribeiro2016}W. L. Ribeiro, G. T. Landi, and F. L. Semião,
Quantum thermodynamics and work fluctuations with applications to
magnetic resonance, Am. J. Phys. \textbf{84}, 948 (2016).

\bibitem{Batalhao2014}T. B. Batalhão, A. M. Souza, L. Mazzola, R.
Auccaise, R. S. Sarthour, I. S. Oliveira, J. Goold, G. De Chiara,
M. Paternostro, and R. M. Serra, Experimental Reconstruction of Work
Distribution and Study of Fluctuation Relations in a Closed Quantum
System, Phys. Rev. Lett. \textbf{113}, 140601 (2014).

\bibitem{Batalhao2015}T.\LyXThinSpace B. Batalhão, A.\LyXThinSpace M.
Souza, R.\LyXThinSpace S. Sarthour, I.\LyXThinSpace S. Oliveira, M.
Paternostro, E. Lutz, and R.\LyXThinSpace M. Serra,  Irreversibility
and the Arrow of Time in a Quenched Quantum System, Phys. Rev. Lett.
\textbf{115}, 190601 (2015).

\bibitem{An2014a}S. An, J.-N. Zhang, M. Um, D. Lv, Y. Lu, J. Zhang,
Z.-Q. Yin, H. T. Quan, and K. Kim, Experimental test of the quantum
Jarzynski equality with a trapped-ion system, Nat. Phys. \textbf{11},
193 (2015).

\bibitem{Naghiloo2017}M. Naghiloo, D. Tan, P. M. Harrington, J. J.
Alonso, E. Lutz, A. Romito, and K. W. Murch, Thermodynamics along
individual trajectories of a quantum bit, arXiv:1703.05885 (2017).

\bibitem{Smith2018}A. Smith, Y. Lu, S. An, X. Zhang, J.-N. Zhang,
Z. Gong, H. T. Quan, C. Jarzynski, and K. Kim, Verification of the
quantum nonequilibrium work relation in the presence of decoherence,
New J. Phys. \textbf{20}, 013008 (2018).

\bibitem{Medeiros2018}R. Medeiros de Araújo, T. Häffner, R. Bernardi,
D. S. Tasca, M. P. J. Lavery, M. J. Padgett, A. Kanaan, L. C. Céleri,
and P. H. Souto Ribeiro, Experimental study of quantum thermodynamics
using optical vortices, J. Phys. Commun. \textbf{2}, 035012 (2018).

\bibitem{Lloyd1997}S. Lloyd, Quantum-mechanical Maxwell\textquoteright s
demon, Phys. Rev. A \textbf{56}, 3374 (1997).

\bibitem{Sagawa2012-1}T. Sagawa, \emph{Thermodynamics of Information
Processing in Small Systems} (Springer Theses, Springer, New York,
2012).

\bibitem{Parrondo2015}J. M. R. Parrondo, J. M. Horowitz, and T. Sagawa,
Thermodynamics of information, Nat. Phys. \textbf{11}, 131 (2015).

\bibitem{Chapman2015}A. Chapman and A. Miyake, How an autonomous
quantum Maxwell demon can harness correlated information, Phys. Rev.
E \textbf{92}, 062125 (2015).

\bibitem{Brandner2015}K. Brandner, M. Bauer, M. T Schmid, and U.
Seifert, Coherence-enhanced efficiency of feedback-driven quantum
engines, New J. Phys. \textbf{17}, 065006 (2015).

\bibitem{Girolami2015}D. Girolami, R. Schmidt, and G. Adesso, Towards
quantum cybernetics, Ann. Phys. \textbf{527}, 757 (2015).

\bibitem{Kammerlander2016}P. Kammerlander and J. Anders, Coherence
and measurement in quantum thermodynamics, Sci. Rep. \textbf{6}, 22174
(2016).

\bibitem{Lebedev2016}A. V. Lebedev, D. Oehri, G. B. Lesovik, and
G. Blatter, Trading coherence and entropy by a quantum Maxwell demon,
Phys. Rev. A \textbf{94}, 052133 (2016).

\bibitem{Kutvonen2016}A. Kutvonen, T. Sagawa, and T. Ala-Nissila,
Thermodynamics of information exchange between two coupled quantum
dots, Phys. Rev. E \textbf{93}, 032147 (2016).

\bibitem{Weilenmann2016}M. Weilenmann, L. Kraemer, P. Faist, and
R. Renner, Axiomatic Relation between Thermodynamic and Information-Theoretic
Entropies, Phys. Rev. Lett. \textbf{117}, 260601 (2016).

\bibitem{Elouard2016} C. Elouard, D. Herrera-Martí, M. Clusel, and
A. Auffèves, The role of quantum measurement in stochastic thermodynamics,
npj Quantum Inf. \textbf{3}, 1 (2017).

\bibitem{Strasberg2017}P. Strasberg, G. Schaller, T. Brandes, and
M. Esposito, Quantum and Information Thermodynamics: A Unifying Framework
based on Repeated Interactions, Phys. Rev. X \textbf{7}, 021003 (2017).

\bibitem{Elouard2017}C. Elouard, D. Herrera-Martí, B. Huard, and
Alexia Auffèves, Extracting work from quantum measurements in Maxwell's
demon engines, Phys. Rev. Lett. \textbf{118}, 260603 (2017).

\bibitem{Shu2017}A. Shu, J. Dai, and V. Scarani, Power of an optical
Maxwell's demon in the presence of photon-number correlations, Phys.
Rev. A \textbf{95}, 022123 (2017).

\bibitem{Sagawa2006}T. Sagawa and M. Ueda, Jarzynski Equality with
Maxwell's Demon, arXiv:cond-mat/0609085v3 (2006).

\bibitem{Sagawa2008}T. Sagawa and M. Ueda, Second Law of Thermodynamics
with Discrete Quantum Feedback Control, Phys. Rev. Lett. \textbf{100},
080403 (2008).

\bibitem{Sagawa2010}T. Sagawa and M. Ueda, Generalized Jarzynski
Equality under Nonequilibrium Feedback Control, Phys. Rev. Lett. \textbf{104},
090602 (2010).

\bibitem{Morikuni2011}Y. Morikuni and H. Tasaki, Quantum Jarzynski-Sagawa-Ueda
Relations, J. Stat. Phys. \textbf{143}, 1 (2011).

\bibitem{Sagawa2012}T. Sagawa and M. Ueda, Fluctuation Theorem with
Information Exchange: Role of Correlations in Stochastic Thermodynamics,
Phys. Rev. Lett. \textbf{109}, 180602 (2012).

\bibitem{Lahiri2012}S. Lahiri, S. Rana, and A. M. Jayannavar, Fluctuation
theorems in the presence of information gain and feedback, J. Phys.
A: Math. Theor. \textbf{45}, 065002 (2012).

\bibitem{Funo2013}K. Funo, Y. Watanabe, and M. Ueda, Integral quantum
fluctuation theorems under measurement and feedback control, Phys.
Rev. E \textbf{88}, 052121 (2013).

\bibitem{Fuchs2001}C. A. Fuchs and K. Jacobs, Information-tradeoff
relations for finite-strength quantum measurements, Phys. Rev. A \textbf{63},
062305 (2001).

\bibitem{Buscemi2008}F. Buscemi, M. Hayashi, and M. Horodecki, Global
information balance in quantum measurements, Phys. Rev. Lett. \textbf{100},
210504 (2008).

\bibitem{Buscemi2009}F. Buscemi, and M. Horodecki, Towards a Unified
Approach to Information-Disturbance Tradeoffs in Quantum Measurements,
Open Syst. Inf. Dyn. \textbf{16}, 29 (2009).

\bibitem{Berta2014}M. Berta, J. M. Renes, and M. M. Wilde, Identifying
the Information Gain of a Quantum Measurement, IEEE Trans. Inf. Theory
\textbf{60}, 7987 (2014).

\bibitem{Jacobs2014}K. Jacobs, \emph{Quantum Measurement Theory and
its Application} (Cambridge University Press, Cambridge, 2014).

\bibitem{Kosloff2013}R. Kosloff, Quantum thermodynamics: A dynamical
viewpoint, Entropy \textbf{15}, 2100 (2013).

\bibitem{Camati2016}P. A. Camati, J. P. S. Peterson, T. B. Batalhão,
K. Micadei, A. M. Souza, R. S. Sarthour, I. S. Oliveira, and R. M.
Serra, Experimental rectification of entropy production by a Maxwell's
Demon in a quantum system, Phys. Rev. Lett. \textbf{117}, 240502 (2016).

\bibitem{Cover2006}T. M. Cover, and J. A. Thomas, \emph{Elements
of Information Theory} (John Wiley \& Songs, Inc., Hoboken, 2006),
2nd edition.

\bibitem{Tajima2013}H. Tajima, Second law of information thermodynamics
with entanglement transfer, Phys. Rev. E \textbf{88}, 042143 (2013).

\bibitem{Mazzola2013}L. Mazzola, G. De Chiara, and M. Paternostro,
Measuring the Characteristic Function of the Work Distribution, Phys.
Rev. Lett. \textbf{110}, 230602 (2013).

\bibitem{Dorner2013a}R. Dorner, S. R. Clark, L. Heaney, R. Fazio,
J. Goold, and V. Vedral, Extracting Quantum Work Statistics and Fluctuation
Theorems by Single-Qubit Interferometry, Phys. Rev. Lett. \textbf{110},
230601 (2013).

\bibitem{Toyabe2010}S. Toyabe, T. Sagawa, M. Ueda, E. Muneyuki, and
M. Sano, Experimental demonstration of information-to-energy conversion
and validation of the generalized Jarzynski equality, Nat. Phys. \textbf{6},
988 (2010).

\bibitem{B=0000E9rut2012}A. Bérut, A. Arakelyan, A. Petrosyan, S.
Ciliberto, R. Dillenschneider, and E. Lutz, Experimental verification
of Landauer\textquoteright s principle linking information and thermodynamics,
Nature (London) \textbf{483}, 187 (2012).

\bibitem{Rold=0000E1n2014}E. Roldán, I. A. Martínez, J.M. R. Parrondo,
and D. Petrov, Universal features in the energetics of symmetry breaking,
Nat. Phys. \textbf{10}, 457 (2014).

\bibitem{Koski2014}J. V. Koski, V. Maisi, T. Sagawa, and J. P. Pekola,
Experimental observation of the role of mutual information in the
nonequilibrium dynamics of a Maxwell demon, Phys. Rev. Lett. \textbf{113},
030601 (2014).

\bibitem{Koski2014-1}J. V. Koski, V. F. Maisi, J. P. Pekola, and
D. V. Averin, Experimental realization of a Szilard engine with a
single electron, Proc. Natl. Acad. Sci. U.S.A. \textbf{111}, 13786
(2014).

\bibitem{Koski2015}J. V. Koski, A. Kutvonen, I. M. Khaymovich, T.
Ala-Nissila, and J. P. Pekola, On-Chip Maxwell\textquoteright s Demon
as an Information- Powered Refrigerator, Phys. Rev. Lett. \textbf{115},
260602 (2015).

\bibitem{vedral2016}M. D. Vidrighin, O. Dahlsten, M. Barbieri, M.
S. Kim, V. Vedral, and I. A. Walmsley, Photonic Maxwell\textquoteright s
Demon, Phys. Rev. Lett. \textbf{116}, 050401 (2016).

\bibitem{Goold2015}J. Goold, M. Huber, A. Riera, L. del Rio, and
P. Skrzypczyk, The role of quantum information in thermodynamics\textemdash A
topical review, J. Phys. A \textbf{49}, 143001 (2016).

\bibitem{Peterson2016}J. P. S. Peterson, R. S. Sarthour, A.M. Souza,
I. S. Oliveira, J. Goold, K. Modi, D. O. Soares-Pinto, and L. C. Céleri,
Experimental demonstration of information to energy conversion in
a quantum system at the Landauer limit, Proc. R. Soc. A \textbf{472},
20150813 (2016).

\bibitem{Mancino2017}L. Mancino, M. Sbroscia, E. Roccia, I. Gianani,
F. Somma, P. Mataloni, M. Paternostro, and M. Barbieri, Information-thermodynamics
of Quantum Generalized Measurements, arXiv:1702.07164v1 (2017).

\bibitem{Ciampini2016}M. A. Ciampini, L. Mancino, A. Orieux, C. Vigliar,
P. Mataloni, M. Paternostro, and M. Barbieri, Experimental extractable
work-based multipartite separability criteria, npj Quantum Inf. \textbf{3},
10 (2017).

\bibitem{Cottet2017}N. Cottet, S. Jezouin, L. Bretheau, P. Campagne-Ibarcq,
Q. Ficheux, J. Anders, A. Auffèves, R. Azouit, P. Rouchon, and B.
Huard, Observing a quantum Maxwell demon at work, PNAS \textbf{114},
7561 (2017).

\bibitem{Masuyama2017}Y. Masuyama, K. Funo, Y. Murashita, A. Noguchi,
S. Kono, Y. Tabuchi, R. Yamazaki, M. Ueda, and Y. Nakamura, Information-to-work
conversion by Maxwell\textquoteright s demon in a superconducting
circuit quantum electrodynamical system, Nat. Comm. \textbf{9}, 1291
(2018).

\bibitem{Naghiloo2018}M. Naghiloo, J. J. Alonso, A. Romito, E. Lutz,
and K. W. Murch, Information gain and loss for a quantum Maxwell's
demon, arXiv:1802.07205 (2018).

\bibitem{Gong2015}Z. Gong and H. T. Quan, Jarzynski equality, Crooks
fluctuation theorem, and the fluctuation theorems of heat for arbitrary
initial states, Phys. Rev. E \textbf{92}, 012131 (2015).

\bibitem{Rana2012}S. Rana, S. Lahiri, and A M Jayannavar, Quantum
Jarzynski equality with multiple measurement and feedback for isolated
system, Pramana \textbf{79}, 233 (2012).

\bibitem{Liu2012}F. Liu, Derivation of quantum work equalities using
a quantum Feynman-Kac formula, Phys. Rev. E \textbf{86}, 010103(R)
(2012).

\bibitem{Liu2014}F. Liu, Equivalence of two Bochkov-Kuzovlev equalities
in quantum two-level systems, Phys. Rev. E \textbf{89}, 042122 (2014).

\bibitem{Wigner1959}E. P. Wigner, \emph{Group Theory and its Application
to the Quantum Mechanics of Atomic Spectra}, translation from German
by J. J. Griffin. (Academic Press Inc., New York, 1959). 

\bibitem{Messiah1962}A. Messiah, \emph{Quantum Mechanics} (North
Holland Publishing Company, Amsterdam, 1962), Vol 2.

\bibitem{Ballentine2000}L. E. Ballentine, \emph{Quantum Mechanics
A Modern Development}, (World Scientific Publishing Co. Pte. Ltd.,
Singapore, 2000) reprint.

\bibitem{Haake2010}F. Haake, \emph{Quantum Signatures of Chao}s,
(Springer, Heidelberg, 2010) 3rd edition.

\bibitem{Talkner2007}P. Talkner, E. Lutz, and P. Hänggi, Fluctuation
theorems: Work is not an observable, Phys. Rev. E \textbf{75}, 050102
(2007).

\bibitem{Talkner2007-1}P. Talkner and P. Hänggi , The Tasaki\textendash Crooks
quantum fluctuation theorem, J. Phys. A: Math. Theor. \textbf{40},
F569 (2007).

\bibitem{Andrieux2008}D. Andrieux and P. Gaspard, Quantum Work Relations
and Response Theory, Phys. Rev. Lett. \textbf{100}, 230404 (2008).

\bibitem{Campisi2010-1}M. Campisi, P. Talkner, and P. Hänggi, Quantum
Bochkov\textendash Kuzovlev work fluctuation theorems, Philos. Trans.
Royal Soc. A \textbf{369}, 291 (2010).

\bibitem{Venkatesh2014}B P. Venkatesh, G. Watanabe, and P. Talkner,
Transient quantum fluctuation theorems and generalized measurements,
New J. Phys. \textbf{16}, 015032 (2014).

\bibitem{Watanabe2014}G. Watanabe, B. P. Venkatesh, P. Talkner, M.
Campisi, and P. Hänggi, Quantum fluctuation theorems and generalized
measurements during the force protocol, Phys. Rev. E \textbf{89},
032114 (2014).

\bibitem{Campisi2011-1}M. Campisi, P. Talkner, and P. Hänggi, Influence
of measurements on the statistics of work performed on a quantum system,
Phys. Rev. E \textbf{83}, 041114 (2011).

\bibitem{Campisi2010}M. Campisi, P. Talkner, and P. Hänggi, Fluctuation
Theorems for Continuously Monitored Quantum Fluxes, Phys. Rev. Lett.
\textbf{105}, 140601 (2010).

\bibitem{Horowitz2010}J. M. Horowitz and S. Vaikuntanathan, Nonequilibrium
detailed fluctuation theorem for repeated discrete feedback, Phys.
Rev. E \textbf{82}, 061120 (2010).

\bibitem{Kaonan-2017}K. Micadei, J. P. S. Peterson, A. M. Souza,
R. S. Sarthour, I. S. Oliveira, G. T. Landi, T. B. Batalhão, R. M.
Serra, and E. Lutz, Reversing the thermodynamic arrow of time using
quantum correlations, arXiv:1711.03323 (2017).

\bibitem{Peterson2018}J. P. S. Peterson, T. B. Batalhão, M. Herrera,
A. M. Souza, R. S. Sarthour, I. S. Oliveira, and R. M. Serra, Experimental
characterization of a spin quantum heat engine, arXiv:1803.06021 (2018).
\end{thebibliography}
\end{document}